\documentclass[onecolumn,aps,prd,preprintnumbers,showpacs,superscriptaddress,nofootinbib,amsmath,amssymb,floats,floatfix,showkeys,notitlepage]{revtex4-1}

\usepackage{hyperref}
\usepackage{natbib}
\usepackage{orcidlink}
\usepackage{palatino}
\usepackage{hyperref}
\hypersetup{
  colorlinks=true,        
  linkcolor=blue,         
  citecolor=blue,         
}

\newcommand{\fe}{\end{equation}}
\newcommand{\se}{\begin{eqnarray}}
\newcommand{\ff}{\end{eqnarray}}

\usepackage{graphicx}
\usepackage{dcolumn}
\usepackage{bm}
\usepackage{enumitem}
\usepackage{amsmath}
\usepackage{amssymb}

\newcommand{\beq}{\begin{equation}}
\newcommand{\eeq}{\end{equation}}
\newcommand{\bea}{\begin{eqnarray}}
\newcommand{\eea}{\end{eqnarray}}

\usepackage{bbm}
\usepackage{amsfonts}
\usepackage{mathrsfs}
\usepackage{latexsym}
\usepackage{epsfig}
\usepackage{epstopdf}
\usepackage{graphicx}
\usepackage{subcaption}
\usepackage{ragged2e}
\usepackage{bm}
\usepackage{color}
\usepackage{comment}

\newcommand{\bbe}{\boldsymbol{\mathrm{e}}}

\newcommand{\bomega}{\boldsymbol{\omega}}

\newcommand{\bR}{\boldsymbol{R}}

\newcommand{\bdiff}{\boldsymbol{\mathrm{d}}}
\newcommand{\bDiff}{\boldsymbol{\mathrm{D}}}

\newcommand{\ph}{\mathrm{ph}}

\begin{document}
\title{Shedding Light on Gravity: Black Hole Shadows and Lensing Signatures in Lorentz Gauge Theory }

\author{Ali \"Ovg\"un \orcidlink{0000-0002-9889-342X}}
\email{ali.ovgun@emu.edu.tr}
\affiliation{Physics Department, Eastern Mediterranean University, Famagusta, 99628 North
Cyprus via Mersin 10, Turkiye.}

\author{Mohsen Fathi
\orcidlink{0000-0002-1602-0722}}
\email{mohsen.fathi@ucentral.cl}
\affiliation{Centro de Investigaci\'{o}n en Ciencias del Espacio y F\'{i}sica Te\'{o}rica, Universidad Central de Chile, La Serena 1710164, Chile.}

\begin{abstract}

Recent advances, including gravitational wave detections and imaging of black hole shadows, have strongly validated general relativity. Nevertheless, ongoing cosmological observations suggest potential limitations of general relativity, spurring interest in modified theories of gravity. This work investigates the Lorentz-gauge formulation of gravity-a novel framework that addresses key conceptual challenges in quantum gravity and cosmology by leveraging the recent black‐hole solutions presented in [10.1103/PhysRevD.111.064008]
\cite{Koivisto:2024asr}. By analyzing black hole shadow structures and gravitational lensing effects—both weak and strong deflection regimes—we highlight unique observational signatures of Lorentz gauge gravity. Our findings provide valuable tools for future observational tests, potentially distinguishing these modified gravity models from general relativity and advancing our understanding of spacetime geometry and fundamental gravitational interactions.

\end{abstract}

\maketitle


\section{Introduction}
General relativity (GR), Einstein's revolutionary theory, interprets gravity as a manifestation of curved spacetime, dynamically described by a metric tensor. Among its most extraordinary predictions is the existence of black holes—regions in spacetime where gravity is so intense that nothing, not even light, can escape \cite{wald1984,mtn1973,Virbhadra:1999nm,Virbhadra:2002ju,Claudel:2000yi,Adler:2022qtb}. The validity and precision of GR have been spectacularly confirmed through recent landmark observations, including gravitational wave detections by LIGO and Virgo collaborations \cite{abbott2016,abbott2019} and high-resolution imaging of black hole shadows by the Event Horizon Telescope (EHT) collaboration \cite{EventHorizonTelescope:2019dse,EventHorizonTelescope:2019ggy,EventHorizonTelescope:2021bee,Akiyama:2022}.

{
The recent advent of very‐long‐baseline interferometry (VLBI) at millimeter and sub‐millimeter wavelengths has opened an unprecedented window into the strong‐field regime of gravity near supermassive black holes.  Early proposals to image the shadow of SgrA* using space–ground VLBI were pioneered by Zakharov et. al, who demonstrated that RADIOASTRON could in principle measure black‐hole parameters from the Galactic center shadow \cite{2005NewA...10..479Z}.  Subsequent theoretical and numerical studies have refined this concept, showing how the shadow can be reconstructed and used as a precision test of general relativity (GR) and its alternatives \cite{Zakharov:2023yjl}.  In parallel, monitoring of stellar orbits—most notably the relativistic motion of the S2 star-provides complementary tests of gravity in the same region \cite{Bambhaniya:2022xbz}. On the theoretical side, analytic expressions for the shadow radius in charged spacetimes have been derived for both Reissner-Nordström and tidal‐charge metrics.  Direct measurements of black‐hole charge parameters were first explored in \cite{Zakharov:2005ek}, and later generalized to include higher‐dimensional and dilatonic extensions \cite{ZAKHAROV201264,Zakharov:2014lqa}.  More recent work has exploited the EHT observations to constrain both electric and tidal charges in M87\* and SgrA* shadows \cite{Zakharov:2021gbg,Zakharov:2022gwk}.  These studies underline the power of shadow measurements to place stringent bounds on deviations from the Kerr geometry. Despite the close relation between photon spheres and shadow boundaries in Kerr and Reissner–Nordström spacetimes, it has been shown that the existence of circular photon orbits alone does not guarantee an observable shadow in more general spherically symmetric metrics \cite{Zakharov:2025cnq,Zakharov:2025mbx}.  This distinction highlights the necessity of a careful analysis of the causal structure and lensing properties within any modified‐gravity framework. Finally, comprehensive reviews of the Galactic Centre as a laboratory for gravity and dark‐matter theories \cite{deLaurentis:2022oqa} and of the current status and future prospects of black‐hole imaging experiments \cite{Genzel:2024vou} provide essential context for ongoing and forthcoming observations.  Recent EHT images have been used to place stringent observational bounds on a variety of modified gravity theories, excluding compact objects in baseline mimetic gravity \cite{khodadi_event_2024}, constraining standard‐model extension coefficients in Lorentz‐violating frameworks \cite{khodadi_probing_2022}, and testing Einstein‐Æther gravity against the shadow of M87* \cite{khodadi_einstein-aether_2021}.  In this work, we build upon these theoretical foundations and observational breakthroughs to examine the shadows and lensing signatures arising in Lorentz Gauge Theory (LGT).
}

Nevertheless, current cosmological observations revealing accelerated cosmic expansion and notable tensions in cosmological parameters have motivated significant interest in exploring modifications to the standard metric dynamics \cite{clifton2012}. Such theoretical developments encompass alternative geometric frameworks and formulations of gravity, moving beyond the traditional metric-based description \cite{sotiriou2010}. A critical conceptual challenge for classical GR is its inherent reliance on a non-degenerate metric; notably, the theory lacks a natural zero-metric "ground state," a deficiency linked to significant theoretical difficulties, particularly its non-renormalizability in quantum formulations \cite{weinberg1979}.

In addressing these fundamental issues, alternative formulations of gravity, such as Plebański's chiral description, provide compelling insights \cite{plebanski1977}. Plebański's framework expresses Einstein’s equations solely through chiral components of the Lorentz group, employing a triple of two-forms and a chiral connection, thus offering a metric-independent initial description of gravity. By imposing suitable conditions, this approach reconstructs spacetime metrics dynamically, underpinning prominent quantum gravity proposals \cite{rovelli2004,ashtekar2004}.

Expanding upon these conceptual advancements, the LGT emerges by gauging the Lorentz invariance of special relativistic parameterized fields, identifying a canonical clock field fundamental in defining spacetime geometry \cite{zlosnik2017,zlosnik2006,Zlosnik:2018qvg,Koivisto:2022uvd,Nikjoo:2023flm}. This innovative approach inherently resolves several longstanding conceptual problems of time in quantum gravity, embedding the metric construction within the chiral Lorentz connection itself. The resulting theory elegantly integrates matter fields, naturally including spinor fields and Yang-Mills interactions, without extraneous constraints \cite{alexandrov2013}. Recent Hamiltonian analyses have identified that only the purely chiral curvature action extends effectively to GR, whereas the action incorporating total curvature represents a topological theory devoid of local degrees of freedom \cite{smolin2004}. This intriguing result points towards promising pathways for quantum gravity formulations.

{The primary motivation for investigating LGT stems from its attempt to address several foundational problems in gravitational theory from a "pregeometric" standpoint. Unlike GR, where the metric is a fundamental field postulated a priori, LGT begins with more basic entities: a gauge connection for the Lorentz group and a scalar field. The spacetime metric, and thus the entire geometric framework of GR, is not fundamental but rather an emergent, composite structure that arises in a symmetry-broken phase of the theory.

This approach offers several distinct advantages and potential solutions to long-standing issues:  \textbf{Addressing the Metric's Origin and Non-Renormalisability:} By treating spacetime geometry as emergent, LGT directly confronts the question of why a metric exists at all. The theory possesses a "topological phase" where no local gravitational degrees of freedom propagate. This phase is described by a topological field theory which is expected to be more manageable at the quantum level. GR, with its dynamical degrees of freedom, is then proposed to emerge from this simpler, topological foundation via a mechanism of symmetry breaking \cite{Mielke:2017nwt}. This provides a conceptual pathway towards a quantum theory of gravity that sidesteps the problems associated with the non-renormalisability of the metric in standard quantisation approaches. \textbf{A Dynamical Solution to the "Problem of Time":} The scalar field $\phi^a$ in the fundamental representation of the Lorentz group is not an arbitrary addition. As originally proposed in \cite{Zlosnik:2018qvg}, this field can be identified with a canonical "clock," which we term the \textit{khronon} in its time-like phase. Its gradient, $\bbe^a = \bDiff\phi^a$, defines the emergent coframe. This mechanism dynamically introduces a physical reference frame into the theory. Consequently, it provides a natural, built-in resolution to the "problem of time" that plagues canonical quantum gravity, without needing to postulate external reference matter fields like the Brown-Kuchar dust \cite{Brown:1994py}. In our framework, such a dust-like energy component emerges naturally from the theory's structure. \textbf{Natural Coupling to Matter:} The first-order, gauge-theoretic nature of LGT allows for a remarkably natural and consistent coupling to matter fields. Spinor fields and other Yang-Mills gauge fields can be incorporated into the pregeometric framework in a way that is often more straightforward than in other formulations of GR or its modifications \cite{Gallagher:2023ghl}. \textbf{Connection to Yang-Mills and Topological Field Theories:} The action for LGT can be formulated as a "curvature-squared" theory, drawing a direct and powerful analogy with the successful Yang-Mills theories of the Standard Model. This structural similarity, combined with its origin as a chiral projection of a topological field theory \cite{Nikjoo:2023flm}, places LGT on a solid theoretical foundation and opens promising avenues for its quantization.

Therefore, Lorentz Gauge Gravity is not merely another alternative to GR, but a comprehensive framework that aims to reconstruct gravity from more fundamental gauge principles. It offers potential resolutions to deep conceptual problems while maintaining a close structural relationship with other well-understood field theories. This motivates our detailed investigation into its classical solutions, such as the black hole spacetimes explored in this work.}

Given these compelling theoretical motivations, further detailed studies of classical solutions are vital. Gravitational lensing phenomena, particularly weak and strong deflection angles, provide powerful observational tests for these gravity theories \cite{weinberg1972,Virbhadra:1999nm}. The weak deflection angle regime, typically involving minor bending of light around massive objects, serves as an important testbed for GR and its alternatives \cite{keeton2005}. Conversely, the strong deflection angle regime, encountered in proximity to compact massive objects like black holes, amplifies gravitational lensing effects, offering critical opportunities to detect subtle deviations from standard GR predictions \cite{Bozza:2002zj,tsukamoto2012}. Gravitational lensing effects around black holes have been extensively studied, incorporating corrections from topological charges, holonomy, loop quantum gravity, and polymerized models, as explored in various theoretical frameworks \cite{Soares:2024rhp, Soares:2023uup, Soares:2023err, Soares:2025hpy, KumarWalia:2022ddq}.

Additionally, the observational characterization of black hole shadows, as impressively demonstrated by the EHT's imaging of supermassive black holes M87* and Sgr A* \cite{EventHorizonTelescope:2019dse,Akiyama:2022}, has opened a novel observational window into spacetime structure and gravity theories. These shadows result from photon spheres—regions where gravitational fields compel photons into unstable circular orbits, producing a distinctive dark silhouette against surrounding luminous emission \cite{cunha2018,cardoso2019}. Precision measurements of shadow size and shape thus serve as sensitive probes of gravity theories, potentially revealing deviations indicative of new physics \cite{bambi2009,johannsen2010}. Observations from the EHT, notably imaging the shadow of supermassive black holes M87* and Sgr A*, have stimulated considerable research interest in testing gravity theories, investigating black holes' scalar hair, and probing the no-hair theorem \cite{Vagnozzi:2019apd,Cardoso:2016ryw,Khodadi:2020jij,Khodadi:2021gbc,Afrin:2021imp,johannsen2010}. Additionally, recent analytical studies have extensively examined how environmental factors like plasma distribution influence black hole shadows and gravitational lensing, yielding critical implications for astrophysical observations \cite{Perlick:2015vta,Bisnovatyi-Kogan:2010flt,Bisnovatyi-Kogan:2018vxl,perlick_calculating_2022}. Recent studies have explored the optical properties and shadows of various black hole models, revealing distinct holographic and shadow images influenced by factors such as Lorentz symmetry breaking, rotation, and surrounding matter \cite{Zeng:2024ptv, Yang:2024nin, He:2024amh, He:2025rjq, Li:2020drn, Hu:2020usx, Guo:2020zmf}.

Modified gravity theories, higher-dimensional models, and nonlinear electrodynamics have also been extensively explored through black hole shadow observations, constraining theoretical parameters effectively \cite{Allahyari:2019jqz,Belhaj:2020okh,khodadi_einstein-aether_2021,Okyay:2021nnh,Ovgun:2021ttv,Kuang:2022ojj,khodadi_probing_2022,Pantig:2022gih,Pulice:2023dqw,khodadi_event_2024}. Investigations into exotic objects such as naked singularities and rotating traversable wormholes further highlight the diverse phenomenological consequences visible in gravitational lensing and shadow imaging \cite{Shaikh:2018lcc,Joshi:2020tlq,Gyulchev:2018fmd}.

Additionally, studies involving shadows of rotating non-Kerr black holes, Einstein-Gauss-Bonnet gravity models, and various scalar-tensor theories continue to refine our understanding of black hole spacetime geometry and potential deviations from General Relativity \cite{Atamurotov:2013sca,Zeng:2020dco,Cunha:2020azh,Cunha:2016wzk,Wei:2018xks,Jafarzade:2020ova,Anacleto:2021qoe}. Furthermore, recent insights into quasinormal modes and greybody factors provide additional tools for distinguishing theoretical predictions from observational data, crucial for future gravitational wave and electromagnetic observational campaigns \cite{Gogoi:2023lvw,AraujoFilho:2024lsi,Wei:2015dua,Konoplya:2019sns}.

Motivated by these theoretical insights and observational advances, this paper focuses on examining black hole solutions, their shadow structures, and gravitational deflection angles within the LGT framework. Such analyses may illuminate distinctive signatures that differentiate modified theories of gravity from classical GR, thereby contributing significantly to the quest for a deeper understanding of gravitational phenomena.

\section{Black holes in LGT}\label{Sec:2}

{This section outlines a theoretical model based on \textbf{Lorentz gauge theory} \cite{Koivisto:2024asr} . The fundamental components of this theory are a dimensionless scalar field, denoted as $\phi^a$, which exists in the fundamental representation of the Lorentz group, and the canonical connection 1-form, $\bomega^a{}_b$, in the adjoint representation. This connection facilitates the definition of a covariant exterior derivative, $\bDiff$. The curvature of this connection is expressed as \cite{Koivisto:2024asr} $\bR^a{}_b = \bdiff\bomega^a{}_b + \bomega^a{}_c\wedge\bomega^c{}_b$. The entire framework is established on an SO(1,3) bundle over a four-dimensional manifold, $M$.

The action of the theory is constructed as an integral of a polynomial 4-form built from the available fields. When considering only the connection, the most general action is a topological field theory with two parameters, $g_1$ and $g_2$:
\begin{equation}
    I_{(0)} = \int_M \left( g_1\eta_{ab}\eta^{cd} + g_2\epsilon_{ab}{}^{cd} \right) \bR^a{}_c\wedge\bR^b{}_d
\end{equation}
Here, $\eta_{ab}$ and $\epsilon_{ab}{}^{cd}$ represent the invariants of the Lorentz algebra. This action, $I_{(0)}$, describes a \textbf{topological field theory}, which lacks local degrees of freedom in the bulk manifold. Such theories are mathematically significant and offer a potential pathway to quantum gravity, where a theory like General Relativity (GR) could emerge through a process of symmetry breaking.

To introduce dynamics, the scalar field $\phi^a$ is incorporated. The construction of the action is guided by a crucial principle: a \textbf{global shift symmetry}. This symmetry dictates that the theory should only depend on the covariant changes in $\phi^a$, rendering its absolute value physically meaningless. A transformation of the form $\phi^a \rightarrow \phi^a + \xi^a$ (where $\bDiff\xi^a=0$) must leave the action unchanged. Under this constraint, three additional independent 4-forms can be added to the action:
\begin{equation}
    I_{(4)} = \lambda\int_M \epsilon_{abcd}\bDiff\phi^a\wedge\bDiff\phi^b\wedge\bDiff\phi^c\wedge\bDiff\phi^d
\end{equation}
and
\begin{equation}
    I_{(2)} = g_3\int_M \bDiff\phi^a\wedge\bDiff\phi^b\wedge\bR_{ab} + g_4\int_M \epsilon_{abcd}\bDiff\phi^a\wedge\bDiff\phi^b\wedge\bR^{cd}
\end{equation}
If the shift symmetry is not held as exact, and the action is permitted to change by a boundary term, additional terms ($\tilde{I}_{(2)}$ and $\tilde{I}_{(4)}$) can be considered. These terms are equivalent to $I_{(2)}$ and $I_{(4)}$ up to boundary terms and suggest that the theory's dynamics can be viewed as emerging from the topological theory through a deformation of the inner product.

Further analysis reveals that the combined action $I=I_{(0)} + I_{(2)} + I_{(4)}$ remains devoid of local dynamics under the specific condition that the coupling constant \textbf{$g_3=0$}. This is due to an enhanced symmetry in the theory.

To obtain a desirable extension of General Relativity, this symmetry must be broken. This is achieved by selecting specific values for the couplings $g_3$ and $g_4$ that correspond to a \textbf{maximally chiral projection} of the curvature, denoted as $\bR^+_{ab}$. The resulting action is:
\begin{equation}
    I_+ = \int_M \bDiff\phi^a\wedge\bDiff\phi^b\wedge\bR^+_{ab}
\end{equation}
The explicit form of this chiral projection, $\bR^{+a}{}_b = \frac{1}{2}\left( \delta^a_c\delta^d_b - \frac{i}{2}\epsilon^{ad}{}_{bc}\right)\bR^{c}{}_d$, necessitates a complexification of the Lorentz group. While this review adheres to the conventional metric signature, it is noted that a Euclidean formulation might offer a more fundamental perspective, particularly concerning the symmetries of the Standard Model and insights from twistor geometry.

In conclusion, the full theory, described by the action $I=I_{(0)}+I_{(2)}+I_{(4)}$, is governed by five dimensionless coupling constants. The theory transitions from being topologically trivial to dynamically active by adjusting the parameter $g_3$. By setting $g_4=-i/4$ and treating $g_3 = a/2$ as an interpolating parameter, the quadratic part of the action becomes:
\begin{equation}
    I_{(2)} = \frac{1}{2}\int_M \bDiff\phi_a\wedge\bDiff\phi^b\wedge\left( a\delta^a_c\delta^d_b - \frac{i}{2}\epsilon^{ad}{}_{bc}\right)\bR^{c}{}_d
\end{equation}
Here, the parameter $a$ (termed the "dymaxion") can be conceptualized as a field that drives the transition between phases: $a=0$ corresponds to the topological phase, while $a=1$ represents the fully dynamical, extended chiral GR phase. This framework provides a mechanism for the emergence of a dynamical spacetime from an underlying topological structure. For the remainder of this work, the theory is analyzed in its dynamical phase where $a=1$.
}

In this framework, the dynamical fields include a scalar field \(\phi^\alpha\), which transforms under the fundamental representation of the Lorentz group, and a connection 1‐form \(\omega^a{}_b\) in the adjoint representation. The connection naturally introduces the covariant exterior derivative \(D\), and its associated curvature 2‐form is defined as $R^a{}_b = d\omega^a{}_b + \omega^a{}_c \wedge \omega^c{}_b.$ This expression encapsulates the nontrivial geometry of the SO(1,3) principal bundle over the 4-dimensional manifold \(M\) \cite{2023IJGMM..2050040K,Koivisto:2024asr,2019CQGra..36f5015W}. In essence, the curvature \(R^a{}_b\) measures the failure of the connection to be locally pure gauge, and it plays a central role in the gravitational dynamics of the theory. {Within LGT}, static and spherically symmetric spacetimes are characterized by metric functions \(f(r)\) and \(g(r)\), which are determined by solving the field equations. An interesting feature of these equations is the appearance of an extra parameter \(A_0\), which originates from the structure of the connection and reflects the choice of the observer’s frame of reference. This parameter is crucial since it adjusts the scale of both the temporal and spatial parts of the metric. For vacuum solutions, the field equations simplify considerably, leading to explicit forms for the metric functions \cite{Koivisto:2024asr} 

\begin{eqnarray}
ds^2 = -\left(\frac{1}{A_0^2}-\frac{m_S}{4\pi m_P^2\, r}\right) dt^2 + \left(\frac{1}{A_0^2}-\frac{m_S}{4\pi m_P^2\, r}\right)^{-1} dr^2 + r^2\left(d\theta^2+\sin^2\theta\, d\phi^2\right),
  \label{eq:metr_1}
\end{eqnarray}
where \(m_P\) and \(m_S\) are mass parameters, and \(A_0\) is the connection coefficient. This metric clearly illustrates how the parameter \(A_0\) not only scales the time component but also rescales the radial part of the geometry. In effect, the connection coefficient \(A_0\) plays a dual role, modulating the gravitational redshift and the spatial curvature simultaneously. A particularly illuminating case arises when one sets \(A_0=\pm1\). In this situation, the metric reduces to the familiar Schwarzschild solution, thereby recovering the standard event horizon at \(r=2M\) (after redefining the combination \(m_S/(4\pi m_P^2)\) as \(2M\), with \(M\) representing the mass of the black hole). This equivalence underscores that \(A_0\) acts as a scaling parameter: deviations from \(|A_0|=1\) yield modified spacetimes. Specifically, for \(A_0>1\), the effective radius of the event horizon increases, and the lapse function \(f(r)\) decreases, suggesting that the black hole appears “inflated” compared to its Schwarzschild counterpart. Conversely, if \(A_0<1\), the event horizon contracts, potentially leading to radical changes in the black hole’s structure and, in extreme cases, to the complete disappearance of the horizon. The equations not only define the geometric structure of the theory but also highlight the significant role played by the connection parameter \(A_0\) in determining the physical properties of static, spherically symmetric spacetimes. The interplay between the gravitational potential (via \(f(r)\)) and the scaling parameter \(A_0\) provides deeper insights into how modifications in the connection can lead to deviations from classical solutions, such as the Schwarzschild metric. The work in \cite{Abdelqader:2014vaa} introduces a set of curvature scalars designed to identify key geometric features of black hole spacetimes—such as the event horizon, the ergosurface—and to provide measures for physical properties like mass and spin \cite{Abdelqader:2014vaa,Tavlayan:2020chf}. These scalars are constructed from the Weyl tensor \(C_{\mu \nu \alpha \beta}\) and its left dual \({}^*C_{\mu \nu \alpha \beta}\), as well as their covariant derivatives. The definitions are as follows:
\begin{equation}
\begin{array}{ll}
I_{1}=C_{\mu \nu \alpha \beta} C^{\mu \nu \alpha \beta}, & I_{2}={}^* C_{\mu \nu \alpha \beta} C^{\mu \nu \alpha \beta},\\[1mm]
I_{3}=\nabla_{\rho} C_{\mu \nu \alpha \beta} \nabla^{\rho} C^{\mu \nu \alpha \beta}, & I_{4}=\nabla_{\rho} C_{\mu \nu \alpha \beta} \nabla^{\rho *} C^{\mu \nu \alpha \beta},\\[1mm]
I_{5}=k_{\mu} k^{\mu}, \quad I_{6}=l_{\mu} l^{\mu}, & I_{7}=k_{\mu} l^{\mu}.
\end{array}
\end{equation}
Here, the covectors \(k_\mu\) and \(l_\mu\) are defined via the gradients of the first two invariants:
 \begin{equation}
k_{\mu}=-\nabla_{\mu} I_{1}, \quad l_{\mu}=-\nabla_{\mu} I_{2}.
\end{equation}
These invariants are powerful because, as shown in \cite{Tavlayan:2020chf}, one may locate the event horizon of Schwarzschild-like black holes by finding where \(I_3\) vanishes. In these spacetimes, \(I_3\) is positive outside the horizon, becomes zero precisely at the horizon, and is negative inside. To illustrate these ideas in a simpler setting, consider the induced metric on a two-dimensional hypersurface with coordinates \((t, r)\). In the induced coordinates, the metric takes the diagonal form
 \begin{equation}
 \gamma_{i j}=\left(\begin{array}{cc}
 f(r) & 0 \\
0 & \frac{1}{ f(r)} \label{induced}
 \end{array}\right).
 \end{equation}
where \(f(r)\) is the lapse function. This metric form is particularly useful when analyzing horizon properties since it retains the essential radial dependence. An important quantity computed from the induced metric is the Kretschmann invariant, which for this two-dimensional geometry is given by \cite{Pantig:2022sjb}
 \begin{equation}
 K={ }^{\Sigma} I_{1}={ }^{\Sigma} R_{i j k l}{ }^{\Sigma} R^{i j k l}= \left( {\frac {{\rm d}^{2}}{{\rm d}{r}^{2}}} f(r) \right) ^{2}.
 \end{equation}
This invariant provides a measure of the curvature intrinsic to the hypersurface. To further probe the horizon, one can define a horizon-detecting invariant based on the gradient of this curvature invariant:
\begin{equation}
{}^{\Sigma}I_5 = \nabla_m {}^{\Sigma}I_1\, \nabla^m {}^{\Sigma}I_1 = 4\,\left(\frac{d^2}{dr^2} f(r)\right)^2\left(\frac{d^3}{dr^3}f(r)\right)^2 f(r).
 \end{equation}
The prescription is that the largest real root of \({}^{\Sigma}I_5\) (i.e. where \({}^{\Sigma}I_5(r_+)=0\)) identifies the event horizon. Similarly, one can consider
 \begin{equation}
{}^{\Sigma}I_3 = \nabla_m {}^{\Sigma}R_{ijkl}\,\nabla^m {}^{\Sigma}R^{ijkl} = \left(\frac{d^3}{dr^3} f(r)\right)^2 f(r),
 \end{equation}
which also vanishes at the horizon. These expressions underline how derivatives of the lapse function encode critical geometric transitions associated with the horizon. Next, by inserting the lapse function from Eq. (\ref{eq:metr_1}) into the induced metric Eq. (\ref{induced}), we can compute explicit forms for these invariants. The Kretschmann invariant associated with the induced metric becomes
\begin{equation}
K = {}^{\Sigma}I_1 = \frac{m_S^2}{4\pi^2 m_P^4\,r^6}.
 \end{equation}
Similarly, the horizon-detecting invariants are found to be
 \begin{equation}
{}^{\Sigma}I_5 = \frac{9 m_S^4}{4\pi^4 m_P^8\,r^{14}}\left(\frac{1}{A_0^2}-\frac{m_S}{4\pi m_P^2\,r}\right),
 \end{equation}
 \begin{equation}
{}^{\Sigma}I_3 = \frac{9 m_S^2}{4\pi^2 m_P^4\,r^8}\left(\frac{1}{A_0^2}-\frac{m_S}{4\pi m_P^2\,r}\right).
 \end{equation}
The common factor \(\left(\frac{1}{A_0^2}-\frac{m_S}{4\pi m_P^2\,r}\right)\) in these expressions is key, as it directly determines the horizon position. Setting either \({}^{\Sigma}I_5\) or \({}^{\Sigma}I_3\) to zero yields the event horizon radius
\begin{equation}
r_h = \frac{A_0^2\,m_S}{4\pi m_P^2}.
 \end{equation}
Thus, the invariants derived from the induced metric not only detect the presence of the event horizon but also give a clear algebraic relation for its location in terms of the parameters \(A_0\), \(m_S\), and \(m_P\). Finally, the Kretschmann scalar for the full four-dimensional spacetime is given by
\begin{equation}
R^{\alpha\beta\gamma\delta}R_{\alpha\beta\gamma\delta} = \frac{m_S^2}{2\pi^2 m_P^4\,r^6} + \frac{4}{r^4}\left[1-\frac{1}{A_0^2}+\frac{m_S}{4\pi m_P^2\,r}\right]^2.
 \end{equation}
Overall, these derivations underscore how curvature invariants—constructed from the Weyl tensor and its derivatives—serve as robust tools for detecting horizons and exploring the geometrical properties of black hole spacetimes. They provide an alternative method to the traditional approach of locating horizons by solving \(g^{rr}=0\), and offer additional insights into the interplay between the spacetime geometry and the parameters that define the gravitational field. {The solution is not gauge-equivalent to the standard Schwarzschild solution. It represents a physically distinct spacetime. Two metrics are gauge-equivalent (i.e., describe the same physical spacetime) if one can be transformed into the other by a coordinate transformation. A definitive way to test this is to compare their scalar curvature invariants, such as the Kretschmann scalar ($K = R^{\alpha\beta\gamma\delta} R_{\alpha\beta\gamma\delta}$), which are independent of the coordinate system. If the invariants are different, the spacetimes are physically distinct. By inspection, $K_{\text{proposed}} \neq K_{\text{Schw}}$ unless $A_0 = 1$. If $A_0 \neq 1$, the second term in $K_{\text{proposed}}$ is non-zero and has a different radial dependence, making it impossible to match $K_{\text{Schw}}$ for all $r$. The main points are as follows:
One may try 
  \begin{eqnarray}
      \tilde t = \frac{t}{A_0}, 
      \quad
      \tilde r = \frac{r}{A_0},
      \quad
      \tilde m_S = \frac{m_S}{A_0}\,,
  \end{eqnarray}
    under which the \(t\!-\!r\) sector becomes
    \(\;-d\tilde t^2+\bigl(1-\tfrac{\tilde m_S}{4\pi m_P^2\tilde r}\bigr)^{-1}d\tilde r^2\).  
    However, the angular part transforms as
\begin{eqnarray}
      r^2\,d\Omega^2
      = A_0^2\,\tilde r^2\,d\Omega^2
      \;\neq\;\tilde r^2\,d\Omega^2
      \quad(\text{unless }A_0^2=1),
  \end{eqnarray}
    showing that in \((t,r,\theta,\varphi)\) coordinates one cannot eliminate \(A_0\) without distorting the two‐sphere. A direct computation gives for metric~(1):
\begin{eqnarray}
      R \;=\;\frac{2\,(1-A_0^2)}{r^2},
      \quad
      R_{\rm Schw}=0.
  \end{eqnarray}
    They coincide only if \(A_0^2=1\), confirming that metric~(1) is not the usual Schwarzschild form in these coordinates.
Since a physical invariant differs, no coordinate transformation can map one metric to the other. Therefore, the solution is not merely a rescaled version of Schwarzschild; it describes a different gravitational field with measurably different curvature.}


\section{Calculation of Hawking Radiation in Jacobi Metric formalism}

\begin{figure}[htp]
\centering
\includegraphics[width=8cm]{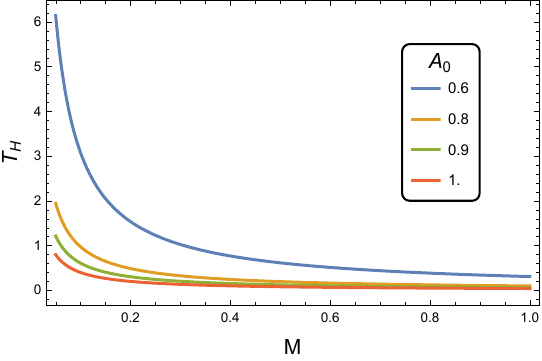}
\caption{Hawking temperature plotted as a function of the mass parameter $m_s$ for various values of $A_0 = 0.5,\,1,\,2,\,3$. Here, we have set $m_p^2 = 1$.}
\label{fig:hawking-temp}
\end{figure}

Recent observational and theoretical advancements have significantly enriched our understanding of black hole physics, particularly concerning intriguing features such as black hole shadows, photon spheres, and gravitational lensing. Hawking radiation, conceptualized via tunneling mechanisms, has provided profound insights into quantum aspects of black holes \cite{Akhmedov:2006pg,
Parikh:1999mf,Shankaranarayanan:2000qv,Angheben:2005rm}. 
To compute the Hawking temperature of a black hole, we adopt a semi-classical tunneling approach that utilizes the Jacobi metric derived from the full four-dimensional covariant metric \cite{Bera:2019oxg}. In this framework, the tunneling probability for a particle crossing the event horizon is related to the imaginary part of the classical action \cite{Akhmedov:2006pg,Kerner:2007rr,Kerner:2006vu,Kerner:2008qv,Yale:2008kx}. We begin by considering the particle’s wave function in the WKB approximation, written as
\begin{equation}\label{eq:psi}
\psi = e^{\frac{i}{\hbar}S},
\end{equation}
where S is the classical action. The key idea is that the tunneling probability is dominated by the exponential of the imaginary part of S. The corresponding Jacobi metric, which encapsulates the effective geometry experienced by the particle, is given by \cite{Gibbons:2015qja,Chanda:2016aph,Das:2016opi}:
\begin{eqnarray}\label{eq:JacobiMetric}
\mathrm{d}s^2 = j_{ij}\mathrm{d}x^{i}\mathrm{d}x^{j} =\Big(E^2 - m^2 f(r)\Big) \left(\frac{\mathrm{d}r^2}{f^2(r)} + \frac{r^2}{f(r)}\Big(\mathrm{d}\theta^2 + \sin^2\theta \mathrm{d}\phi^2\Big)\right).
\end{eqnarray}
Here, $M$ and $m$ are the energy and mass of the tunneling particle, respectively, and $f(r)$ is the lapse function that encodes the gravitational potential. This metric effectively redefines the “distance” in configuration space in such a way that the kinetic term of the particle’s motion is modulated by the combination $E^2 - m^2 f(r)$. In the context of Hawking radiation, the tunneling process occurs predominantly along the radial direction near the horizon. Thus, the action for the particle undergoing radial tunneling can be written as
\begin{equation}\label{eq:Action}
S = -\int \sqrt{j_{ij}\frac{\mathrm{d}x^i}{\mathrm{d}s}\frac{\mathrm{d}x^j}{\mathrm{d}s}}\mathrm{d}s~.
\end{equation}
The integrand represents the proper “speed” of the particle in the effective geometry. Explicitly, for the radial component, this becomes
\begin{equation}\label{eq:Integrand}
\sqrt{j_{ij}\frac{\mathrm{d}x^i}{\mathrm{d}s}\frac{\mathrm{d}x^j}{\mathrm{d}s}} = \pm \left(E^2 - m^2 f(r)\right)^{\frac{1}{2}} f^{-1}(r) \frac{\mathrm{d}r}{\mathrm{d}s}.
\end{equation}
The choice of the $\pm$ sign distinguishes between incoming and outgoing trajectories. From Eq. \eqref{eq:Integrand} and the definition of the action in Eq. \eqref{eq:Action}, the radial momentum of the particle is obtained as \cite{Bera:2019oxg}
\begin{equation}\label{eq:RadialMomentum}
p_r = \partial_r S = \mp \left(E^2 - m^2 f(r)\right)^{\frac{1}{2}} f^{-1}(r).
\end{equation}
Since the tunneling process is confined to the near-horizon region, the condition $f(r) < 0$ is imposed. In our expression for $p_r$, the sign ensures that the outgoing particle (with $p_r>0$) corresponds to the appropriate tunneling direction, in line with conventional treatments such as those found in \cite{Srinivasan:1998ty}.
Near the horizon, the full four-dimensional metric simplifies to an effective (1+1)-dimensional form because the angular components become subdominant \cite{Robinson:2005pd,Iso:2006wa,Majhi:2010onr}. To capture the physics near the horizon, we expand the lapse function $f(r)$ in a Taylor series about the horizon $r = r_H$:
\begin{eqnarray}\label{eq:TaylorExpansion}
f(r) = f(r_H) + f’(r_H)(r - r_H) + \mathcal{O}((r - r_H)^2) \notag \\ \approx 2\tilde{\kappa} (r - r_H) + \mathcal{O}((r - r_H)^2),
\end{eqnarray}
where we have used the fact that $f(r_H)=0$ and defined the surface gravity as
\begin{equation}\label{eq:SurfaceGravity}
\tilde{\kappa} = \frac{1}{2}f’(r_H).
\end{equation}
Substituting the expansion \eqref{eq:TaylorExpansion} into Eq. \eqref{eq:Integrand}, the near-horizon radial action becomes \cite{Bera:2019oxg}	
\begin{eqnarray}\label{eq:NearHorizonAction}
S = \mp \frac{E}{2\tilde{\kappa}} \int_{r_H-\epsilon}^{r_H+\epsilon} \frac{\mathrm{d}r}{(r - r_H)} \pm \frac{m^2}{2E}\int_{r_H-\epsilon}^{r_H+\epsilon} \mathrm{d}r \mp \int_{r_H-\epsilon}^{r_H+\epsilon} \mathcal{O}(r - r_H)\mathrm{d}r,
\end{eqnarray}
where $\epsilon$ is a small positive parameter that spans a region across the horizon. In the first integral, we perform a change of variable $r - r_H = \epsilon e^{i\theta}$ to evaluate the contour integral. This yields
\begin{equation}\label{eq:ContourIntegral}
\int_{r_H-\epsilon}^{r_H+\epsilon} \frac{\mathrm{d}r}{(r - r_H)} = - i\pi.
\end{equation}
The second integral vanishes in the limit $\epsilon \to 0$, and the contributions from higher-order terms (third integral) are negligible. Thus, the dominant contribution to the action is
\begin{equation}\label{eq:ActionFinal}
S = \pm \frac{i\pi E}{2\tilde{\kappa}} + \textrm{(real part)}.
\end{equation}
Here, the positive sign corresponds to the outgoing trajectory, while the negative sign corresponds to the incoming trajectory. The WKB wave functions for the outgoing and incoming particles can now be written as
\begin{equation}\label{eq:WKBOut}
\psi_{\rm out} = A e^{\frac{i}{\hbar}S_{\rm out}}, \quad \psi_{\rm in} = A e^{\frac{i}{\hbar}S_{\rm in}},
\end{equation}
with A being a normalization constant. The associated probabilities are then
\begin{equation}\label{eq:ProbabilityOut}
P_{\rm out} = |\psi_{\rm out}|^2 = |A|^2 e^{-\frac{\pi E}{\hbar \tilde{\kappa}}},
\end{equation}
and
\begin{equation}\label{eq:ProbabilityIn}
P_{\rm in} = |\psi_{\rm in}|^2 = |A|^2 e^{\frac{\pi E}{\hbar \tilde{\kappa}}}.
\end{equation}
Since the real part of the action does not contribute to the probability, the tunneling rate is determined by the ratio
\begin{equation}\label{eq:TunnelingRate}
\Gamma = \frac{P_{\rm out}}{P_{\rm in}} = e^{-\frac{2\pi E}{\hbar \tilde{\kappa}}} \equiv e^{-\frac{E}{T_H}}.
\end{equation}
Comparing Eq. \eqref{eq:TunnelingRate} with the Boltzmann factor, we identify the Hawking temperature as
\begin{equation}\label{eq:HawkingTemp}
T_H = \frac{\tilde{\kappa}}{2\pi}.
\end{equation}
To further clarify, the surface gravity $\tilde{\kappa}$ can also be expressed in terms of the norm of the timelike Killing vector field $\chi^{\mu}$ as
\begin{equation}\label{eq:SurfaceGravityKilling}
\tilde{\kappa} = \left.\sqrt{-\frac{1}{2}\nabla_{\mu}\chi_{\nu}\nabla^{\mu}\chi^{\nu}} \right|_{r=r_H} = \left.\frac{1}{2}\frac{\partial f(r)}{\partial r}\right|_{r=r_H}.
\end{equation}
In our model, using the explicit form of the lapse function, the Hawking temperature can be further determined as
\begin{equation}\label{eq:TemperatureFinal}
T_H = \frac{m_p^2}{A_0^4 m_s},
\end{equation}
which encapsulates the dependence on the parameters $m_p$, $m_s$, and the scale parameter $A_0$. {By considering the Schwarzschild mass, $M = m_s / (8\pi m_p^2)$, and substitute $m_s = 8\pi M m_p^2$ into Eq. \eqref{eq:TemperatureFinal}}
\begin{equation}
T_H \;=\;\frac{1}{8\,\pi\,M\,A_0^4}\,. 
\end{equation}
Figure~\ref{fig:hawking-temp} illustrates how the Hawking temperature $T_H$ varies with the black hole mass parameter $M$ for different values of $A_0$. Figure~\ref{fig:hawking-temp} shows that as $M$ increases, the temperature rapidly decreases. Additionally, smaller values of $A_0$ result in higher temperatures.
In summary, this semi-classical tunneling method based on the Jacobi metric provides a consistent framework to derive the Hawking temperature. The detailed derivation—from the form of the Jacobi metric through the evaluation of the tunneling action and the resulting probability—shows that the Hawking temperature is intimately linked to the surface gravity of the black hole. 

\section{Weak Deflection Angle Using the Gauss-Bonnet Theorem}

In this section, we analyze the weak gravitational lensing phenomena caused by the black hole using the Gauss–Bonnet theorem \cite{Gibbons:2008rj}. This approach involves studying the deflection of photons traveling in the spacetime geometry induced by the black hole. To start, we consider the optical metric projected onto the equatorial plane defined by \( \theta = \pi /2 \). This optical geometry is represented by the metric:
\begin{equation}\label{eq:optical_metric-}
\mathrm{d}t^{2}=\frac{1}{f(r)^2}\mathrm{d}r^2+\frac{r^2}{f(r)}\mathrm{d}\varphi^2.
\end{equation}
The Gaussian optical curvature associated with this metric, which measures the intrinsic curvature affecting photon trajectories, is calculated using the Ricci scalar R as a relation \(\mathcal{K} = R/2\).  A straightforward calculation yields
{\begin{equation}\mathcal{K}=\frac{1}{2} \left(f(r) f''(r)-\frac{1}{2} f'(r)^2\right) \end{equation} where primes denote \(d/dr\).}
Explicitly, Gaussian curvature becomes:
\begin{equation}\label{eq:gaussian_curvature}
\mathcal{K} = \frac{3 m_s^2}{64 \pi^2 m_p^4 r^4} - \frac{m_s}{4 \pi A_0^2 m_p^2 r^3}.
\end{equation}
To derive the weak deflection angle using the Gauss–Bonnet theorem, we first define a non-singular region \(\mathcal{D}_{R}\) bounded by the curve \(\partial\mathcal{D}_{R}=\gamma_{\tilde{g}}\cup C_{R}\). Applying the Gauss–Bonnet theorem yields \cite{Gibbons:2008rj}:
\begin{equation}\label{eq:gauss_bonnet}
\iint\limits_{\mathcal{D}_{R}}\mathcal{K}\,\mathrm{d}S+\oint\limits_{\partial \mathcal{D}_{R}}\kappa \,\mathrm{d}t+\sum_{i}\theta_{i}=2\pi \chi(\mathcal{D}_{R}),
\end{equation}
where \(\kappa\) represents the geodesic curvature of the boundary, \(\theta_{i}\) is the exterior angle at the \(i^{th}\) vertex, and \(\chi(\mathcal{D}_{R})=1\) denotes the Euler characteristic for a simply connected domain.
We choose our region outside the photon trajectory, and since the photon path \(\gamma_{\tilde{g}}\) is geodesic, its geodesic curvature vanishes, i.e., \(\kappa(\gamma_{\tilde{g}})=0\). For the boundary at infinity, represented by the curve \(C_R:=r(\varphi)=R=\text{constant}\), the geodesic curvature is computed via:
\begin{equation}\label{eq:geodesic_curvature}
\kappa(C_{R})=|\nabla_{\dot{C}_{R}}\dot{C}_{R}|.
\end{equation}
Expanding this expression further, we have the radial component:
\begin{equation}\label{eq:radial_part}
(\nabla_{\dot{C}_{R}}\dot{C}_{R})^{r}=\dot{C}_{R}^{\varphi}(\partial_{\varphi}\dot{C}_{R}^{r})+\tilde{\Gamma}_{\varphi\varphi}^{r}(\dot{C}_{R}^{\varphi})^{2}.
\end{equation}
The first term in Eq.~\eqref{eq:radial_part} vanishes due to the constancy of \(R\). Evaluating the second term using the unit-speed condition, we obtain:
\begin{equation}\label{eq:kappa_limit}
\lim_{R\rightarrow \infty}\kappa(C_{R})=\frac{1}{R}.
\end{equation}
At large radial distances, we further approximate:
\begin{equation}\label{eq:large_R_limit}
\lim_{R\rightarrow \infty}\mathrm{d}t \approx R \,\mathrm{d}\varphi,
\end{equation}
thus combining Eqs.~\eqref{eq:kappa_limit} and \eqref{eq:large_R_limit}, we have \(\kappa(C_{R})\mathrm{d}t=\mathrm{d}\varphi\).
Using the straight-line approximation for photon trajectories, the radial coordinate is related to the impact parameter \(b\) as \(r=b/\sin\varphi\). Consequently, the Gauss–Bonnet theorem simplifies considerably, providing a straightforward integral representation for the deflection angle \cite{Gibbons:2008rj}:
\begin{equation}\label{eq:deflection_integral}
\alpha=-\int_{0}^{\pi}\int_{\frac{b}{\sin \varphi}}^{\infty}\mathcal{K}\,\mathrm{d}S,
\end{equation}
with the surface element given explicitly as:
\begin{equation}\label{eq:surface_element}
\mathrm{d}S \approx \left(\frac{3 A_0^5 m_s}{8 \pi m_p^2}+A_0^3 r\right)\mathrm{d}r\,\mathrm{d}\varphi.
\end{equation}
Evaluating the integral \eqref{eq:deflection_integral} using the Gaussian curvature from Eq.~\eqref{eq:gaussian_curvature}, we find the weak deflection angle, including second-order contributions:
\begin{equation}\label{eq:weak_deflection}
\alpha \approx \frac{3 A_0^3 m_s^2}{256 \pi b^2 m_p^4}+\frac{A_0 m_s}{2 \pi b m_p^2}.
\end{equation}

{Using the definition of the Schwarzschild mass, $M = m_s / (8\pi m_p^2)$, we substitute $m_s = 8\pi M m_p^2$ into Eq. \eqref{eq:weak_deflection}. This yields the deflection angle in the desired form:
\begin{equation}
\alpha \approx \frac{4 A_0 M}{b} + \frac{3\pi A_0^3 M^2}{4 b^2}
\label{eq:compact_derived}
\end{equation}
\begin{figure}[htp]
    \centering
    \includegraphics[width=8cm]{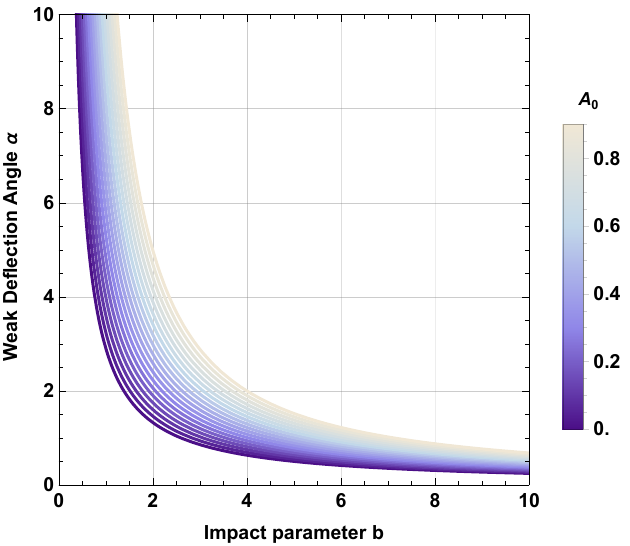}
    \caption{Weak deflection angle plotted as a function of radial distance parameter \( r \) for various values of the scale parameter \( A_0 \). The Planck mass squared is fixed at \( M = 1 \).}
    \label{fig:wda}
\end{figure}
Figure~\ref{fig:wda} illustrates how the weak deflection angle depends on the  impact parameter \( b \) for different values of the parameter \( A_0 \). We constrain the parameter $A_0$ by equating the theoretical prediction in Eq. \eqref{eq:compact_derived} with high-precision observational data from light deflection at the solar limb. The observed angle, $\theta$, is related to the General Relativistic prediction, $\delta_{\rm GR} = 4M_{\odot}/R_{\odot} \approx 1.7516687''$, by the PPN parameter $\gamma$:
\begin{equation}
\theta = \frac{1+\gamma}{2} \delta_{\rm GR}
\end{equation}
The accepted value for $\gamma$ is $0.99992 \pm 0.00012$  \cite{Lambert:2011fk}. Setting $\alpha = \theta$ for a light ray grazing the Sun ($M=M_{\odot}$, $b=R_{\odot}$), we have:
\begin{equation}
A_0 \delta_{\rm GR} + \frac{3\pi A_0^3}{16} \delta_{\rm GR}^2 = \frac{1+\gamma}{2} \delta_{\rm GR}
\end{equation}
Given that $\delta_{\rm GR} \approx 8.492 \times 10^{-6}$ rad, the term proportional to $\delta_{\rm GR}^2$ is negligible. The equation simplifies to $A_0 \approx (1+\gamma)/2$. Using the central value of $\gamma$, we find $A_0 \approx 0.99996$. The uncertainty is dominated by that of $\gamma$, giving $\Delta A_0 \approx \Delta\gamma/2 = 0.00006$.

By re-expressing the deflection angle in terms of the Schwarzschild mass and comparing it with observational data, we derive a stringent constraint on the model parameter $A_0$:
\begin{equation}
A_0 = 0.99996 \pm 0.00006
\end{equation}
This result shows that $A_0$ is consistent with unity, as predicted by General Relativity, to a high degree of precision.}

\subsection{Derivation of deflection angle using geodesics method}

We now independently verify the deflection angle by solving the geodesic equations directly. Introducing the impact parameter \(b \equiv L/E\), we write
\begin{equation}
\dot{t} = \frac{E}{f(r)}, \quad \dot{\phi} = \frac{L}{r^2}\,,
\end{equation}
which follow from symmetries of the spacetime and corresponding constants of motion \(E\) and \(L\).
Substituting these expressions into the null geodesic condition (\(\mathcal{L}=0\)), we have
\begin{equation}
-f(r)\left(\frac{E}{f(r)}\right)^2 + \frac{\dot{r}^2}{f(r)} + r^2\left(\frac{L}{r^2}\right)^2 = 0\,,
\end{equation}
which simplifies to
\begin{equation}
\dot{r}^2 = E^2 - \frac{L^2}{r^2}\,f(r)\,.
\end{equation}
The above equation is the radial geodesic equation, expressing the radial velocity in terms of the energy and angular momentum constants of motion.
Introducing the inverse radial coordinate \( u \equiv 1/r \), we note
\begin{equation}
\frac{dr}{d\phi} = -\frac{1}{u^2}\frac{du}{d\phi}\,, \quad \dot{\phi} = L u^2\,,
\end{equation}
thus giving
\begin{equation}
\dot{r} = \frac{dr}{d\phi}\dot{\phi} = - L \frac{du}{d\phi}\,.
\end{equation}
Substituting this into the radial equation yields
\begin{equation}
L^2\left(\frac{du}{d\phi}\right)^2 = E^2 - L^2 u^2\,f\left(\frac{1}{u}\right)\,.
\end{equation}
Dividing by \(L^2\) and using the impact parameter \(b\), we find
\begin{equation}
\left(\frac{du}{d\phi}\right)^2 = \frac{1}{b^2} - u^2\,f\left(\frac{1}{u}\right)\,.
\end{equation}
Expanding the metric function \(f(1/u)\), we have explicitly
\begin{equation}
f\left(\frac{1}{u}\right) = \frac{1}{A_0^2} - \frac{m_S}{4\pi m_P^2}\, u\,,
\end{equation}
thus yielding the orbital equation
\begin{equation}
\left(\frac{du}{d\phi}\right)^2 = \frac{1}{b^2} - \frac{u^2}{A_0^2} + \frac{m_S}{4\pi m_P^2} u^3\,.
\label{eq:du_dphi}
\end{equation}
This equation explicitly relates the inverse radius \(u\) and the azimuthal angle \(\phi\).
Differentiating Eq.~(\ref{eq:du_dphi}) again gives the second-order orbital equation:
\begin{equation}
\frac{d^2 u}{d\phi^2} + \frac{u}{A_0^2} = \frac{3 m_S}{8\pi m_P^2} u^2\,,
\label{eq:orbit}
\end{equation}
which governs the photon trajectory around the massive source. To solve perturbatively, let \(u(\phi)=u_0(\phi)+\delta u(\phi)\), where \(u_0\) solves the unperturbed equation (with \(m_S=0\)) and \(\delta u\) is a small correction. \paragraph{Unperturbed solution:} For \(m_S=0\), Eq.~(\ref{eq:orbit}) reduces to a harmonic oscillator-type equation
\begin{equation}
\frac{d^2u_0}{d\phi^2} + \frac{u_0}{A_0^2}=0\,,
\end{equation}
whose physically relevant solution is
\begin{equation}
u_0(\phi)=\frac{1}{b}\sin\left(\frac{\phi}{A_0}\right)\,.
\end{equation}
This solution describes a straight trajectory (no deflection), and the closest approach is at \(\phi=\pi A_0/2\).

\paragraph{First-order correction:} The perturbation satisfies
\begin{equation}
\frac{d^2\,\delta u}{d\phi^2}+\frac{\delta u}{A_0^2}=\frac{3 m_S}{8\pi m_P^2}u_0^2(\phi)\,.
\end{equation}
Expanding \(u_0^2(\phi)\) explicitly, we have
\begin{equation}
u_0^2(\phi)=\frac{1}{2b^2}\left[1-\cos\left(\frac{2\phi}{A_0}\right)\right]\,,
\end{equation}
thus yielding
\begin{equation}
\frac{d^2\,\delta u}{d\phi^2}+\frac{\delta u}{A_0^2}=\frac{3 m_S}{16\pi m_P^2 b^2}\left[1-\cos\left(\frac{2\phi}{A_0}\right)\right]\,.
\end{equation}
We try a particular solution of the form
\begin{equation}
\delta u_p(\phi)=C+D\cos\left(\frac{2\phi}{A_0}\right)\,.
\end{equation}
Substituting and matching coefficients yields
\begin{equation}
C=\frac{3 m_S A_0^2}{16\pi m_P^2 b^2}\,,\quad D=\frac{m_S A_0^2}{16\pi m_P^2 b^2}\,.
\end{equation}
Thus, the particular solution is
\begin{equation}
\delta u_p(\phi)=\frac{3 m_S A_0^2}{16\pi m_P^2 b^2}+\frac{m_S A_0^2}{16\pi m_P^2 b^2}\cos\left(\frac{2\phi}{A_0}\right)\,.
\end{equation}
Adding suitable homogeneous solutions and applying physical boundary conditions (that the orbit is asymptotically straight), we obtain the full corrected trajectory. Finally, by examining the asymptotic limit \(u\to 0\), the total deflection angle \(\alpha\) can be extracted. Performing this carefully (as in standard methods from the literature~\cite{dolan2023einstein}), the final deflection angle up to first order in perturbation is obtained:
\begin{equation}
\alpha\approx \frac{3 A_0^3 m_S^2}{256\pi b^2 m_P^4}+\frac{A_0 m_S}{2\pi b m_P^2}\,.
\label{eq:deflection}
\end{equation}
The first term represents a second-order correction, while the second term is the leading-order deflection angle, reproducing known weak-field lensing results.


\section{Shadow of the Black Hole}
\label{sec:shadow}

The photon sphere around a black hole is defined by unstable circular photon orbits that are significantly influenced by the gravitational field. Photons in these orbits either spiral into the black hole's event horizon or escape to infinity, forming a luminous photon ring. This ring delineates the observable boundary of the black hole shadow for distant observers \cite{Synge:1966,Cunningham:1972,Bardeen:1973a,Luminet:1979nyg}. Historically, different terminologies such as \textit{escape cone} \cite{Synge:1966}, \textit{cone of gravitational radiation capture} \cite{zeldovich_relativistic_1966}, \textit{optical appearance of black holes}, and \textit{black hole image} \cite{Bardeen:1973b,Chandrasekhar:1998,Luminet:1979nyg,luminet_seeing_2018} have been utilized. However, the contemporary and widely adopted term \textit{black hole shadow}, introduced by Falcke et al. \cite{Falcke_2000}, refers explicitly to the dark region surrounded by the photon ring \cite{johannsen2010,johnson_universal_2020}. In this section, we investigate the properties of the black hole shadow, focusing particularly on its dependence on the underlying model parameters, such as \( A_0 \), \( m_s \), and \( m_p \). Given the spherical symmetry of the spacetime considered, the analysis is simplified by restricting null geodesic trajectories to the equatorial plane (\( \theta = \pi/2 \)). The metric in this case reduces to:
\begin{equation}
\mathrm{d}s^2=-A(r)\mathrm{d}t^2+B(r)\mathrm{d}r^2+C(r)\mathrm{d}\phi^2,
\label{eq:metric_0}
\end{equation}
where \( B(r)=A(r)^{-1} \) and \( C(r)=r^2 \). The geodesic motion of photons can be described through the Lagrangian:
\begin{equation}
2\mathcal{L} = -A(r)\dot{t}^2 + B(r)\dot{r}^2 + C(r)\dot{\phi}^2,
\label{eq:Lag}
\end{equation}
where the dot indicates differentiation with respect to an affine parameter \(\lambda\). Constants of motion, namely the energy \(E\) and angular momentum \(L\), arise from symmetry: $
E = A(r)\dot{t}, \quad L = C(r)\dot{\phi}.$ Using these constants, the critical impact parameter \(b\), which quantifies the photon trajectory, is defined as:
\begin{equation}
b = \frac{L}{E} = \frac{C(r)}{A(r)}\frac{\mathrm{d}\phi}{\mathrm{d}t}.
\label{eq:Def_b}
\end{equation}
Applying the null condition \(ds^2=0\), the radial equation for photon trajectories becomes:
\begin{equation}
\left(\frac{\mathrm{d}r}{\mathrm{d}\phi}\right)^2 = \frac{C(r)}{B(r)}\left[\frac{h(r)^2}{b^2}-1\right],
\label{eq:drdphi}
\end{equation}
where
\begin{equation}
h(r)^2 = \frac{C(r)}{A(r)}.
\label{eq:h(r)}
\end{equation}
The photon sphere is determined by the condition \(\frac{\mathrm{d}}{\mathrm{d}r}h^2(r)=0\), explicitly:
\begin{equation}
C'(r_{\mathrm{ph}})A(r_{\mathrm{ph}})-C(r_{\mathrm{ph}})A'(r_{\mathrm{ph}})=0,
\label{eq:photonspherecomplete}
\end{equation}
yielding the photon sphere radius:
\begin{equation}
r_{\mathrm{ph}} = \frac{3 A_0^2 m_S}{8 \pi m_P^2}.
\label{eq:rph_explicit}
\end{equation}
The black hole shadow radius \(r_{\mathrm{sh}}\), as seen by an observer at radial coordinate \(r_O\), is:
\begin{equation}
r_{\mathrm{sh}} = r_O \sin\beta = r_{\mathrm{ph}} \sqrt{\frac{1}{A(r_{\mathrm{ph}})}},
\label{eq:rshcomplete1}
\end{equation}
leading explicitly to:
\begin{equation}
r_{\mathrm{sh}} = \frac{3\sqrt{3} A_0^3 m_s}{8\pi m_p^2}.
\label{eq:rshcomplete2}
\end{equation}

\begin{figure}[htp]
    \centering
    \includegraphics[width=8cm]{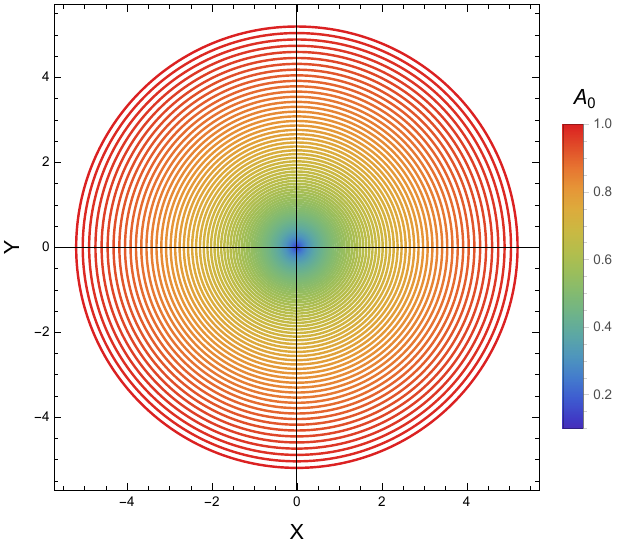}
    \caption{Shadow radius as functions of the parameter \(A_0\), assuming \(M=1\).}
    \label{fig:shadow}
\end{figure}

{Another way to interpret the physical impact of the parameter $A_0$ and opens a direct path to confronting our model with observational data from the EHT \cite{EventHorizonTelescope:2019dse, EventHorizonTelescope:2021bee}. We define the Schwarzschild mass as $M = m_s / (8\pi m_p^2)$. The standard Schwarzschild photon sphere and shadow radii are $r_{\text{ph-Sch}} = 3M$ and $r_{\text{sh-Sch}} = 3\sqrt{3}M$, respectively. We can now re-express our results from Eqs.~\eqref{eq:rph_explicit} and \eqref{eq:rshcomplete2} in these terms.
Photon sphere radius is obtained as
\begin{equation}
r_{\mathrm{ph}} = \frac{3 A_0^2 m_S}{8 \pi m_P^2} = 3 A_0^2 M = A_0^2 \, r_{\text{ph-Sch}},
\end{equation}
then we calculate black hole shadow radius
\begin{equation}
r_{\mathrm{sh}} = \frac{3\sqrt{3} A_0^3 m_s}{8\pi m_p^2} = 3\sqrt{3} A_0^3 M = A_0^3r_{\text{sh-Sch}}.
\end{equation}
The shadow size is directly modulated by $A_0^3$: if $A_0 > 1$, the shadow is larger than predicted by General Relativity, while for $A_0 < 1$, the reverse holds. Figures~\ref{fig:shadow} highlight the sensitivity of the shadow radii to the parameter \(A_0\). The shadow radius increases with the parameter \(A_0\). This result underscores the potential observational impact of different model parameters in gravitational lensing phenomena.

This provides a powerful test for our model. The angular diameter of the shadow as seen from Earth is given by $\theta_{\text{sh}} = 2 r_{\text{sh}} / D$, where $D$ is the distance to the black hole. Using our derived expression, we can solve for $A_0$:
\begin{equation}
\theta_{\text{sh}} = \frac{2 A_0^3 r_{\text{sh-Sch}}}{D} = \frac{6\sqrt{3} A_0^3 M}{D} \quad \implies \quad A_0 = \left( \frac{\theta_{\text{sh}} D}{6\sqrt{3} M} \right)^{1/3}
\end{equation}

We now apply this to the EHT observations for M87* and Sgr A*:
\begin{enumerate}
    \item \textbf{For M87*:} Using the observational values of $\theta_{\text{sh}} = (42 \pm 3) \, \mu\text{as}$, $M = 6.5 \times 10^9 \, M_{\odot}$, and $D = 16.8 \, \text{Mpc}$, we find a constraint of $A_0 \approx 1.04 \pm 0.08$.

    \item \textbf{For Sagittarius A* (Sgr A*):} Using $\theta_{\text{sh}} = (48.7 \pm 7) \, \mu\text{as}$, $M = 4.3 \times 10^6 \, M_{\odot}$, and $D \approx 8.2 \, \text{kpc}$, we find a constraint of $A_0 \approx 0.98 \pm 0.14$.
\end{enumerate}

Both results are remarkably consistent with General Relativity ($A_0=1$) within their respective uncertainties. }

\section{Strong lensing by the black hole}\label{sec:sLensing}

\begin{figure}[htp]
    \centering
    \includegraphics[width=8cm]{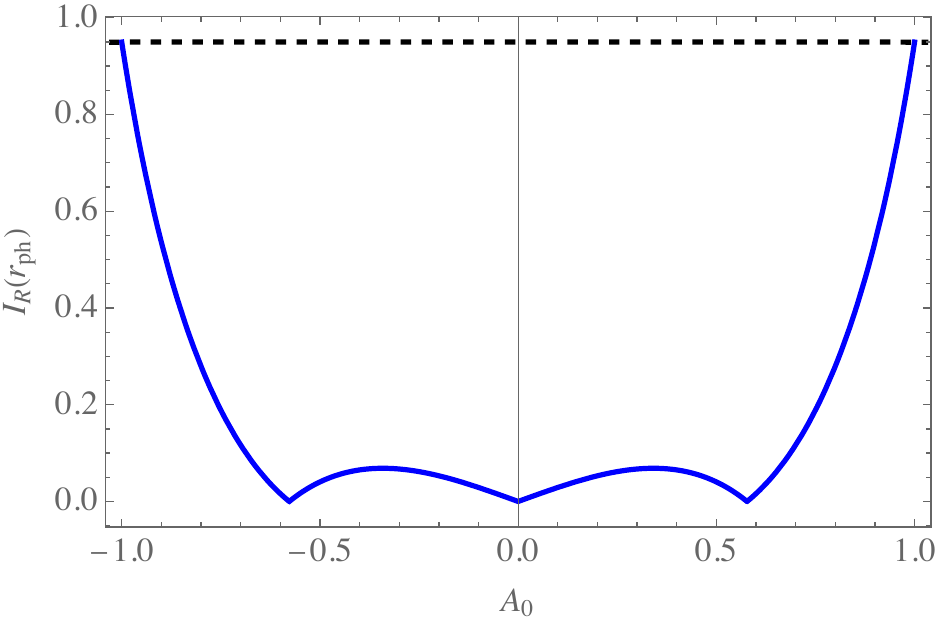}
    \caption{The behavior of $I_R(r_\ph)$ as a function of $A_0$. The dashed line corresponds to $I_R = 0.9496$, which is the value for the Schwarzschild black hole lensing.}
    \label{fig:IR}
\end{figure}

\begin{figure}[htp]
    \centering
    \includegraphics[width=8cm]{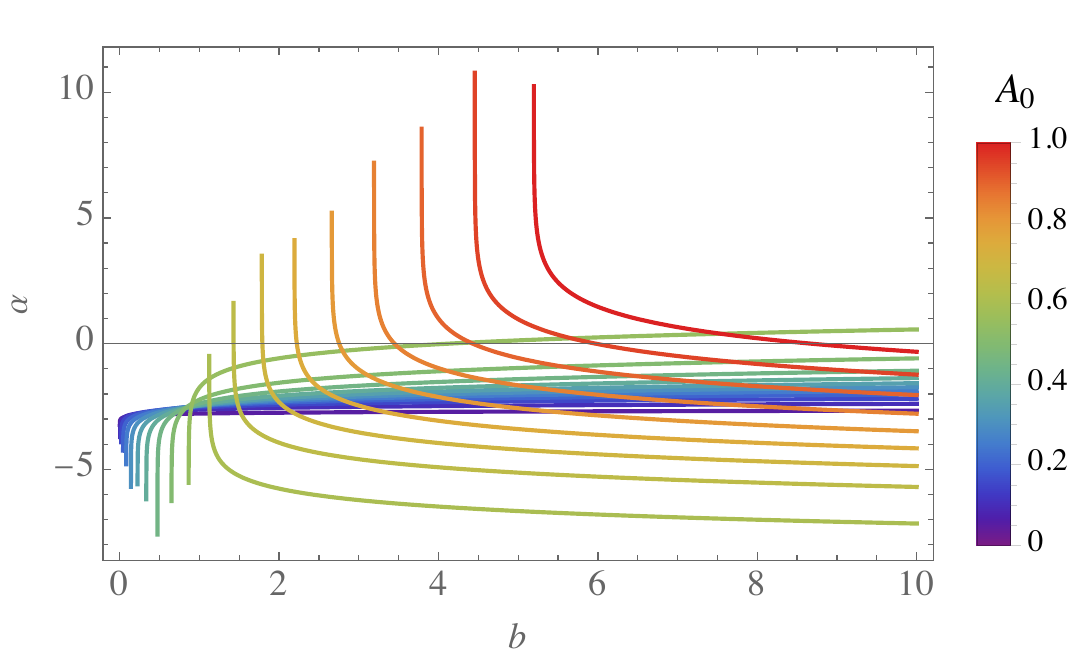}
    \caption{The $b$-profile of the strong deflection angle $\alpha$, plotted for $0< A_0\leq1$.}
    \label{fig:alpha_strong}
\end{figure}

Recalling the form in Eq. \eqref{eq:Lag} for null geodesics (i.e., with $\mathcal{L}=0$), we assume that light rays approach the black hole at the radial distance $r=r_0$. This simplifies the trajectory equation at this turning point to
\begin{equation}
A(r_0) \dot{t}_0^2=C(r_0) \dot{\phi}_0^2,
    \label{eq:A0C0}
\end{equation}
Using the definition of the impact parameter in Eq. \eqref{eq:Def_b} at $r=r_0$, the function $R(r)$ in Eq. \eqref{eq:drdphi} can be rewritten as
\begin{equation}
R(r)=\frac{A(r_0) C(r)}{A(r) C(r_0)}-1.
    \label{eq:R_b0}
\end{equation}
We define the critical impact parameter as \cite{Bozza:2002zj}
\begin{equation}
b_c\equiv b(r_\ph) = \lim_{r_0\rightarrow r_\ph}\sqrt{\frac{C(r_0)}{A(r_0)}}.
    \label{eq:bc}
\end{equation}
In Ref. \cite{Bozza:2002zj}, the deflection angle of light rays passing at the radial distance $r_0$ from a static asymptotically flat black hole is given as
\begin{equation}
\alpha(r_0)=I(r_0)-\pi = 2\int_{r_0}^\infty \frac{\sqrt{B(r)}}{\sqrt{C(r) R(r)}}\,dr-\pi :=2\int_{r_0}^\infty f(r)\, dr-\pi,
    \label{eq:alpha_0}
\end{equation}
where $R(r)$ is given in Eq. \eqref{eq:R_b0}. Introducing the variable $z\equiv 1-r_0/r$, the integral in Eq. \eqref{eq:alpha_0} can be separated into the divergent part, $I_D(r_0)$, and the regular part, $I_R(r_0)$. The divergent part is expressed as \cite{Bozza:2002zj,Tsukamoto:2017fxq,ali_testing_2025}
\begin{equation}
I_D(r_0) = \int_0^1 f_0(z,r_0)\,dz,
    \label{eq:ID_0}
\end{equation}
where
\begin{equation}
f_0(z,r_0) = \frac{2r_0}{\sqrt{x_1(r_0) z+x_2 (r_0) z^2}},
    \label{eq:f0}
\end{equation}
and
\begin{widetext}
\begin{subequations}
    \begin{align}
      &  x_1(r_0) = \frac{1-A(r_0)}{C(r_0) A'(r_0)}\Big[C'(r_0) A(r_0)-C(r_0) A'(r_0)\Big],\\
      &  x_2(r_0) = \frac{\bigl(1-A(r_0)\bigr)^2}{2C(r_0)^2 A'(r_0)^3}\Big[
      2C(r_0) C'(r_0)A'(r_0)^2+\bigl(C(r_0) C''(r_0)-2C'(r_0)^2\bigr)A(r_0)A'(r_0)-C(r_0) C'(r_0)A(r_0)A''(r_0)
      \Big].
    \end{align}
    \label{eq:x12}
\end{subequations}
\end{widetext}
This integral can be evaluated analytically, yielding
\begin{equation}
I_D(r_0)=\frac{4r_0}{\sqrt{x_2(r_0)}}\ln\left(\frac{\sqrt{x_2(r_0)}+\sqrt{x_1(r_0)+x_2(r_0)}}{\sqrt{x_1(r_0)}}\right).
    \label{eq:ID_1}
\end{equation}
In the strong lensing regime, i.e., when $r_0\rightarrow r_\ph$, expanding $x_1(r_0)$ around $r_0-r_\ph$ gives
\begin{equation}
     x_1(r_0) = \frac{C(r_\ph) r_\ph \mathfrak{L'(r_\ph)}}{B(r_\ph)}\left(r_0-r_\ph\right)+\mathcal{O}(r_0-r_\ph)^2,
     \label{eq:x1_0c}
\end{equation}
where $\mathfrak{L}(r) = C'(r)/C(r)-A'(r)/A(r)$. The impact parameter is similarly expanded as
\begin{equation}
b(r_0)=b_c+\frac{1}{4}\sqrt{\frac{C(r_\ph)}{A(r_\ph)}}\,\mathfrak{L}'(r_\ph)\left(r_0-r_\ph\right)^2+\mathcal{O}(r_0-r_\ph)^2.
    \label{eq:b_0c}
\end{equation}
\begin{widetext}
Thus, we obtain
\begin{equation}
\lim_{r_0\rightarrow r_\ph} x_1(r_0) = \lim_{b\rightarrow b_c} \frac{2 C(r_\ph) r_\ph \sqrt{\mathfrak{L}'(r_\ph)}}{B(r_\ph)}\left(\frac{b}{b_c}-1\right)^{1/2}.
    \label{eq:x1_0c_1}
\end{equation}
This provides
\begin{equation}
I_D(b) = -\frac{r_\ph}{\sqrt{x_2(r_\ph)}}\ln\left(\frac{b}{b_c}-1\right)+\frac{r_\ph}{\sqrt{x_2(r_\ph)}}\ln\Bigl(r_\ph^2 \mathfrak{L}'(r
_\ph)\Bigr)+\mathcal{O}(b-b_c),
    \label{eq:ID_2}
\end{equation}
where
\begin{equation}
x_2(r_\ph) = \frac{C(r_\ph)\bigl(1-A(r_\ph)\bigr)^2\Bigl[C''(r_\ph) A(r_\ph)-C(r_\ph) A''(r_\ph)\Bigr]}{2A(r_\ph)^2 C'(r_\ph)}.
    \label{eq:x2ph}
\end{equation}
\end{widetext}
The regular part of the integral in Eq. \eqref{eq:alpha_0}, $I_R$, is expressed as
\begin{equation}
I_R(b) = \int_0^1 \Bigl[f(z,r_0)-f_0(z,r_0)\Bigr]\,dz.
    \label{eq:IR_0}
\end{equation}
Thus, the deflection angle in the strong lensing regime is given by
\begin{equation}
\alpha(b) = -\bar{a} \ln\left(\frac{b}{b_c}-1\right)+\bar{k}+\mathcal{O}(b-b_c),
    \label{eq:alpha_1}
\end{equation}
where
\begin{subequations}
    \begin{align}
        & \bar{a} = \sqrt{\frac{2 B(r_\ph) A(r_\ph)}{C''(r_\ph) A(r_\ph)-C(r_\ph) A''(r_\ph)}},\\
        & \bar{k} = \bar{a}\ln\Bigl(r_\ph^2 \mathfrak{L}'(r_\ph)\Bigr) + I_R(r_\ph) - \pi.
    \end{align}
    \label{eq:babk}
\end{subequations}
Now, using the spacetime metric functions from the line element \eqref{eq:metr_1} for the black hole in LGT and comparing them with those in Eq. \eqref{eq:metric_0}, while considering the photon sphere radius given in Eq. \eqref{eq:rph_explicit}, we obtain the critical impact parameter as
\begin{equation}
b_c = \sqrt{\frac{C(r_\ph)}{A(r_\ph)}} = \frac{3\sqrt{3} A_0^3 m_S}{8\pi m_P^2},
\end{equation}
which coincides with the shadow radius presented in Eq. \eqref{eq:rshcomplete2}. By utilizing the relations derived in this section, we further obtain
\begin{eqnarray}
    && \bar{a} = A_0,\label{eq:bara_1}\\
    && \bar{k} = A_0\ln\left(2+\frac{256 \pi ^2 m_P^4 }{27 A_0^6 m_S^2}\right) + I_R(r_\ph) - \pi.\label{eq:kb_1}
\end{eqnarray}
The integrand of Eq. \eqref{eq:IR_0} is given by
\begin{equation}
\frac{2\left(3A_0^2-1\right)\left[\sqrt{-18+\frac{3}{A_0^2}+9A_0^2}-\sqrt{-\frac{\left(3A_0^2-1\right)^2\left(3A_0^2z-3-z\right)}{A_0^2}}\right]}{\sqrt{\frac{\left(3A_0^2-1\right)^4\left[3-\left(3A_0^2-1\right)z\right]}{A_0^4}}}.
    \label{eq:IR_0_integ}
\end{equation}
Since this integral cannot be computed analytically for general $A_0$, we note that for the Schwarzschild black hole case ($A_0 = \pm 1$), the integral evaluates to $I_R^{\mathrm{SBH}} \approx 0.9496$, which matches the result reported in Ref. \cite{Bozza:2002zj}. More generally, as $A_0^2 \to 1$, the values of $I_R$ approach this Schwarzschild black hole value (see Fig. \ref{fig:IR}).
Additionally, for $A_0^2 = 1$, we find $\bar{k} = -0.40023$, which corresponds to the Schwarzschild black hole lensing result reported in Ref. \cite{Bozza:2002zj}. Finally, by incorporating these relations, we present in Fig. \ref{fig:alpha_strong} the $b$-profile of the strong deflection angle as a function of $A_0$.
%

\subsection{The lens equation and the EHT constraints}

\begin{figure}[htp]
    \centering
    \includegraphics[width=7cm]{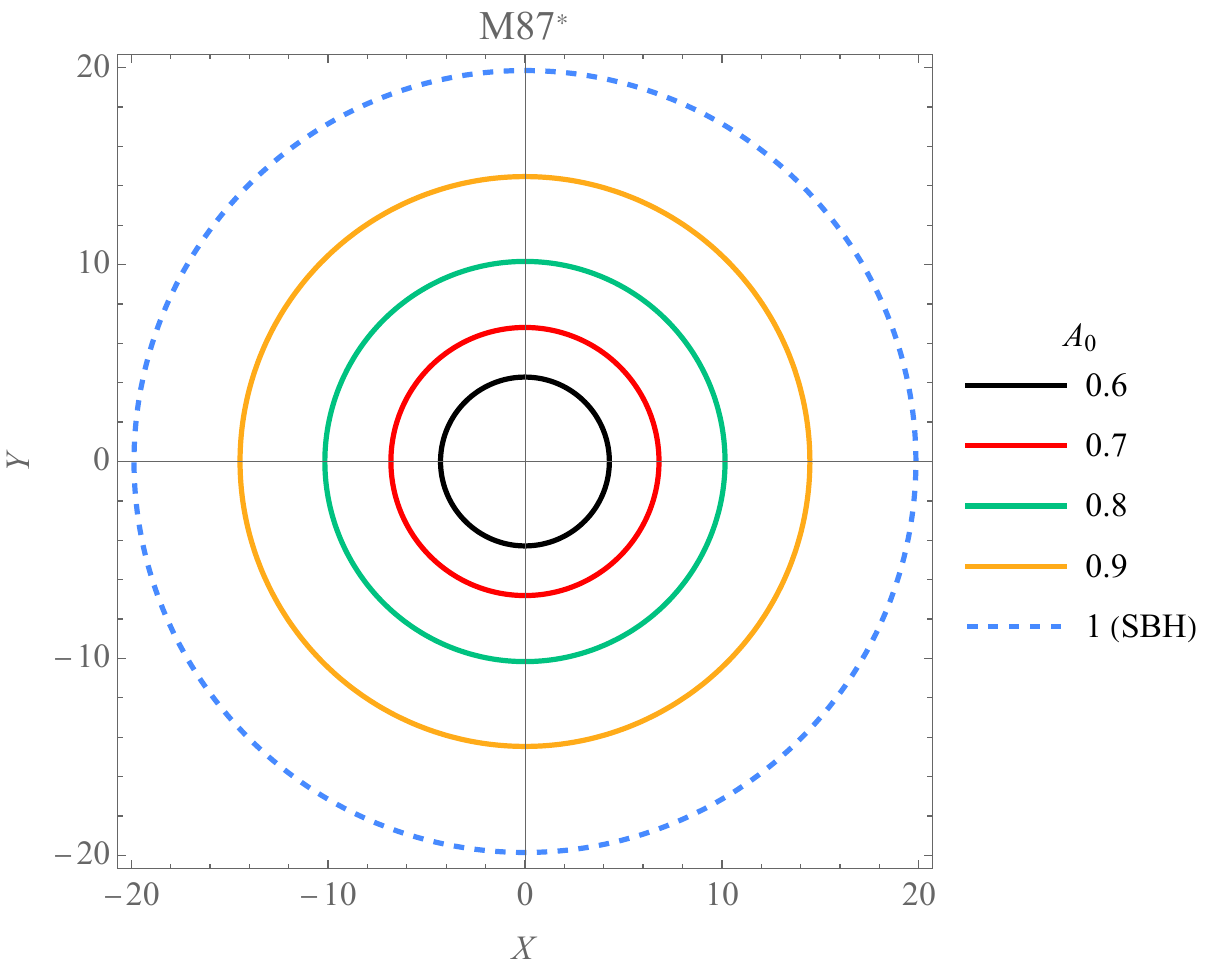} (a)
    \includegraphics[width=7cm]{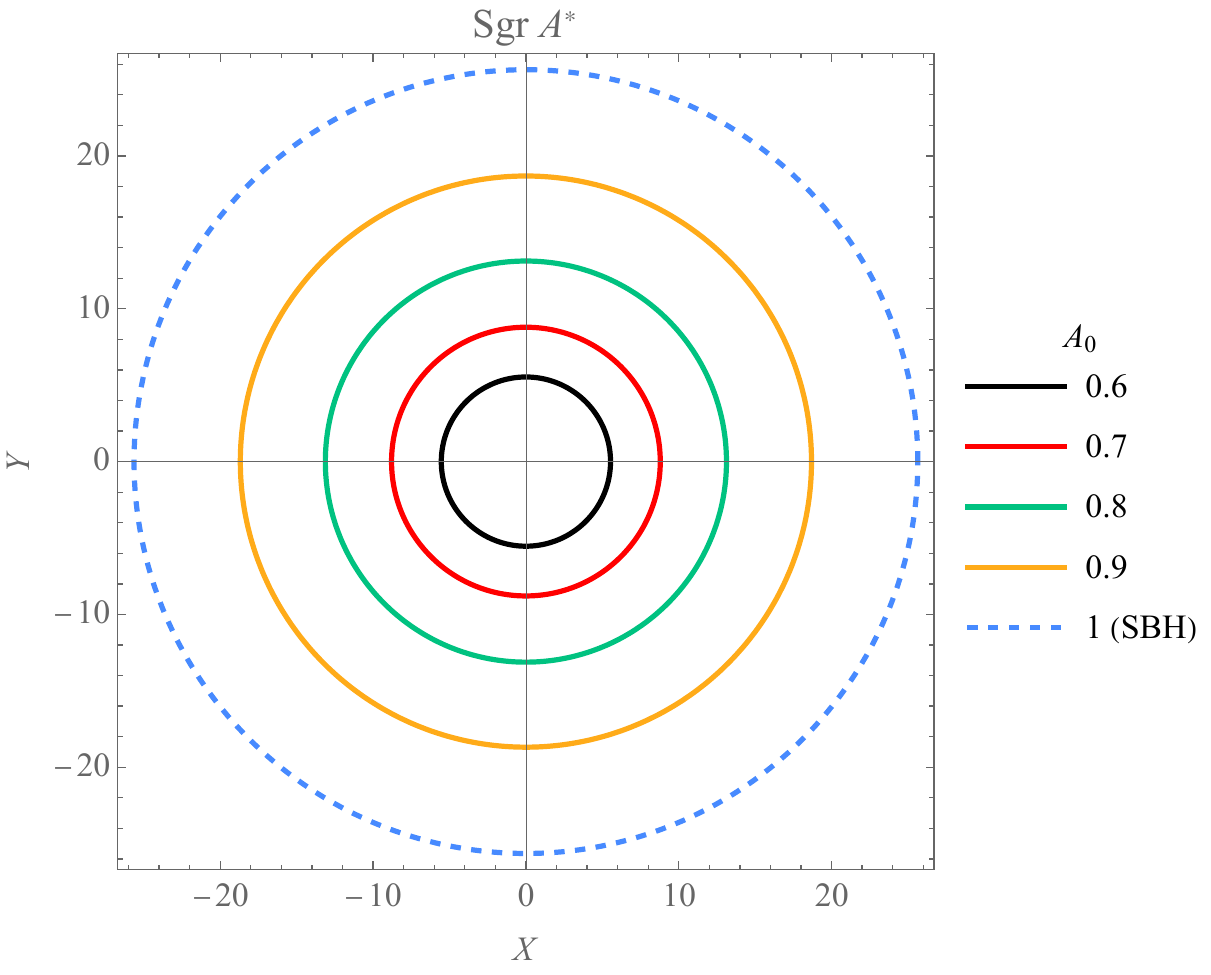} (b)
    \caption{The Einstein rings of M87* and Sgr A* considered as LGT black holes.}
    \label{fig:Erings}
\end{figure}

Following the discussion in Ref.~\cite{Bozza:2001xd}, we consider a scenario in which a light source, denoted as $S$, is nearly perfectly aligned with a black hole that serves as a gravitational lens, labeled as $L$. In this configuration, relativistic images are expected to form. The lens equation describing this setup is given by  
\begin{equation}
    \psi = \theta - \frac{D_{LS}}{D_{OS}} \Delta\alpha_n,
    \label{eq:lensEq_0}
\end{equation}
where $D_{LS}$ represents the distance between the lens and the light source, while $D_{OS}$ denotes the distance between the observer and the light source. The angular positions of the source and the image are given by $\psi$ and $\theta$, respectively. Additionally, the quantity $\Delta\alpha_n = \alpha(\theta) - 2n\pi$ accounts for the deviation of the deflection angle after considering the total number of loops completed by the photons. For $\alpha(\theta_n^0) = 2n\pi$, we obtain  
\begin{equation}
    \theta_n^0 = \frac{b_c}{D_{OL}} \left(1 + \epsilon_n\right),
    \label{eq:thetan0}
\end{equation}
where  
\begin{equation}
    \epsilon_n = \exp\left(\frac{\bar{k}-2 n\pi}{\bar{a}}\right).
    \label{eq:en}
\end{equation}
Expanding $\alpha(\theta)$ around $\theta_n^0$ and introducing $\Delta\theta_n = \theta - \theta_n^0$, the deviation simplifies to  
\begin{equation}
    \Delta\alpha_n = -\frac{\bar{a}D_{OL}}{b_c \epsilon_n} \Delta\theta_n.
    \label{eq:Deltalpha_n}
\end{equation}
Thus, the lens equation can be rewritten as  
\begin{equation}
    \psi = \theta + \left(\frac{\bar{a}D_{LS} D_{OL}}{b_c \epsilon_n D_{OS}}\right) \Delta\theta_n.
    \label{eq:lensEq_2}
\end{equation}
Assuming that $D_{OL} \gg b_c$, the angular position of the $n$th relativistic image is given by \cite{Bozza:2002zj}  
\begin{equation}
    \theta_n = \theta_n^0 + \frac{b_c \epsilon_n \left(\psi - \theta_n^0\right) D_{OS}}{\bar{a}D_{LS} D_{OL}}.
    \label{eq:theta_n}
\end{equation}
From this equation, it follows that when $\psi = \theta_n^0$, the image coincides with the source. The sign of $\psi$ determines whether the image forms on the same side ($\psi > 0$) or the opposite side ($\psi < 0$) of the lens.  In cases where the black hole, source, and observer are nearly aligned ($\psi \approx 0$), and assuming that the observer and lens are equidistant from the source ($D_{OS} = D_{LS} = 2 D_{OL}$), light deflection occurs symmetrically, leading to the formation of Einstein rings \cite{einstein_lens-like_1936, mellier_probing_1999, bartelmann_weak_2001, schmidt_weak_2008, guzik_tests_2010}. Under these conditions, Eq.~\eqref{eq:theta_n} simplifies to \cite{bozza_time_2004}  
\begin{equation}
    \theta_n^{E} = \left(1 - \frac{b_c \epsilon_n D_{OS}}{\bar{a}D_{LS} D_{OL}}\right) \theta_n^0.
    \label{eq:theta_nE_0}
\end{equation}
Furthermore, for $D_{OL} \gg b_c$, the angular radius of the $n$th relativistic Einstein ring is given by  
\begin{equation}
    \theta_n^{E} = \frac{b_c \left( 1 + \epsilon_n \right)}{D_{OL}}.
    \label{eq:thetanE_1}
\end{equation}
To apply these theoretical results to astrophysical observations, we consider the supermassive black holes M87* and Sgr A*. The black hole M87* has a mass of $(6.5 \pm 0.7) \times 10^9 \, M_\odot$ and is located at a distance of $D_{OL} = 16.8 \, \mathrm{Mpc}$ from Earth \cite{EventHorizonTelescope:2019dse, EventHorizonTelescope:2019ggy}. Meanwhile, Sgr A* has a mass of $4^{+1.1}_{-0.6} \times 10^6 \, M_\odot$ and is situated at a distance of $7.97 \, \mathrm{kpc}$ from Earth \cite{Akiyama:2022}. Substituting these parameters into Eq.~\eqref{eq:thetanE_1}, Fig.~\ref{fig:Erings} illustrates the outermost Einstein rings ($n = 1$) for M87* and Sgr A*, assuming they are black holes within the framework of the LGT. These rings are depicted in the celestial coordinate system of an observer on Earth, with coordinates $X$ and $Y$, and are computed for different values of $A_0$.  
As observed, an increase in $A_0$ leads to a significant enlargement of the Einstein rings. Consequently, the Schwarzschild black hole (SBH) exhibits the largest rings when $A_0^2 \leq 1$.

A key quantity in strong lensing is the magnification of the $n$th relativistic image, defined as \cite{Virbhadra:1999nm, Bozza:2002zj}
\begin{equation}
    \mu_n = \left. \left( \frac{\psi}{\theta} \frac{d \psi}{d \theta} \right)^{-1} \right|_{\theta_n^0} = \frac{b_c^2 \epsilon_n \left( 1 + \epsilon_n \right) D_{OS}}{\bar{a} \psi D_{LS} D_{OL}^2}.
    \label{eq:mun}
\end{equation}
Since the magnification is inversely proportional to $D_{OL}^2$, relativistic images are typically faint. The outermost image appears the brightest, while the brightness of higher-order images diminishes exponentially. However, in the limiting case where $\psi$ approaches zero, implying near-perfect alignment of the source and the lens, the images undergo significant magnification. It is noteworthy that while the outermost image, corresponding to $\theta_1$, remains distinct, higher-order images tend to cluster around $\theta_\infty \equiv \theta_n |_{n \to \infty}$ \cite{Bozza:2002zj}, which is identified as $\theta_\infty = \theta_c$, representing the innermost relativistic image.

Astrophysical applications of strong lensing also involve two important observables. The first is the angular separation between the outermost and innermost relativistic images, given by
\begin{equation}
    s = \theta_1 - \theta_\infty \approx \theta_\infty \epsilon_1,
    \label{eq:s_0}
\end{equation}
while the second is the relative magnification between the outermost relativistic image and the collective group of inner relativistic images, expressed as \cite{Bozza:2002zj}
\begin{equation}
    r_{\mathrm{mag}} = \frac{\mu_1}{\sum_{n=2}^\infty \mu_n} = 2.5 \log_{10} \left( \exp \left[ \frac{2\pi}{\bar{a}} \right] \right),
    \label{eq:rmag_0}
\end{equation}
which notably remains independent of the observer's distance, $D_{OL}$.

Before concluding this section, it is crucial to establish constraints on the scalar hair parameter of black holes based on recent observations of M87* and Sgr A*.

In 2019, the EHT collaboration published the first horizon-scale image of the supermassive black hole M87*, providing strong observational evidence for black holes. Their analysis revealed that the compact emission region had an angular diameter of $\theta_d = (42 \pm 3) \, \mu\text{as}$, along with a central flux depression exceeding a factor of $\gtrsim 10$, corresponding to the black hole's shadow \cite{EventHorizonTelescope:2019dse, EventHorizonTelescope:2019ggy}. Subsequently, in 2022, the EHT collaboration captured an image of Sgr A*, the supermassive black hole at the center of the Milky Way. This image displayed a distinctive ring structure with an angular diameter of $\theta_d = (48.7 \pm 7) \, \mu\text{as}$ and a deviation from the Schwarzschild shadow characterized by $\delta = -0.08^{+0.09}_{-0.09}$ (VLTI) and $\delta = -0.04^{+0.09}_{-0.10}$ (Keck). Additionally, EHT observations of Sgr A* estimated the angular diameter of the emission ring to be $\theta_d = (51.8 \pm 2.3) \, \mu\text{as}$ \cite{Akiyama:2022}.

\begin{figure}[htp]
    \centering
    \includegraphics[width=7cm]{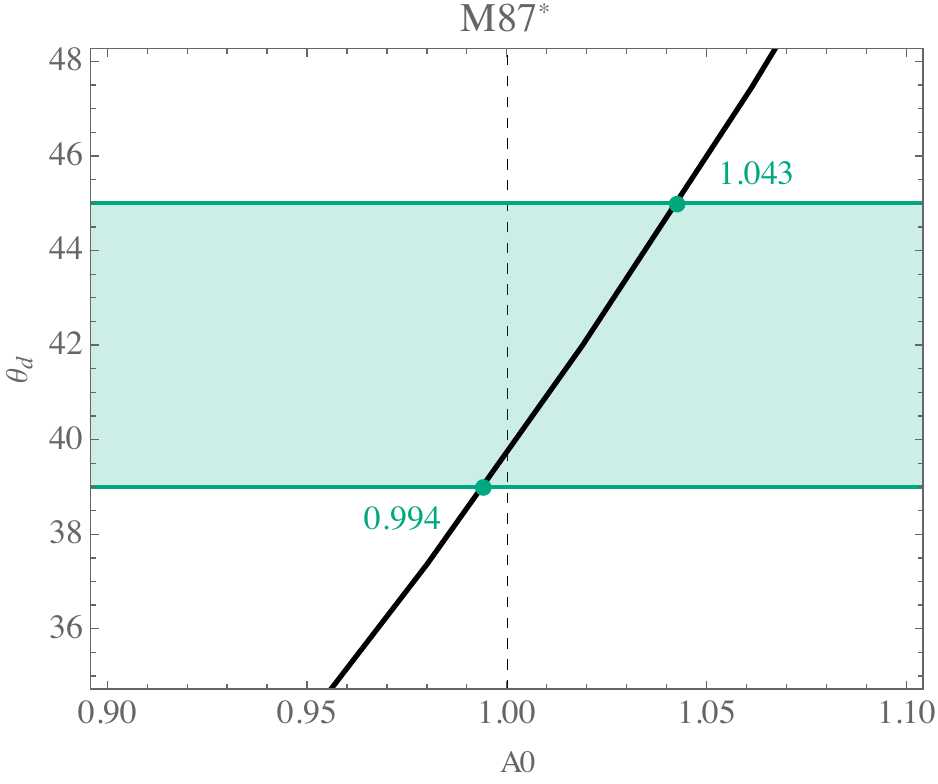} (a)
    \includegraphics[width=7cm]{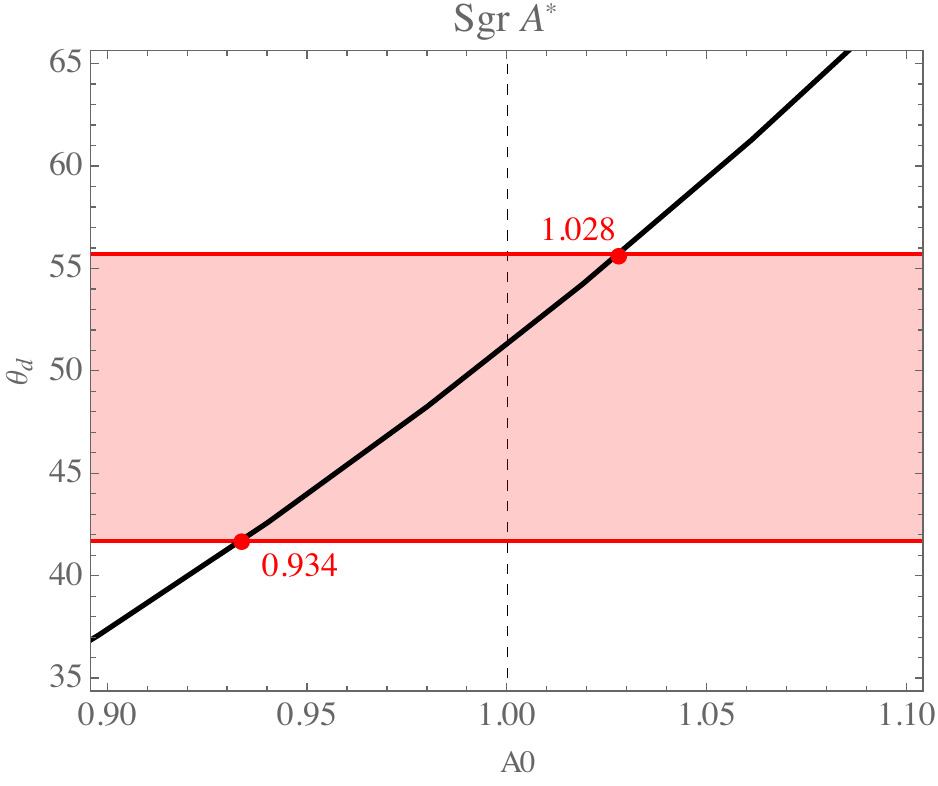} (b)
    \caption{The $A_0$-profile of the theoretical shadow angular diameter ($2\theta_\infty$) of the LGT black hole, within the observed shadow diameters of (a) M87* and (b) Sgr A*, at the $1\sigma$ confidence level.}
    \label{fig:EHTconstraints}
\end{figure}

Using these EHT observational data, we model M87* and Sgr A* as LGT black holes to constrain the parameter $A_0$. By considering the apparent radius of the photon sphere, $\theta_{\infty}$, as the angular size of the black hole shadow, we derive constraints on $A_0$ at the $1\sigma$ confidence level. To establish this relationship, we use $\theta_d = 2 \theta_\infty$, linking the observed angular diameter of the shadows to the theoretical prediction. In Fig.~\ref{fig:EHTconstraints}(a), we present the $A_0$-profile of $2\theta_\infty$ for M87*.
For Sgr A*, we consider the averaged shadow angular diameter, derived from independent algorithms analyzing EHT observations. These studies suggest that $\theta_d$ for Sgr A* lies within the range $46.9 \, \mu\text{as}$ to $50 \, \mu\text{as}$. Including the $1\sigma$ uncertainty, this range extends from $41.7 \, \mu\text{as}$ to $55.6 \, \mu\text{as}$ \cite{Akiyama:2022}. In Fig.~\ref{fig:EHTconstraints}(b), we incorporate this confidence interval to compare the theoretical shadow diameter with the observational results, thereby constraining $A_0$. Our findings suggest that the LGT black hole aligns most closely with the EHT data when $A_0$ approaches the Schwarzschild limit, as $A_0$ remains nearly unity within its uncertainty range.


\section{Conclusion}\label{sec:conclusion}

In this work, we have investigated black hole solutions and their observational signatures within the framework of Lorentz gauge gravity, an intriguing alternative to General Relativity motivated by recent cosmological observations and foundational issues in quantum gravity. We have systematically studied the gravitational lensing effects and shadow structures of static, spherically symmetric black hole solutions arising from this theory.

Through explicit analytical and numerical analyses, we have shown that the Lorentz gauge parameter \( A_0 \), characterizing the scale of the connection, significantly influences black hole properties, including horizon radius, photon sphere radius, shadow radius, and gravitational lensing angles. Notably, deviations from the Schwarzschild geometry emerge clearly through variations in \( A_0 \), potentially providing observational discriminators to test the validity of this gravitational framework.

Specifically, we computed weak deflection angles using both Gauss-Bonnet theorem and geodesic methods, confirming consistency between approaches and highlighting modifications introduced by the LGT. In the strong lensing regime, we derived the critical impact parameter and characterized the logarithmic divergence of the deflection angle, identifying clear signatures distinguishing these solutions from classical Schwarzschild black holes.

Furthermore, our analysis of the black hole shadow revealed its explicit dependence on \( A_0 \), exhibiting a pronounced sensitivity that could be tested with high-precision observations such as those from the EHT. The computed shadow radius provides robust criteria for distinguishing Lorentz gauge gravity from standard General Relativity through direct imaging. These findings demonstrate the potential of gravitational lensing and shadow imaging as powerful observational tools to probe modified gravity theories. Our results thus contribute significantly toward establishing precise observational tests that can discriminate between LGT and classical GR, paving the way for deeper insights into the fundamental nature of gravity and spacetime structure.

\section*{Acknowledgements}
A.{\"O}. would like to acknowledge the contribution of the COST Action CA21106 - COSMIC WISPers in the Dark Universe: Theory, astrophysics and experiments (CosmicWISPers), the COST Action CA21136 - Addressing observational tensions in cosmology with systematics and fundamental physics (CosmoVerse), the COST Action CA22113 - Fundamental challenges in theoretical physics (THEORY-CHALLENGES), the COST Action CA23130 - Bridging high and low energies in search of quantum gravity (BridgeQG) and the COST Action CA23115 - Relativistic Quantum Information (RQI) funded by COST (European Cooperation in Science and
Technology). We also thank EMU, TUBITAK, ULAKBIM (Turkiye) and SCOAP3 (Switzerland) for their support.  The work of M.F. is supported by Universidad Central de Chile through the project No. PDUCEN20240008.

\bibliography{references}

\begin{thebibliography}{137}%
\makeatletter
\providecommand \@ifxundefined [1]{%
 \@ifx{#1\undefined}
}%
\providecommand \@ifnum [1]{%
 \ifnum #1\expandafter \@firstoftwo
 \else \expandafter \@secondoftwo
 \fi
}%
\providecommand \@ifx [1]{%
 \ifx #1\expandafter \@firstoftwo
 \else \expandafter \@secondoftwo
 \fi
}%
\providecommand \natexlab [1]{#1}%
\providecommand \enquote  [1]{``#1''}%
\providecommand \bibnamefont  [1]{#1}%
\providecommand \bibfnamefont [1]{#1}%
\providecommand \citenamefont [1]{#1}%
\providecommand \href@noop [0]{\@secondoftwo}%
\providecommand \href [0]{\begingroup \@sanitize@url \@href}%
\providecommand \@href[1]{\@@startlink{#1}\@@href}%
\providecommand \@@href[1]{\endgroup#1\@@endlink}%
\providecommand \@sanitize@url [0]{\catcode `\\12\catcode `\$12\catcode `\&12\catcode `\#12\catcode `\^12\catcode `\_12\catcode `\%12\relax}%
\providecommand \@@startlink[1]{}%
\providecommand \@@endlink[0]{}%
\providecommand \url  [0]{\begingroup\@sanitize@url \@url }%
\providecommand \@url [1]{\endgroup\@href {#1}{\urlprefix }}%
\providecommand \urlprefix  [0]{URL }%
\providecommand \Eprint [0]{\href }%
\providecommand \doibase [0]{http://dx.doi.org/}%
\providecommand \selectlanguage [0]{\@gobble}%
\providecommand \bibinfo  [0]{\@secondoftwo}%
\providecommand \bibfield  [0]{\@secondoftwo}%
\providecommand \translation [1]{[#1]}%
\providecommand \BibitemOpen [0]{}%
\providecommand \bibitemStop [0]{}%
\providecommand \bibitemNoStop [0]{.\EOS\space}%
\providecommand \EOS [0]{\spacefactor3000\relax}%
\providecommand \BibitemShut  [1]{\csname bibitem#1\endcsname}%
\let\auto@bib@innerbib\@empty
\bibitem [{\citenamefont {Koivisto}\ and\ \citenamefont {Zheng}(2025)}]{Koivisto:2024asr}%
  \BibitemOpen
  \bibfield  {author} {\bibinfo {author} {\bibfnamefont {T.~S.}\ \bibnamefont {Koivisto}}\ and\ \bibinfo {author} {\bibfnamefont {L.}~\bibnamefont {Zheng}},\ }\href {\doibase 10.1103/PhysRevD.111.064008} {\bibfield  {journal} {\bibinfo  {journal} {Phys. Rev. D}\ }\textbf {\bibinfo {volume} {111}},\ \bibinfo {pages} {064008} (\bibinfo {year} {2025})}\BibitemShut {NoStop}%
\bibitem [{\citenamefont {Wald}(1984)}]{wald1984}%
  \BibitemOpen
  \bibfield  {author} {\bibinfo {author} {\bibfnamefont {R.~M.}\ \bibnamefont {Wald}},\ }\href@noop {} {\emph {\bibinfo {title} {General Relativity}}}\ (\bibinfo  {publisher} {University of Chicago Press},\ \bibinfo {year} {1984})\BibitemShut {NoStop}%
\bibitem [{\citenamefont {Misner}\ \emph {et~al.}(1973)\citenamefont {Misner}, \citenamefont {Thorne},\ and\ \citenamefont {Wheeler}}]{mtn1973}%
  \BibitemOpen
  \bibfield  {author} {\bibinfo {author} {\bibfnamefont {C.~W.}\ \bibnamefont {Misner}}, \bibinfo {author} {\bibfnamefont {K.~S.}\ \bibnamefont {Thorne}}, \ and\ \bibinfo {author} {\bibfnamefont {J.~A.}\ \bibnamefont {Wheeler}},\ }\href@noop {} {\emph {\bibinfo {title} {Gravitation}}}\ (\bibinfo  {publisher} {Freeman},\ \bibinfo {year} {1973})\BibitemShut {NoStop}%
\bibitem [{\citenamefont {Virbhadra}\ and\ \citenamefont {Ellis}(2000)}]{Virbhadra:1999nm}%
  \BibitemOpen
  \bibfield  {author} {\bibinfo {author} {\bibfnamefont {K.~S.}\ \bibnamefont {Virbhadra}}\ and\ \bibinfo {author} {\bibfnamefont {G.~F.~R.}\ \bibnamefont {Ellis}},\ }\href {\doibase 10.1103/PhysRevD.62.084003} {\bibfield  {journal} {\bibinfo  {journal} {Phys. Rev. D}\ }\textbf {\bibinfo {volume} {62}},\ \bibinfo {pages} {084003} (\bibinfo {year} {2000})},\ \Eprint {http://arxiv.org/abs/astro-ph/9904193} {arXiv:astro-ph/9904193} \BibitemShut {NoStop}%
\bibitem [{\citenamefont {Virbhadra}\ and\ \citenamefont {Ellis}(2002)}]{Virbhadra:2002ju}%
  \BibitemOpen
  \bibfield  {author} {\bibinfo {author} {\bibfnamefont {K.~S.}\ \bibnamefont {Virbhadra}}\ and\ \bibinfo {author} {\bibfnamefont {G.~F.~R.}\ \bibnamefont {Ellis}},\ }\href {\doibase 10.1103/PhysRevD.65.103004} {\bibfield  {journal} {\bibinfo  {journal} {Phys. Rev. D}\ }\textbf {\bibinfo {volume} {65}},\ \bibinfo {pages} {103004} (\bibinfo {year} {2002})}\BibitemShut {NoStop}%
\bibitem [{\citenamefont {Claudel}\ \emph {et~al.}(2001)\citenamefont {Claudel}, \citenamefont {Virbhadra},\ and\ \citenamefont {Ellis}}]{Claudel:2000yi}%
  \BibitemOpen
  \bibfield  {author} {\bibinfo {author} {\bibfnamefont {C.-M.}\ \bibnamefont {Claudel}}, \bibinfo {author} {\bibfnamefont {K.~S.}\ \bibnamefont {Virbhadra}}, \ and\ \bibinfo {author} {\bibfnamefont {G.~F.~R.}\ \bibnamefont {Ellis}},\ }\href {\doibase 10.1063/1.1308507} {\bibfield  {journal} {\bibinfo  {journal} {J. Math. Phys.}\ }\textbf {\bibinfo {volume} {42}},\ \bibinfo {pages} {818} (\bibinfo {year} {2001})},\ \Eprint {http://arxiv.org/abs/gr-qc/0005050} {arXiv:gr-qc/0005050} \BibitemShut {NoStop}%
\bibitem [{\citenamefont {Adler}\ and\ \citenamefont {Virbhadra}(2022)}]{Adler:2022qtb}%
  \BibitemOpen
  \bibfield  {author} {\bibinfo {author} {\bibfnamefont {S.~L.}\ \bibnamefont {Adler}}\ and\ \bibinfo {author} {\bibfnamefont {K.~S.}\ \bibnamefont {Virbhadra}},\ }\href {\doibase 10.1007/s10714-022-02976-7} {\bibfield  {journal} {\bibinfo  {journal} {Gen. Rel. Grav.}\ }\textbf {\bibinfo {volume} {54}},\ \bibinfo {pages} {93} (\bibinfo {year} {2022})},\ \Eprint {http://arxiv.org/abs/2205.04628} {arXiv:2205.04628 [gr-qc]} \BibitemShut {NoStop}%
\bibitem [{\citenamefont {Abbott}\ \emph {et~al.}(2016)\citenamefont {Abbott} \emph {et~al.}}]{abbott2016}%
  \BibitemOpen
  \bibfield  {author} {\bibinfo {author} {\bibfnamefont {B.~P.}\ \bibnamefont {Abbott}} \emph {et~al.} (\bibinfo {collaboration} {LIGO Scientific, Virgo}),\ }\href {\doibase 10.1103/PhysRevLett.116.061102} {\bibfield  {journal} {\bibinfo  {journal} {Phys. Rev. Lett.}\ }\textbf {\bibinfo {volume} {116}},\ \bibinfo {pages} {061102} (\bibinfo {year} {2016})},\ \Eprint {http://arxiv.org/abs/1602.03837} {arXiv:1602.03837 [gr-qc]} \BibitemShut {NoStop}%
\bibitem [{\citenamefont {Abbott}\ \emph {et~al.}(2019)\citenamefont {Abbott} \emph {et~al.}}]{abbott2019}%
  \BibitemOpen
  \bibfield  {author} {\bibinfo {author} {\bibfnamefont {B.~P.}\ \bibnamefont {Abbott}} \emph {et~al.} (\bibinfo {collaboration} {LIGO Scientific, Virgo}),\ }\href {\doibase 10.1103/PhysRevX.9.031040} {\bibfield  {journal} {\bibinfo  {journal} {Phys. Rev. X}\ }\textbf {\bibinfo {volume} {9}},\ \bibinfo {pages} {031040} (\bibinfo {year} {2019})},\ \Eprint {http://arxiv.org/abs/1811.12907} {arXiv:1811.12907 [astro-ph.HE]} \BibitemShut {NoStop}%
\bibitem [{\citenamefont {{[The Event Horizon Telescope Collaboration]}}\ \emph {et~al.}(2019{\natexlab{a}})\citenamefont {{[The Event Horizon Telescope Collaboration]}}, \citenamefont {Akiyama} \emph {et~al.}}]{EventHorizonTelescope:2019dse}%
  \BibitemOpen
  \bibfield  {author} {\bibinfo {author} {\bibnamefont {{[The Event Horizon Telescope Collaboration]}}}, \bibinfo {author} {\bibfnamefont {K.}~\bibnamefont {Akiyama}},  \emph {et~al.},\ }\href {\doibase 10.3847/2041-8213/ab0ec7} {\bibfield  {journal} {\bibinfo  {journal} {Astrophys. J. Lett.}\ }\textbf {\bibinfo {volume} {875}},\ \bibinfo {pages} {L1} (\bibinfo {year} {2019}{\natexlab{a}})},\ \Eprint {http://arxiv.org/abs/1906.11238} {arXiv:1906.11238 [astro-ph.GA]} \BibitemShut {NoStop}%
\bibitem [{\citenamefont {{[The Event Horizon Telescope Collaboration]}}\ \emph {et~al.}(2019{\natexlab{b}})\citenamefont {{[The Event Horizon Telescope Collaboration]}}, \citenamefont {Akiyama} \emph {et~al.}}]{EventHorizonTelescope:2019ggy}%
  \BibitemOpen
  \bibfield  {author} {\bibinfo {author} {\bibnamefont {{[The Event Horizon Telescope Collaboration]}}}, \bibinfo {author} {\bibfnamefont {K.}~\bibnamefont {Akiyama}},  \emph {et~al.},\ }\href {\doibase 10.3847/2041-8213/ab1141} {\bibfield  {journal} {\bibinfo  {journal} {Astrophys. J. Lett.}\ }\textbf {\bibinfo {volume} {875}},\ \bibinfo {pages} {L6} (\bibinfo {year} {2019}{\natexlab{b}})},\ \Eprint {http://arxiv.org/abs/1906.11243} {arXiv:1906.11243 [astro-ph.GA]} \BibitemShut {NoStop}%
\bibitem [{\citenamefont {{[The Event Horizon Telescope Collaboration]}}\ \emph {et~al.}(2021)\citenamefont {{[The Event Horizon Telescope Collaboration]}}, \citenamefont {Akiyama} \emph {et~al.}}]{EventHorizonTelescope:2021bee}%
  \BibitemOpen
  \bibfield  {author} {\bibinfo {author} {\bibnamefont {{[The Event Horizon Telescope Collaboration]}}}, \bibinfo {author} {\bibfnamefont {K.}~\bibnamefont {Akiyama}},  \emph {et~al.},\ }\href {\doibase 10.3847/2041-8213/abe71d} {\bibfield  {journal} {\bibinfo  {journal} {Astrophys. J. Lett.}\ }\textbf {\bibinfo {volume} {910}},\ \bibinfo {pages} {L12} (\bibinfo {year} {2021})},\ \Eprint {http://arxiv.org/abs/2105.01169} {arXiv:2105.01169 [astro-ph.HE]} \BibitemShut {NoStop}%
\bibitem [{\citenamefont {{[The Event Horizon Telescope Collaboration]}}\ \emph {et~al.}(2022)\citenamefont {{[The Event Horizon Telescope Collaboration]}}, \citenamefont {Akiyama} \emph {et~al.}}]{Akiyama:2022}%
  \BibitemOpen
  \bibfield  {author} {\bibinfo {author} {\bibnamefont {{[The Event Horizon Telescope Collaboration]}}}, \bibinfo {author} {\bibfnamefont {K.}~\bibnamefont {Akiyama}},  \emph {et~al.},\ }\href {\doibase 10.3847/2041-8213/ac6674} {\bibfield  {journal} {\bibinfo  {journal} {Astrophys. J. Lett.}\ }\textbf {\bibinfo {volume} {930}},\ \bibinfo {pages} {L12} (\bibinfo {year} {2022})},\ \Eprint {http://arxiv.org/abs/2311.08680} {arXiv:2311.08680 [astro-ph.HE]} \BibitemShut {NoStop}%
\bibitem [{\citenamefont {{Zakharov}}\ \emph {et~al.}(2005)\citenamefont {{Zakharov}}, \citenamefont {{Nucita}}, \citenamefont {{De Paolis}},\ and\ \citenamefont {{Ingrosso}}}]{2005NewA...10..479Z}%
  \BibitemOpen
  \bibfield  {author} {\bibinfo {author} {\bibfnamefont {A.~F.}\ \bibnamefont {{Zakharov}}}, \bibinfo {author} {\bibfnamefont {A.~A.}\ \bibnamefont {{Nucita}}}, \bibinfo {author} {\bibfnamefont {F.}~\bibnamefont {{De Paolis}}}, \ and\ \bibinfo {author} {\bibfnamefont {G.}~\bibnamefont {{Ingrosso}}},\ }\href {\doibase 10.1016/j.newast.2005.02.007} {\bibfield  {journal} {\bibinfo  {journal} {New Astronomy}\ }\textbf {\bibinfo {volume} {10}},\ \bibinfo {pages} {479} (\bibinfo {year} {2005})},\ \Eprint {http://arxiv.org/abs/astro-ph/0411511} {arXiv:astro-ph/0411511 [astro-ph]} \BibitemShut {NoStop}%
\bibitem [{\citenamefont {Zakharov}(2024)}]{Zakharov:2023yjl}%
  \BibitemOpen
  \bibfield  {author} {\bibinfo {author} {\bibfnamefont {A.~F.}\ \bibnamefont {Zakharov}},\ }\href {\doibase 10.1142/S0218271823400047} {\bibfield  {journal} {\bibinfo  {journal} {Int. J. Mod. Phys. D}\ }\textbf {\bibinfo {volume} {33}},\ \bibinfo {pages} {2340004} (\bibinfo {year} {2024})},\ \Eprint {http://arxiv.org/abs/2308.01301} {arXiv:2308.01301 [gr-qc]} \BibitemShut {NoStop}%
\bibitem [{\citenamefont {Bambhaniya}\ \emph {et~al.}(2024)\citenamefont {Bambhaniya}, \citenamefont {Joshi}, \citenamefont {Dey}, \citenamefont {Joshi}, \citenamefont {Mazumdar}, \citenamefont {Harada},\ and\ \citenamefont {Nakao}}]{Bambhaniya:2022xbz}%
  \BibitemOpen
  \bibfield  {author} {\bibinfo {author} {\bibfnamefont {P.}~\bibnamefont {Bambhaniya}}, \bibinfo {author} {\bibfnamefont {A.~B.}\ \bibnamefont {Joshi}}, \bibinfo {author} {\bibfnamefont {D.}~\bibnamefont {Dey}}, \bibinfo {author} {\bibfnamefont {P.~S.}\ \bibnamefont {Joshi}}, \bibinfo {author} {\bibfnamefont {A.}~\bibnamefont {Mazumdar}}, \bibinfo {author} {\bibfnamefont {T.}~\bibnamefont {Harada}}, \ and\ \bibinfo {author} {\bibfnamefont {K.-i.}\ \bibnamefont {Nakao}},\ }\href {\doibase 10.1140/epjc/s10052-024-12477-3} {\bibfield  {journal} {\bibinfo  {journal} {Eur. Phys. J. C}\ }\textbf {\bibinfo {volume} {84}},\ \bibinfo {pages} {124} (\bibinfo {year} {2024})},\ \Eprint {http://arxiv.org/abs/2209.12610} {arXiv:2209.12610 [gr-qc]} \BibitemShut {NoStop}%
\bibitem [{\citenamefont {Zakharov}\ \emph {et~al.}(2005)\citenamefont {Zakharov}, \citenamefont {De~Paolis}, \citenamefont {Ingrosso},\ and\ \citenamefont {Nucita}}]{Zakharov:2005ek}%
  \BibitemOpen
  \bibfield  {author} {\bibinfo {author} {\bibfnamefont {A.~F.}\ \bibnamefont {Zakharov}}, \bibinfo {author} {\bibfnamefont {F.}~\bibnamefont {De~Paolis}}, \bibinfo {author} {\bibfnamefont {G.}~\bibnamefont {Ingrosso}}, \ and\ \bibinfo {author} {\bibfnamefont {A.~A.}\ \bibnamefont {Nucita}},\ }\href {\doibase 10.1051/0004-6361:20053432} {\bibfield  {journal} {\bibinfo  {journal} {Astron. Astrophys.}\ }\textbf {\bibinfo {volume} {442}},\ \bibinfo {pages} {795} (\bibinfo {year} {2005})},\ \Eprint {http://arxiv.org/abs/astro-ph/0505286} {arXiv:astro-ph/0505286} \BibitemShut {NoStop}%
\bibitem [{\citenamefont {Zakharov}\ \emph {et~al.}(2012)\citenamefont {Zakharov}, \citenamefont {Paolis}, \citenamefont {Ingrosso},\ and\ \citenamefont {Nucita}}]{ZAKHAROV201264}%
  \BibitemOpen
  \bibfield  {author} {\bibinfo {author} {\bibfnamefont {A.~F.}\ \bibnamefont {Zakharov}}, \bibinfo {author} {\bibfnamefont {F.~D.}\ \bibnamefont {Paolis}}, \bibinfo {author} {\bibfnamefont {G.}~\bibnamefont {Ingrosso}}, \ and\ \bibinfo {author} {\bibfnamefont {A.~A.}\ \bibnamefont {Nucita}},\ }\href {\doibase https://doi.org/10.1016/j.newar.2011.09.002} {\bibfield  {journal} {\bibinfo  {journal} {New Astronomy Reviews}\ }\textbf {\bibinfo {volume} {56}},\ \bibinfo {pages} {64} (\bibinfo {year} {2012})},\ \bibinfo {note} {supermassive Black Holes and Spectral Lines}\BibitemShut {NoStop}%
\bibitem [{\citenamefont {Zakharov}(2014)}]{Zakharov:2014lqa}%
  \BibitemOpen
  \bibfield  {author} {\bibinfo {author} {\bibfnamefont {A.~F.}\ \bibnamefont {Zakharov}},\ }\href {\doibase 10.1103/PhysRevD.90.062007} {\bibfield  {journal} {\bibinfo  {journal} {Phys. Rev. D}\ }\textbf {\bibinfo {volume} {90}},\ \bibinfo {pages} {062007} (\bibinfo {year} {2014})},\ \Eprint {http://arxiv.org/abs/1407.7457} {arXiv:1407.7457 [gr-qc]} \BibitemShut {NoStop}%
\bibitem [{\citenamefont {Zakharov}(2022{\natexlab{a}})}]{Zakharov:2021gbg}%
  \BibitemOpen
  \bibfield  {author} {\bibinfo {author} {\bibfnamefont {A.~F.}\ \bibnamefont {Zakharov}},\ }\href {\doibase 10.3390/universe8030141} {\bibfield  {journal} {\bibinfo  {journal} {Universe}\ }\textbf {\bibinfo {volume} {8}},\ \bibinfo {pages} {141} (\bibinfo {year} {2022}{\natexlab{a}})},\ \Eprint {http://arxiv.org/abs/2108.01533} {arXiv:2108.01533 [gr-qc]} \BibitemShut {NoStop}%
\bibitem [{\citenamefont {Zakharov}(2022{\natexlab{b}})}]{Zakharov:2022gwk}%
  \BibitemOpen
  \bibfield  {author} {\bibinfo {author} {\bibfnamefont {A.~F.}\ \bibnamefont {Zakharov}},\ }in\ \href {\doibase 10.17184/eac.7531} {\emph {\bibinfo {booktitle} {{ICRANet-ISFAHAN Astronomy Meeting}: {From the Ancient Persian Astronomy to Recent Developments in Theoretical and Experimental Physics, Astrophysics and General Relativity}}}}\ (\bibinfo {year} {2022})\ \Eprint {http://arxiv.org/abs/2208.06805} {arXiv:2208.06805 [astro-ph.GA]} \BibitemShut {NoStop}%
\bibitem [{\citenamefont {Zakharov}(2025{\natexlab{a}})}]{Zakharov:2025cnq}%
  \BibitemOpen
  \bibfield  {author} {\bibinfo {author} {\bibfnamefont {A.~F.}\ \bibnamefont {Zakharov}},\ }\href {\doibase 10.1134/S106377882570019X} {\bibfield  {journal} {\bibinfo  {journal} {Phys. Atom. Nucl.}\ }\textbf {\bibinfo {volume} {88}},\ \bibinfo {pages} {154} (\bibinfo {year} {2025}{\natexlab{a}})}\BibitemShut {NoStop}%
\bibitem [{\citenamefont {Zakharov}(2025{\natexlab{b}})}]{Zakharov:2025mbx}%
  \BibitemOpen
  \bibfield  {author} {\bibinfo {author} {\bibfnamefont {A.~F.}\ \bibnamefont {Zakharov}},\ }\href {\doibase 10.1134/S1547477125700177} {\bibfield  {journal} {\bibinfo  {journal} {Phys. Part. Nucl. Lett.}\ }\textbf {\bibinfo {volume} {22}},\ \bibinfo {pages} {568} (\bibinfo {year} {2025}{\natexlab{b}})}\BibitemShut {NoStop}%
\bibitem [{\citenamefont {de~Laurentis}\ \emph {et~al.}(2023)\citenamefont {de~Laurentis}, \citenamefont {De~Martino},\ and\ \citenamefont {Della~Monica}}]{deLaurentis:2022oqa}%
  \BibitemOpen
  \bibfield  {author} {\bibinfo {author} {\bibfnamefont {M.}~\bibnamefont {de~Laurentis}}, \bibinfo {author} {\bibfnamefont {I.}~\bibnamefont {De~Martino}}, \ and\ \bibinfo {author} {\bibfnamefont {R.}~\bibnamefont {Della~Monica}},\ }\href {\doibase 10.1088/1361-6633/ace91b} {\bibfield  {journal} {\bibinfo  {journal} {Rept. Prog. Phys.}\ }\textbf {\bibinfo {volume} {86}},\ \bibinfo {pages} {104901} (\bibinfo {year} {2023})},\ \Eprint {http://arxiv.org/abs/2211.07008} {arXiv:2211.07008 [astro-ph.GA]} \BibitemShut {NoStop}%
\bibitem [{\citenamefont {Genzel}\ \emph {et~al.}(2024)\citenamefont {Genzel}, \citenamefont {Eisenhauer},\ and\ \citenamefont {Gillessen}}]{Genzel:2024vou}%
  \BibitemOpen
  \bibfield  {author} {\bibinfo {author} {\bibfnamefont {R.}~\bibnamefont {Genzel}}, \bibinfo {author} {\bibfnamefont {F.}~\bibnamefont {Eisenhauer}}, \ and\ \bibinfo {author} {\bibfnamefont {S.}~\bibnamefont {Gillessen}},\ }\href {\doibase 10.1007/s00159-024-00154-z} {\bibfield  {journal} {\bibinfo  {journal} {Astron. Astrophys. Rev.}\ }\textbf {\bibinfo {volume} {32}},\ \bibinfo {pages} {3} (\bibinfo {year} {2024})},\ \Eprint {http://arxiv.org/abs/2404.03522} {arXiv:2404.03522 [astro-ph.GA]} \BibitemShut {NoStop}%
\bibitem [{\citenamefont {Khodadi}\ \emph {et~al.}(2024)\citenamefont {Khodadi}, \citenamefont {Vagnozzi},\ and\ \citenamefont {Firouzjaee}}]{khodadi_event_2024}%
  \BibitemOpen
  \bibfield  {author} {\bibinfo {author} {\bibfnamefont {M.}~\bibnamefont {Khodadi}}, \bibinfo {author} {\bibfnamefont {S.}~\bibnamefont {Vagnozzi}}, \ and\ \bibinfo {author} {\bibfnamefont {J.~T.}\ \bibnamefont {Firouzjaee}},\ }\href {\doibase 10.1038/s41598-024-78264-y} {\bibfield  {journal} {\bibinfo  {journal} {Scientific Reports}\ }\textbf {\bibinfo {volume} {14}},\ \bibinfo {pages} {26932} (\bibinfo {year} {2024})}\BibitemShut {NoStop}%
\bibitem [{\citenamefont {Khodadi}\ and\ \citenamefont {Lambiase}(2022)}]{khodadi_probing_2022}%
  \BibitemOpen
  \bibfield  {author} {\bibinfo {author} {\bibfnamefont {M.}~\bibnamefont {Khodadi}}\ and\ \bibinfo {author} {\bibfnamefont {G.}~\bibnamefont {Lambiase}},\ }\href {\doibase 10.1103/PhysRevD.106.104050} {\bibfield  {journal} {\bibinfo  {journal} {Physical Review D}\ }\textbf {\bibinfo {volume} {106}},\ \bibinfo {pages} {104050} (\bibinfo {year} {2022})}\BibitemShut {NoStop}%
\bibitem [{\citenamefont {Khodadi}\ and\ \citenamefont {Saridakis}(2021)}]{khodadi_einstein-aether_2021}%
  \BibitemOpen
  \bibfield  {author} {\bibinfo {author} {\bibfnamefont {M.}~\bibnamefont {Khodadi}}\ and\ \bibinfo {author} {\bibfnamefont {E.~N.}\ \bibnamefont {Saridakis}},\ }\href {\doibase 10.1016/j.dark.2021.100835} {\bibfield  {journal} {\bibinfo  {journal} {Physics of the Dark Universe}\ }\textbf {\bibinfo {volume} {32}},\ \bibinfo {pages} {100835} (\bibinfo {year} {2021})}\BibitemShut {NoStop}%
\bibitem [{\citenamefont {Clifton}\ \emph {et~al.}(2012)\citenamefont {Clifton}, \citenamefont {Ferreira}, \citenamefont {Padilla},\ and\ \citenamefont {Skordis}}]{clifton2012}%
  \BibitemOpen
  \bibfield  {author} {\bibinfo {author} {\bibfnamefont {T.}~\bibnamefont {Clifton}}, \bibinfo {author} {\bibfnamefont {P.~G.}\ \bibnamefont {Ferreira}}, \bibinfo {author} {\bibfnamefont {A.}~\bibnamefont {Padilla}}, \ and\ \bibinfo {author} {\bibfnamefont {C.}~\bibnamefont {Skordis}},\ }\href {\doibase 10.1016/j.physrep.2012.01.001} {\bibfield  {journal} {\bibinfo  {journal} {Phys. Rept.}\ }\textbf {\bibinfo {volume} {513}},\ \bibinfo {pages} {1} (\bibinfo {year} {2012})},\ \Eprint {http://arxiv.org/abs/1106.2476} {arXiv:1106.2476 [astro-ph.CO]} \BibitemShut {NoStop}%
\bibitem [{\citenamefont {Sotiriou}\ and\ \citenamefont {Faraoni}(2010)}]{sotiriou2010}%
  \BibitemOpen
  \bibfield  {author} {\bibinfo {author} {\bibfnamefont {T.~P.}\ \bibnamefont {Sotiriou}}\ and\ \bibinfo {author} {\bibfnamefont {V.}~\bibnamefont {Faraoni}},\ }\href {\doibase 10.1103/RevModPhys.82.451} {\bibfield  {journal} {\bibinfo  {journal} {Rev. Mod. Phys.}\ }\textbf {\bibinfo {volume} {82}},\ \bibinfo {pages} {451} (\bibinfo {year} {2010})},\ \Eprint {http://arxiv.org/abs/0805.1726} {arXiv:0805.1726 [gr-qc]} \BibitemShut {NoStop}%
\bibitem [{\citenamefont {Weinberg}(1979)}]{weinberg1979}%
  \BibitemOpen
  \bibfield  {author} {\bibinfo {author} {\bibfnamefont {S.}~\bibnamefont {Weinberg}},\ }in\ \href@noop {} {\emph {\bibinfo {booktitle} {General Relativity: An Einstein Centenary Survey}}}\ (\bibinfo  {publisher} {Cambridge University Press},\ \bibinfo {year} {1979})\BibitemShut {NoStop}%
\bibitem [{\citenamefont {Plebański}(1977)}]{plebanski1977}%
  \BibitemOpen
  \bibfield  {author} {\bibinfo {author} {\bibfnamefont {J.~F.}\ \bibnamefont {Plebański}},\ }\href@noop {} {\bibfield  {journal} {\bibinfo  {journal} {J. Math. Phys.}\ }\textbf {\bibinfo {volume} {18}},\ \bibinfo {pages} {2511} (\bibinfo {year} {1977})}\BibitemShut {NoStop}%
\bibitem [{\citenamefont {Rovelli}(2004)}]{rovelli2004}%
  \BibitemOpen
  \bibfield  {author} {\bibinfo {author} {\bibfnamefont {C.}~\bibnamefont {Rovelli}},\ }\href@noop {} {\emph {\bibinfo {title} {Quantum Gravity}}}\ (\bibinfo  {publisher} {Cambridge University Press},\ \bibinfo {year} {2004})\BibitemShut {NoStop}%
\bibitem [{\citenamefont {Ashtekar}\ and\ \citenamefont {Lewandowski}(2004)}]{ashtekar2004}%
  \BibitemOpen
  \bibfield  {author} {\bibinfo {author} {\bibfnamefont {A.}~\bibnamefont {Ashtekar}}\ and\ \bibinfo {author} {\bibfnamefont {J.}~\bibnamefont {Lewandowski}},\ }\href {\doibase 10.1088/0264-9381/21/15/R01} {\bibfield  {journal} {\bibinfo  {journal} {Class. Quant. Grav.}\ }\textbf {\bibinfo {volume} {21}},\ \bibinfo {pages} {R53} (\bibinfo {year} {2004})},\ \Eprint {http://arxiv.org/abs/gr-qc/0404018} {arXiv:gr-qc/0404018} \BibitemShut {NoStop}%
\bibitem [{\citenamefont {Z{\l}o{\'s}nik}\ and\ \citenamefont {Westman}(2017)}]{zlosnik2017}%
  \BibitemOpen
  \bibfield  {author} {\bibinfo {author} {\bibfnamefont {T.~G.}\ \bibnamefont {Z{\l}o{\'s}nik}}\ and\ \bibinfo {author} {\bibfnamefont {H.~F.}\ \bibnamefont {Westman}},\ }\href {\doibase 10.1088/1361-6382/aa944f} {\bibfield  {journal} {\bibinfo  {journal} {Class. Quant. Grav.}\ }\textbf {\bibinfo {volume} {34}},\ \bibinfo {pages} {245001} (\bibinfo {year} {2017})},\ \Eprint {http://arxiv.org/abs/1601.00567} {arXiv:1601.00567 [gr-qc]} \BibitemShut {NoStop}%
\bibitem [{\citenamefont {Zlosnik}\ \emph {et~al.}(2006)\citenamefont {Zlosnik}, \citenamefont {Ferreira},\ and\ \citenamefont {Starkman}}]{zlosnik2006}%
  \BibitemOpen
  \bibfield  {author} {\bibinfo {author} {\bibfnamefont {T.~G.}\ \bibnamefont {Zlosnik}}, \bibinfo {author} {\bibfnamefont {P.~G.}\ \bibnamefont {Ferreira}}, \ and\ \bibinfo {author} {\bibfnamefont {G.~D.}\ \bibnamefont {Starkman}},\ }\href {\doibase 10.1103/PhysRevD.74.044037} {\bibfield  {journal} {\bibinfo  {journal} {Phys. Rev. D}\ }\textbf {\bibinfo {volume} {74}},\ \bibinfo {pages} {044037} (\bibinfo {year} {2006})},\ \Eprint {http://arxiv.org/abs/gr-qc/0606039} {arXiv:gr-qc/0606039} \BibitemShut {NoStop}%
\bibitem [{\citenamefont {Z{\l}o{\'s}nik}\ \emph {et~al.}(2018)\citenamefont {Z{\l}o{\'s}nik}, \citenamefont {Urban}, \citenamefont {Marzola},\ and\ \citenamefont {Koivisto}}]{Zlosnik:2018qvg}%
  \BibitemOpen
  \bibfield  {author} {\bibinfo {author} {\bibfnamefont {T.}~\bibnamefont {Z{\l}o{\'s}nik}}, \bibinfo {author} {\bibfnamefont {F.}~\bibnamefont {Urban}}, \bibinfo {author} {\bibfnamefont {L.}~\bibnamefont {Marzola}}, \ and\ \bibinfo {author} {\bibfnamefont {T.}~\bibnamefont {Koivisto}},\ }\href {\doibase 10.1088/1361-6382/aaea96} {\bibfield  {journal} {\bibinfo  {journal} {Class. Quant. Grav.}\ }\textbf {\bibinfo {volume} {35}},\ \bibinfo {pages} {235003} (\bibinfo {year} {2018})},\ \Eprint {http://arxiv.org/abs/1807.01100} {arXiv:1807.01100 [gr-qc]} \BibitemShut {NoStop}%
\bibitem [{\citenamefont {Koivisto}\ and\ \citenamefont {Zlosnik}(2023)}]{Koivisto:2022uvd}%
  \BibitemOpen
  \bibfield  {author} {\bibinfo {author} {\bibfnamefont {T.~S.}\ \bibnamefont {Koivisto}}\ and\ \bibinfo {author} {\bibfnamefont {T.}~\bibnamefont {Zlosnik}},\ }\href {\doibase 10.1103/PhysRevD.107.124013} {\bibfield  {journal} {\bibinfo  {journal} {Phys. Rev. D}\ }\textbf {\bibinfo {volume} {107}},\ \bibinfo {pages} {124013} (\bibinfo {year} {2023})},\ \Eprint {http://arxiv.org/abs/2212.04562} {arXiv:2212.04562 [gr-qc]} \BibitemShut {NoStop}%
\bibitem [{\citenamefont {Nikjoo}\ and\ \citenamefont {Zlosnik}(2024)}]{Nikjoo:2023flm}%
  \BibitemOpen
  \bibfield  {author} {\bibinfo {author} {\bibfnamefont {M.}~\bibnamefont {Nikjoo}}\ and\ \bibinfo {author} {\bibfnamefont {T.}~\bibnamefont {Zlosnik}},\ }\href {\doibase 10.1088/1361-6382/ad1c84} {\bibfield  {journal} {\bibinfo  {journal} {Class. Quant. Grav.}\ }\textbf {\bibinfo {volume} {41}},\ \bibinfo {pages} {045005} (\bibinfo {year} {2024})},\ \Eprint {http://arxiv.org/abs/2308.01108} {arXiv:2308.01108 [gr-qc]} \BibitemShut {NoStop}%
\bibitem [{\citenamefont {Alexandrov}(2014)}]{alexandrov2013}%
  \BibitemOpen
  \bibfield  {author} {\bibinfo {author} {\bibfnamefont {S.}~\bibnamefont {Alexandrov}},\ }\href {\doibase 10.1007/s10714-013-1639-1} {\bibfield  {journal} {\bibinfo  {journal} {Gen. Rel. Grav.}\ }\textbf {\bibinfo {volume} {46}},\ \bibinfo {pages} {1639} (\bibinfo {year} {2014})},\ \Eprint {http://arxiv.org/abs/1308.6586} {arXiv:1308.6586 [hep-th]} \BibitemShut {NoStop}%
\bibitem [{\citenamefont {Smolin}(2004)}]{smolin2004}%
  \BibitemOpen
  \bibfield  {author} {\bibinfo {author} {\bibfnamefont {L.}~\bibnamefont {Smolin}},\ }\bibfield  {booktitle} {\emph {\bibinfo {booktitle} {{3rd International Symposium on Quantum Theory and Symmetries}}},\ }\href {\doibase 10.1142/9789812702340_0078} {\ ,\ \bibinfo {pages} {655} (\bibinfo {year} {2004})},\ \Eprint {http://arxiv.org/abs/hep-th/0408048} {arXiv:hep-th/0408048} \BibitemShut {NoStop}%
\bibitem [{\citenamefont {Mielke}(2017)}]{Mielke:2017nwt}%
  \BibitemOpen
  \bibfield  {author} {\bibinfo {author} {\bibfnamefont {E.~W.}\ \bibnamefont {Mielke}},\ }\href {\doibase 10.1007/978-3-319-29734-7} {\emph {\bibinfo {title} {{Geometrodynamics of Gauge Fields. On the Geometry of Yang-Mills and Gravitational Gauge Theories}}}},\ Mathematical Physics Studies\ (\bibinfo  {publisher} {Springer},\ \bibinfo {year} {2017})\BibitemShut {NoStop}%
\bibitem [{\citenamefont {Brown}\ and\ \citenamefont {Kuchar}(1995)}]{Brown:1994py}%
  \BibitemOpen
  \bibfield  {author} {\bibinfo {author} {\bibfnamefont {J.~D.}\ \bibnamefont {Brown}}\ and\ \bibinfo {author} {\bibfnamefont {K.~V.}\ \bibnamefont {Kuchar}},\ }\href {\doibase 10.1103/PhysRevD.51.5600} {\bibfield  {journal} {\bibinfo  {journal} {Phys. Rev. D}\ }\textbf {\bibinfo {volume} {51}},\ \bibinfo {pages} {5600} (\bibinfo {year} {1995})},\ \Eprint {http://arxiv.org/abs/gr-qc/9409001} {arXiv:gr-qc/9409001} \BibitemShut {NoStop}%
\bibitem [{\citenamefont {Gallagher}\ \emph {et~al.}(2024)\citenamefont {Gallagher}, \citenamefont {Koivisto}, \citenamefont {Marzola}, \citenamefont {Varrin},\ and\ \citenamefont {Zlosnik}}]{Gallagher:2023ghl}%
  \BibitemOpen
  \bibfield  {author} {\bibinfo {author} {\bibfnamefont {P.}~\bibnamefont {Gallagher}}, \bibinfo {author} {\bibfnamefont {T.~S.}\ \bibnamefont {Koivisto}}, \bibinfo {author} {\bibfnamefont {L.}~\bibnamefont {Marzola}}, \bibinfo {author} {\bibfnamefont {L.}~\bibnamefont {Varrin}}, \ and\ \bibinfo {author} {\bibfnamefont {T.}~\bibnamefont {Zlosnik}},\ }\href {\doibase 10.1103/PhysRevD.109.L061503} {\bibfield  {journal} {\bibinfo  {journal} {Phys. Rev. D}\ }\textbf {\bibinfo {volume} {109}},\ \bibinfo {pages} {L061503} (\bibinfo {year} {2024})},\ \Eprint {http://arxiv.org/abs/2311.07464} {arXiv:2311.07464 [hep-th]} \BibitemShut {NoStop}%
\bibitem [{\citenamefont {Weinberg}(1972)}]{weinberg1972}%
  \BibitemOpen
  \bibfield  {author} {\bibinfo {author} {\bibfnamefont {S.}~\bibnamefont {Weinberg}},\ }\href@noop {} {\emph {\bibinfo {title} {Gravitation and Cosmology}}}\ (\bibinfo  {publisher} {Wiley},\ \bibinfo {year} {1972})\BibitemShut {NoStop}%
\bibitem [{\citenamefont {Keeton}\ and\ \citenamefont {Petters}(2005)}]{keeton2005}%
  \BibitemOpen
  \bibfield  {author} {\bibinfo {author} {\bibfnamefont {C.~R.}\ \bibnamefont {Keeton}}\ and\ \bibinfo {author} {\bibfnamefont {A.~O.}\ \bibnamefont {Petters}},\ }\href {\doibase 10.1103/PhysRevD.72.104006} {\bibfield  {journal} {\bibinfo  {journal} {Phys. Rev. D}\ }\textbf {\bibinfo {volume} {72}},\ \bibinfo {pages} {104006} (\bibinfo {year} {2005})},\ \Eprint {http://arxiv.org/abs/gr-qc/0511019} {arXiv:gr-qc/0511019} \BibitemShut {NoStop}%
\bibitem [{\citenamefont {Bozza}(2002)}]{Bozza:2002zj}%
  \BibitemOpen
  \bibfield  {author} {\bibinfo {author} {\bibfnamefont {V.}~\bibnamefont {Bozza}},\ }\href {\doibase 10.1103/PhysRevD.66.103001} {\bibfield  {journal} {\bibinfo  {journal} {Phys. Rev. D}\ }\textbf {\bibinfo {volume} {66}},\ \bibinfo {pages} {103001} (\bibinfo {year} {2002})},\ \Eprint {http://arxiv.org/abs/gr-qc/0208075} {arXiv:gr-qc/0208075} \BibitemShut {NoStop}%
\bibitem [{\citenamefont {Tsukamoto}(2018{\natexlab{a}})}]{tsukamoto2012}%
  \BibitemOpen
  \bibfield  {author} {\bibinfo {author} {\bibfnamefont {N.}~\bibnamefont {Tsukamoto}},\ }\href {\doibase 10.1103/PhysRevD.97.064021} {\bibfield  {journal} {\bibinfo  {journal} {Phys. Rev. D}\ }\textbf {\bibinfo {volume} {97}},\ \bibinfo {pages} {064021} (\bibinfo {year} {2018}{\natexlab{a}})},\ \Eprint {http://arxiv.org/abs/1708.07427} {arXiv:1708.07427 [gr-qc]} \BibitemShut {NoStop}%
\bibitem [{\citenamefont {Soares}\ \emph {et~al.}(2024)\citenamefont {Soares}, \citenamefont {Vit\'oria},\ and\ \citenamefont {Pereira}}]{Soares:2024rhp}%
  \BibitemOpen
  \bibfield  {author} {\bibinfo {author} {\bibfnamefont {A.~R.}\ \bibnamefont {Soares}}, \bibinfo {author} {\bibfnamefont {R.~L.~L.}\ \bibnamefont {Vit\'oria}}, \ and\ \bibinfo {author} {\bibfnamefont {C.~F.~S.}\ \bibnamefont {Pereira}},\ }\href {\doibase 10.1103/PhysRevD.110.084004} {\bibfield  {journal} {\bibinfo  {journal} {Phys. Rev. D}\ }\textbf {\bibinfo {volume} {110}},\ \bibinfo {pages} {084004} (\bibinfo {year} {2024})},\ \Eprint {http://arxiv.org/abs/2408.03217} {arXiv:2408.03217 [gr-qc]} \BibitemShut {NoStop}%
\bibitem [{\citenamefont {Soares}\ \emph {et~al.}(2023{\natexlab{a}})\citenamefont {Soares}, \citenamefont {Pereira}, \citenamefont {Vit\'oria},\ and\ \citenamefont {Rocha}}]{Soares:2023uup}%
  \BibitemOpen
  \bibfield  {author} {\bibinfo {author} {\bibfnamefont {A.~R.}\ \bibnamefont {Soares}}, \bibinfo {author} {\bibfnamefont {C.~F.~S.}\ \bibnamefont {Pereira}}, \bibinfo {author} {\bibfnamefont {R.~L.~L.}\ \bibnamefont {Vit\'oria}}, \ and\ \bibinfo {author} {\bibfnamefont {E.~M.}\ \bibnamefont {Rocha}},\ }\href {\doibase 10.1103/PhysRevD.108.124024} {\bibfield  {journal} {\bibinfo  {journal} {Phys. Rev. D}\ }\textbf {\bibinfo {volume} {108}},\ \bibinfo {pages} {124024} (\bibinfo {year} {2023}{\natexlab{a}})},\ \Eprint {http://arxiv.org/abs/2309.05106} {arXiv:2309.05106 [gr-qc]} \BibitemShut {NoStop}%
\bibitem [{\citenamefont {Soares}\ \emph {et~al.}(2023{\natexlab{b}})\citenamefont {Soares}, \citenamefont {Vit\'oria},\ and\ \citenamefont {Pereira}}]{Soares:2023err}%
  \BibitemOpen
  \bibfield  {author} {\bibinfo {author} {\bibfnamefont {A.~R.}\ \bibnamefont {Soares}}, \bibinfo {author} {\bibfnamefont {R.~L.~L.}\ \bibnamefont {Vit\'oria}}, \ and\ \bibinfo {author} {\bibfnamefont {C.~F.~S.}\ \bibnamefont {Pereira}},\ }\href {\doibase 10.1140/epjc/s10052-023-12071-z} {\bibfield  {journal} {\bibinfo  {journal} {Eur. Phys. J. C}\ }\textbf {\bibinfo {volume} {83}},\ \bibinfo {pages} {903} (\bibinfo {year} {2023}{\natexlab{b}})},\ \Eprint {http://arxiv.org/abs/2305.11105} {arXiv:2305.11105 [gr-qc]} \BibitemShut {NoStop}%
\bibitem [{\citenamefont {Soares}\ \emph {et~al.}(2025)\citenamefont {Soares}, \citenamefont {Pereira}, \citenamefont {Vit\'oria}, \citenamefont {Silva},\ and\ \citenamefont {Belich}}]{Soares:2025hpy}%
  \BibitemOpen
  \bibfield  {author} {\bibinfo {author} {\bibfnamefont {A.~R.}\ \bibnamefont {Soares}}, \bibinfo {author} {\bibfnamefont {C.~F.~S.}\ \bibnamefont {Pereira}}, \bibinfo {author} {\bibfnamefont {R.~L.~L.}\ \bibnamefont {Vit\'oria}}, \bibinfo {author} {\bibfnamefont {M.~V. d.~S.}\ \bibnamefont {Silva}}, \ and\ \bibinfo {author} {\bibfnamefont {H.}~\bibnamefont {Belich}},\ }\href@noop {} {\  (\bibinfo {year} {2025})},\ \Eprint {http://arxiv.org/abs/2503.06373} {arXiv:2503.06373 [gr-qc]} \BibitemShut {NoStop}%
\bibitem [{\citenamefont {Kumar~Walia}(2023)}]{KumarWalia:2022ddq}%
  \BibitemOpen
  \bibfield  {author} {\bibinfo {author} {\bibfnamefont {R.}~\bibnamefont {Kumar~Walia}},\ }\href {\doibase 10.1088/1475-7516/2023/03/029} {\bibfield  {journal} {\bibinfo  {journal} {JCAP}\ }\textbf {\bibinfo {volume} {03}},\ \bibinfo {pages} {029} (\bibinfo {year} {2023})},\ \Eprint {http://arxiv.org/abs/2207.02106} {arXiv:2207.02106 [gr-qc]} \BibitemShut {NoStop}%
\bibitem [{\citenamefont {Cunha}\ and\ \citenamefont {Herdeiro}(2018)}]{cunha2018}%
  \BibitemOpen
  \bibfield  {author} {\bibinfo {author} {\bibfnamefont {P.~V.~P.}\ \bibnamefont {Cunha}}\ and\ \bibinfo {author} {\bibfnamefont {C.~A.~R.}\ \bibnamefont {Herdeiro}},\ }\href {\doibase 10.1007/s10714-018-2361-9} {\bibfield  {journal} {\bibinfo  {journal} {Gen. Rel. Grav.}\ }\textbf {\bibinfo {volume} {50}},\ \bibinfo {pages} {42} (\bibinfo {year} {2018})},\ \Eprint {http://arxiv.org/abs/1801.00860} {arXiv:1801.00860 [gr-qc]} \BibitemShut {NoStop}%
\bibitem [{\citenamefont {Cardoso}\ and\ \citenamefont {Pani}(2019)}]{cardoso2019}%
  \BibitemOpen
  \bibfield  {author} {\bibinfo {author} {\bibfnamefont {V.}~\bibnamefont {Cardoso}}\ and\ \bibinfo {author} {\bibfnamefont {P.}~\bibnamefont {Pani}},\ }\href {\doibase 10.1007/s41114-019-0020-4} {\bibfield  {journal} {\bibinfo  {journal} {Living Rev. Rel.}\ }\textbf {\bibinfo {volume} {22}},\ \bibinfo {pages} {4} (\bibinfo {year} {2019})},\ \Eprint {http://arxiv.org/abs/1904.05363} {arXiv:1904.05363 [gr-qc]} \BibitemShut {NoStop}%
\bibitem [{\citenamefont {Bambi}\ and\ \citenamefont {Freese}(2009)}]{bambi2009}%
  \BibitemOpen
  \bibfield  {author} {\bibinfo {author} {\bibfnamefont {C.}~\bibnamefont {Bambi}}\ and\ \bibinfo {author} {\bibfnamefont {K.}~\bibnamefont {Freese}},\ }\href {\doibase 10.1103/PhysRevD.79.043002} {\bibfield  {journal} {\bibinfo  {journal} {Phys. Rev. D}\ }\textbf {\bibinfo {volume} {79}},\ \bibinfo {pages} {043002} (\bibinfo {year} {2009})},\ \Eprint {http://arxiv.org/abs/0812.1328} {arXiv:0812.1328 [astro-ph]} \BibitemShut {NoStop}%
\bibitem [{\citenamefont {Johannsen}\ and\ \citenamefont {Psaltis}(2010)}]{johannsen2010}%
  \BibitemOpen
  \bibfield  {author} {\bibinfo {author} {\bibfnamefont {T.}~\bibnamefont {Johannsen}}\ and\ \bibinfo {author} {\bibfnamefont {D.}~\bibnamefont {Psaltis}},\ }\href {\doibase 10.1088/0004-637X/718/1/446} {\bibfield  {journal} {\bibinfo  {journal} {Astrophys. J.}\ }\textbf {\bibinfo {volume} {718}},\ \bibinfo {pages} {446} (\bibinfo {year} {2010})},\ \Eprint {http://arxiv.org/abs/1005.1931} {arXiv:1005.1931 [astro-ph.HE]} \BibitemShut {NoStop}%
\bibitem [{\citenamefont {Vagnozzi}\ and\ \citenamefont {Visinelli}(2019)}]{Vagnozzi:2019apd}%
  \BibitemOpen
  \bibfield  {author} {\bibinfo {author} {\bibfnamefont {S.}~\bibnamefont {Vagnozzi}}\ and\ \bibinfo {author} {\bibfnamefont {L.}~\bibnamefont {Visinelli}},\ }\href {\doibase 10.1103/PhysRevD.100.024020} {\bibfield  {journal} {\bibinfo  {journal} {Phys. Rev. D}\ }\textbf {\bibinfo {volume} {100}},\ \bibinfo {pages} {024020} (\bibinfo {year} {2019})},\ \Eprint {http://arxiv.org/abs/1905.12421} {arXiv:1905.12421 [gr-qc]} \BibitemShut {NoStop}%
\bibitem [{\citenamefont {Cardoso}\ and\ \citenamefont {Gualtieri}(2016)}]{Cardoso:2016ryw}%
  \BibitemOpen
  \bibfield  {author} {\bibinfo {author} {\bibfnamefont {V.}~\bibnamefont {Cardoso}}\ and\ \bibinfo {author} {\bibfnamefont {L.}~\bibnamefont {Gualtieri}},\ }\href {\doibase 10.1088/0264-9381/33/17/174001} {\bibfield  {journal} {\bibinfo  {journal} {Class. Quant. Grav.}\ }\textbf {\bibinfo {volume} {33}},\ \bibinfo {pages} {174001} (\bibinfo {year} {2016})},\ \Eprint {http://arxiv.org/abs/1607.03133} {arXiv:1607.03133 [gr-qc]} \BibitemShut {NoStop}%
\bibitem [{\citenamefont {Khodadi}\ \emph {et~al.}(2020)\citenamefont {Khodadi}, \citenamefont {Allahyari}, \citenamefont {Vagnozzi},\ and\ \citenamefont {Mota}}]{Khodadi:2020jij}%
  \BibitemOpen
  \bibfield  {author} {\bibinfo {author} {\bibfnamefont {M.}~\bibnamefont {Khodadi}}, \bibinfo {author} {\bibfnamefont {A.}~\bibnamefont {Allahyari}}, \bibinfo {author} {\bibfnamefont {S.}~\bibnamefont {Vagnozzi}}, \ and\ \bibinfo {author} {\bibfnamefont {D.~F.}\ \bibnamefont {Mota}},\ }\href {\doibase 10.1088/1475-7516/2020/09/026} {\bibfield  {journal} {\bibinfo  {journal} {JCAP}\ }\textbf {\bibinfo {volume} {09}},\ \bibinfo {pages} {026} (\bibinfo {year} {2020})},\ \Eprint {http://arxiv.org/abs/2005.05992} {arXiv:2005.05992 [gr-qc]} \BibitemShut {NoStop}%
\bibitem [{\citenamefont {Khodadi}\ \emph {et~al.}(2021)\citenamefont {Khodadi}, \citenamefont {Lambiase},\ and\ \citenamefont {Mota}}]{Khodadi:2021gbc}%
  \BibitemOpen
  \bibfield  {author} {\bibinfo {author} {\bibfnamefont {M.}~\bibnamefont {Khodadi}}, \bibinfo {author} {\bibfnamefont {G.}~\bibnamefont {Lambiase}}, \ and\ \bibinfo {author} {\bibfnamefont {D.~F.}\ \bibnamefont {Mota}},\ }\href {\doibase 10.1088/1475-7516/2021/09/028} {\bibfield  {journal} {\bibinfo  {journal} {JCAP}\ }\textbf {\bibinfo {volume} {09}},\ \bibinfo {pages} {028} (\bibinfo {year} {2021})},\ \Eprint {http://arxiv.org/abs/2107.00834} {arXiv:2107.00834 [gr-qc]} \BibitemShut {NoStop}%
\bibitem [{\citenamefont {Afrin}\ \emph {et~al.}(2021)\citenamefont {Afrin}, \citenamefont {Kumar},\ and\ \citenamefont {Ghosh}}]{Afrin:2021imp}%
  \BibitemOpen
  \bibfield  {author} {\bibinfo {author} {\bibfnamefont {M.}~\bibnamefont {Afrin}}, \bibinfo {author} {\bibfnamefont {R.}~\bibnamefont {Kumar}}, \ and\ \bibinfo {author} {\bibfnamefont {S.~G.}\ \bibnamefont {Ghosh}},\ }\href {\doibase 10.1093/mnras/stab1260} {\bibfield  {journal} {\bibinfo  {journal} {Mon. Not. Roy. Astron. Soc.}\ }\textbf {\bibinfo {volume} {504}},\ \bibinfo {pages} {5927} (\bibinfo {year} {2021})},\ \Eprint {http://arxiv.org/abs/2103.11417} {arXiv:2103.11417 [gr-qc]} \BibitemShut {NoStop}%
\bibitem [{\citenamefont {Perlick}\ \emph {et~al.}(2015)\citenamefont {Perlick}, \citenamefont {Tsupko},\ and\ \citenamefont {Bisnovatyi-Kogan}}]{Perlick:2015vta}%
  \BibitemOpen
  \bibfield  {author} {\bibinfo {author} {\bibfnamefont {V.}~\bibnamefont {Perlick}}, \bibinfo {author} {\bibfnamefont {O.~Y.}\ \bibnamefont {Tsupko}}, \ and\ \bibinfo {author} {\bibfnamefont {G.~S.}\ \bibnamefont {Bisnovatyi-Kogan}},\ }\href {\doibase 10.1103/PhysRevD.92.104031} {\bibfield  {journal} {\bibinfo  {journal} {Phys. Rev. D}\ }\textbf {\bibinfo {volume} {92}},\ \bibinfo {pages} {104031} (\bibinfo {year} {2015})},\ \Eprint {http://arxiv.org/abs/1507.04217} {arXiv:1507.04217 [gr-qc]} \BibitemShut {NoStop}%
\bibitem [{\citenamefont {Bisnovatyi-Kogan}\ and\ \citenamefont {Tsupko}(2010)}]{Bisnovatyi-Kogan:2010flt}%
  \BibitemOpen
  \bibfield  {author} {\bibinfo {author} {\bibfnamefont {G.~S.}\ \bibnamefont {Bisnovatyi-Kogan}}\ and\ \bibinfo {author} {\bibfnamefont {O.~Y.}\ \bibnamefont {Tsupko}},\ }\href {\doibase 10.1111/j.1365-2966.2010.16290.x} {\bibfield  {journal} {\bibinfo  {journal} {Mon. Not. Roy. Astron. Soc.}\ }\textbf {\bibinfo {volume} {404}},\ \bibinfo {pages} {1790} (\bibinfo {year} {2010})},\ \Eprint {http://arxiv.org/abs/1006.2321} {arXiv:1006.2321 [astro-ph.CO]} \BibitemShut {NoStop}%
\bibitem [{\citenamefont {Bisnovatyi-Kogan}\ and\ \citenamefont {Tsupko}(2018)}]{Bisnovatyi-Kogan:2018vxl}%
  \BibitemOpen
  \bibfield  {author} {\bibinfo {author} {\bibfnamefont {G.~S.}\ \bibnamefont {Bisnovatyi-Kogan}}\ and\ \bibinfo {author} {\bibfnamefont {O.~Y.}\ \bibnamefont {Tsupko}},\ }\href {\doibase 10.1103/PhysRevD.98.084020} {\bibfield  {journal} {\bibinfo  {journal} {Phys. Rev. D}\ }\textbf {\bibinfo {volume} {98}},\ \bibinfo {pages} {084020} (\bibinfo {year} {2018})},\ \Eprint {http://arxiv.org/abs/1805.03311} {arXiv:1805.03311 [gr-qc]} \BibitemShut {NoStop}%
\bibitem [{\citenamefont {Perlick}\ and\ \citenamefont {Tsupko}(2022)}]{perlick_calculating_2022}%
  \BibitemOpen
  \bibfield  {author} {\bibinfo {author} {\bibfnamefont {V.}~\bibnamefont {Perlick}}\ and\ \bibinfo {author} {\bibfnamefont {O.~Y.}\ \bibnamefont {Tsupko}},\ }\href {\doibase 10.1016/j.physrep.2021.10.004} {\bibfield  {journal} {\bibinfo  {journal} {Phys. Rept.}\ }\textbf {\bibinfo {volume} {947}},\ \bibinfo {pages} {1} (\bibinfo {year} {2022})},\ \Eprint {http://arxiv.org/abs/2105.07101} {arXiv:2105.07101 [gr-qc]} \BibitemShut {NoStop}%
\bibitem [{\citenamefont {Zeng}\ \emph {et~al.}(2025)\citenamefont {Zeng}, \citenamefont {Li}, \citenamefont {Li}, \citenamefont {Liang},\ and\ \citenamefont {Xu}}]{Zeng:2024ptv}%
  \BibitemOpen
  \bibfield  {author} {\bibinfo {author} {\bibfnamefont {X.-X.}\ \bibnamefont {Zeng}}, \bibinfo {author} {\bibfnamefont {L.-F.}\ \bibnamefont {Li}}, \bibinfo {author} {\bibfnamefont {P.}~\bibnamefont {Li}}, \bibinfo {author} {\bibfnamefont {B.}~\bibnamefont {Liang}}, \ and\ \bibinfo {author} {\bibfnamefont {P.}~\bibnamefont {Xu}},\ }\href {\doibase 10.1007/s11433-024-2526-4} {\bibfield  {journal} {\bibinfo  {journal} {Sci. China Phys. Mech. Astron.}\ }\textbf {\bibinfo {volume} {68}},\ \bibinfo {pages} {220412} (\bibinfo {year} {2025})},\ \Eprint {http://arxiv.org/abs/2411.12528} {arXiv:2411.12528 [gr-qc]} \BibitemShut {NoStop}%
\bibitem [{\citenamefont {Yang}\ \emph {et~al.}(2025)\citenamefont {Yang}, \citenamefont {Aslam}, \citenamefont {Zeng},\ and\ \citenamefont {Saleem}}]{Yang:2024nin}%
  \BibitemOpen
  \bibfield  {author} {\bibinfo {author} {\bibfnamefont {C.-Y.}\ \bibnamefont {Yang}}, \bibinfo {author} {\bibfnamefont {M.~I.}\ \bibnamefont {Aslam}}, \bibinfo {author} {\bibfnamefont {X.-X.}\ \bibnamefont {Zeng}}, \ and\ \bibinfo {author} {\bibfnamefont {R.}~\bibnamefont {Saleem}},\ }\href {\doibase 10.1016/j.jheap.2025.01.017} {\bibfield  {journal} {\bibinfo  {journal} {JHEAp}\ }\textbf {\bibinfo {volume} {46}},\ \bibinfo {pages} {345} (\bibinfo {year} {2025})},\ \Eprint {http://arxiv.org/abs/2411.11807} {arXiv:2411.11807 [astro-ph.HE]} \BibitemShut {NoStop}%
\bibitem [{\citenamefont {He}\ \emph {et~al.}(2024)\citenamefont {He}, \citenamefont {Li}, \citenamefont {Yang},\ and\ \citenamefont {Zeng}}]{He:2024amh}%
  \BibitemOpen
  \bibfield  {author} {\bibinfo {author} {\bibfnamefont {K.-J.}\ \bibnamefont {He}}, \bibinfo {author} {\bibfnamefont {G.-P.}\ \bibnamefont {Li}}, \bibinfo {author} {\bibfnamefont {C.-Y.}\ \bibnamefont {Yang}}, \ and\ \bibinfo {author} {\bibfnamefont {X.-X.}\ \bibnamefont {Zeng}},\ }\href@noop {} {\  (\bibinfo {year} {2024})},\ \Eprint {http://arxiv.org/abs/2411.11680} {arXiv:2411.11680 [astro-ph.HE]} \BibitemShut {NoStop}%
\bibitem [{\citenamefont {He}\ \emph {et~al.}(2025)\citenamefont {He}, \citenamefont {Yang},\ and\ \citenamefont {Zeng}}]{He:2025rjq}%
  \BibitemOpen
  \bibfield  {author} {\bibinfo {author} {\bibfnamefont {K.-J.}\ \bibnamefont {He}}, \bibinfo {author} {\bibfnamefont {C.-Y.}\ \bibnamefont {Yang}}, \ and\ \bibinfo {author} {\bibfnamefont {X.-X.}\ \bibnamefont {Zeng}},\ }\href@noop {} {\  (\bibinfo {year} {2025})},\ \Eprint {http://arxiv.org/abs/2501.06778} {arXiv:2501.06778 [astro-ph.HE]} \BibitemShut {NoStop}%
\bibitem [{\citenamefont {Li}\ \emph {et~al.}(2020)\citenamefont {Li}, \citenamefont {Guo},\ and\ \citenamefont {Chen}}]{Li:2020drn}%
  \BibitemOpen
  \bibfield  {author} {\bibinfo {author} {\bibfnamefont {P.-C.}\ \bibnamefont {Li}}, \bibinfo {author} {\bibfnamefont {M.}~\bibnamefont {Guo}}, \ and\ \bibinfo {author} {\bibfnamefont {B.}~\bibnamefont {Chen}},\ }\href {\doibase 10.1103/PhysRevD.101.084041} {\bibfield  {journal} {\bibinfo  {journal} {Phys. Rev. D}\ }\textbf {\bibinfo {volume} {101}},\ \bibinfo {pages} {084041} (\bibinfo {year} {2020})},\ \Eprint {http://arxiv.org/abs/2001.04231} {arXiv:2001.04231 [gr-qc]} \BibitemShut {NoStop}%
\bibitem [{\citenamefont {Hu}\ \emph {et~al.}(2021)\citenamefont {Hu}, \citenamefont {Zhong}, \citenamefont {Li}, \citenamefont {Guo},\ and\ \citenamefont {Chen}}]{Hu:2020usx}%
  \BibitemOpen
  \bibfield  {author} {\bibinfo {author} {\bibfnamefont {Z.}~\bibnamefont {Hu}}, \bibinfo {author} {\bibfnamefont {Z.}~\bibnamefont {Zhong}}, \bibinfo {author} {\bibfnamefont {P.-C.}\ \bibnamefont {Li}}, \bibinfo {author} {\bibfnamefont {M.}~\bibnamefont {Guo}}, \ and\ \bibinfo {author} {\bibfnamefont {B.}~\bibnamefont {Chen}},\ }\href {\doibase 10.1103/PhysRevD.103.044057} {\bibfield  {journal} {\bibinfo  {journal} {Phys. Rev. D}\ }\textbf {\bibinfo {volume} {103}},\ \bibinfo {pages} {044057} (\bibinfo {year} {2021})},\ \Eprint {http://arxiv.org/abs/2012.07022} {arXiv:2012.07022 [gr-qc]} \BibitemShut {NoStop}%
\bibitem [{\citenamefont {Guo}\ and\ \citenamefont {Li}(2020)}]{Guo:2020zmf}%
  \BibitemOpen
  \bibfield  {author} {\bibinfo {author} {\bibfnamefont {M.}~\bibnamefont {Guo}}\ and\ \bibinfo {author} {\bibfnamefont {P.-C.}\ \bibnamefont {Li}},\ }\href {\doibase 10.1140/epjc/s10052-020-8164-7} {\bibfield  {journal} {\bibinfo  {journal} {Eur. Phys. J. C}\ }\textbf {\bibinfo {volume} {80}},\ \bibinfo {pages} {588} (\bibinfo {year} {2020})},\ \Eprint {http://arxiv.org/abs/2003.02523} {arXiv:2003.02523 [gr-qc]} \BibitemShut {NoStop}%
\bibitem [{\citenamefont {Allahyari}\ \emph {et~al.}(2020)\citenamefont {Allahyari}, \citenamefont {Khodadi}, \citenamefont {Vagnozzi},\ and\ \citenamefont {Mota}}]{Allahyari:2019jqz}%
  \BibitemOpen
  \bibfield  {author} {\bibinfo {author} {\bibfnamefont {A.}~\bibnamefont {Allahyari}}, \bibinfo {author} {\bibfnamefont {M.}~\bibnamefont {Khodadi}}, \bibinfo {author} {\bibfnamefont {S.}~\bibnamefont {Vagnozzi}}, \ and\ \bibinfo {author} {\bibfnamefont {D.~F.}\ \bibnamefont {Mota}},\ }\href {\doibase 10.1088/1475-7516/2020/02/003} {\bibfield  {journal} {\bibinfo  {journal} {JCAP}\ }\textbf {\bibinfo {volume} {02}},\ \bibinfo {pages} {003} (\bibinfo {year} {2020})},\ \Eprint {http://arxiv.org/abs/1912.08231} {arXiv:1912.08231 [gr-qc]} \BibitemShut {NoStop}%
\bibitem [{\citenamefont {Belhaj}\ \emph {et~al.}(2021)\citenamefont {Belhaj}, \citenamefont {Belmahi}, \citenamefont {Benali}, \citenamefont {El~Hadri}, \citenamefont {El~Moumni},\ and\ \citenamefont {Torrente-Lujan}}]{Belhaj:2020okh}%
  \BibitemOpen
  \bibfield  {author} {\bibinfo {author} {\bibfnamefont {A.}~\bibnamefont {Belhaj}}, \bibinfo {author} {\bibfnamefont {H.}~\bibnamefont {Belmahi}}, \bibinfo {author} {\bibfnamefont {M.}~\bibnamefont {Benali}}, \bibinfo {author} {\bibfnamefont {W.}~\bibnamefont {El~Hadri}}, \bibinfo {author} {\bibfnamefont {H.}~\bibnamefont {El~Moumni}}, \ and\ \bibinfo {author} {\bibfnamefont {E.}~\bibnamefont {Torrente-Lujan}},\ }\href {\doibase 10.1016/j.physletb.2020.136025} {\bibfield  {journal} {\bibinfo  {journal} {Phys. Lett. B}\ }\textbf {\bibinfo {volume} {812}},\ \bibinfo {pages} {136025} (\bibinfo {year} {2021})},\ \Eprint {http://arxiv.org/abs/2008.13478} {arXiv:2008.13478 [hep-th]} \BibitemShut {NoStop}%
\bibitem [{\citenamefont {Okyay}\ and\ \citenamefont {\"Ovg\"un}(2022)}]{Okyay:2021nnh}%
  \BibitemOpen
  \bibfield  {author} {\bibinfo {author} {\bibfnamefont {M.}~\bibnamefont {Okyay}}\ and\ \bibinfo {author} {\bibfnamefont {A.}~\bibnamefont {\"Ovg\"un}},\ }\href {\doibase 10.1088/1475-7516/2022/01/009} {\bibfield  {journal} {\bibinfo  {journal} {JCAP}\ }\textbf {\bibinfo {volume} {01}},\ \bibinfo {pages} {009} (\bibinfo {year} {2022})},\ \Eprint {http://arxiv.org/abs/2108.07766} {arXiv:2108.07766 [gr-qc]} \BibitemShut {NoStop}%
\bibitem [{\citenamefont {\"Ovg\"un}(2021)}]{Ovgun:2021ttv}%
  \BibitemOpen
  \bibfield  {author} {\bibinfo {author} {\bibfnamefont {A.}~\bibnamefont {\"Ovg\"un}},\ }\href {\doibase 10.1016/j.physletb.2021.136517} {\bibfield  {journal} {\bibinfo  {journal} {Phys. Lett. B}\ }\textbf {\bibinfo {volume} {820}},\ \bibinfo {pages} {136517} (\bibinfo {year} {2021})},\ \Eprint {http://arxiv.org/abs/2105.05035} {arXiv:2105.05035 [gr-qc]} \BibitemShut {NoStop}%
\bibitem [{\citenamefont {Kuang}\ \emph {et~al.}(2022)\citenamefont {Kuang}, \citenamefont {Tang}, \citenamefont {Wang},\ and\ \citenamefont {Wang}}]{Kuang:2022ojj}%
  \BibitemOpen
  \bibfield  {author} {\bibinfo {author} {\bibfnamefont {X.-M.}\ \bibnamefont {Kuang}}, \bibinfo {author} {\bibfnamefont {Z.-Y.}\ \bibnamefont {Tang}}, \bibinfo {author} {\bibfnamefont {B.}~\bibnamefont {Wang}}, \ and\ \bibinfo {author} {\bibfnamefont {A.}~\bibnamefont {Wang}},\ }\href {\doibase 10.1103/PhysRevD.106.064012} {\bibfield  {journal} {\bibinfo  {journal} {Phys. Rev. D}\ }\textbf {\bibinfo {volume} {106}},\ \bibinfo {pages} {064012} (\bibinfo {year} {2022})},\ \Eprint {http://arxiv.org/abs/2206.05878} {arXiv:2206.05878 [gr-qc]} \BibitemShut {NoStop}%
\bibitem [{\citenamefont {Pantig}\ \emph {et~al.}(2022)\citenamefont {Pantig}, \citenamefont {Mastrototaro}, \citenamefont {Lambiase},\ and\ \citenamefont {\"Ovg\"un}}]{Pantig:2022gih}%
  \BibitemOpen
  \bibfield  {author} {\bibinfo {author} {\bibfnamefont {R.~C.}\ \bibnamefont {Pantig}}, \bibinfo {author} {\bibfnamefont {L.}~\bibnamefont {Mastrototaro}}, \bibinfo {author} {\bibfnamefont {G.}~\bibnamefont {Lambiase}}, \ and\ \bibinfo {author} {\bibfnamefont {A.}~\bibnamefont {\"Ovg\"un}},\ }\href {\doibase 10.1140/epjc/s10052-022-11125-y} {\bibfield  {journal} {\bibinfo  {journal} {Eur. Phys. J. C}\ }\textbf {\bibinfo {volume} {82}},\ \bibinfo {pages} {1155} (\bibinfo {year} {2022})},\ \Eprint {http://arxiv.org/abs/2208.06664} {arXiv:2208.06664 [gr-qc]} \BibitemShut {NoStop}%
\bibitem [{\citenamefont {Puli\c{c}e}\ \emph {et~al.}(2023)\citenamefont {Puli\c{c}e}, \citenamefont {Pantig}, \citenamefont {\"Ovg\"un},\ and\ \citenamefont {Demir}}]{Pulice:2023dqw}%
  \BibitemOpen
  \bibfield  {author} {\bibinfo {author} {\bibfnamefont {B.}~\bibnamefont {Puli\c{c}e}}, \bibinfo {author} {\bibfnamefont {R.~C.}\ \bibnamefont {Pantig}}, \bibinfo {author} {\bibfnamefont {A.}~\bibnamefont {\"Ovg\"un}}, \ and\ \bibinfo {author} {\bibfnamefont {D.}~\bibnamefont {Demir}},\ }\href {\doibase 10.1088/1361-6382/acf08c} {\bibfield  {journal} {\bibinfo  {journal} {Class. Quant. Grav.}\ }\textbf {\bibinfo {volume} {40}},\ \bibinfo {pages} {195003} (\bibinfo {year} {2023})},\ \Eprint {http://arxiv.org/abs/2308.08415} {arXiv:2308.08415 [gr-qc]} \BibitemShut {NoStop}%
\bibitem [{\citenamefont {Shaikh}\ \emph {et~al.}(2019)\citenamefont {Shaikh}, \citenamefont {Kocherlakota}, \citenamefont {Narayan},\ and\ \citenamefont {Joshi}}]{Shaikh:2018lcc}%
  \BibitemOpen
  \bibfield  {author} {\bibinfo {author} {\bibfnamefont {R.}~\bibnamefont {Shaikh}}, \bibinfo {author} {\bibfnamefont {P.}~\bibnamefont {Kocherlakota}}, \bibinfo {author} {\bibfnamefont {R.}~\bibnamefont {Narayan}}, \ and\ \bibinfo {author} {\bibfnamefont {P.~S.}\ \bibnamefont {Joshi}},\ }\href {\doibase 10.1093/mnras/sty2624} {\bibfield  {journal} {\bibinfo  {journal} {Mon. Not. Roy. Astron. Soc.}\ }\textbf {\bibinfo {volume} {482}},\ \bibinfo {pages} {52} (\bibinfo {year} {2019})},\ \Eprint {http://arxiv.org/abs/1802.08060} {arXiv:1802.08060 [astro-ph.HE]} \BibitemShut {NoStop}%
\bibitem [{\citenamefont {Joshi}\ \emph {et~al.}(2020)\citenamefont {Joshi}, \citenamefont {Dey}, \citenamefont {Joshi},\ and\ \citenamefont {Bambhaniya}}]{Joshi:2020tlq}%
  \BibitemOpen
  \bibfield  {author} {\bibinfo {author} {\bibfnamefont {A.~B.}\ \bibnamefont {Joshi}}, \bibinfo {author} {\bibfnamefont {D.}~\bibnamefont {Dey}}, \bibinfo {author} {\bibfnamefont {P.~S.}\ \bibnamefont {Joshi}}, \ and\ \bibinfo {author} {\bibfnamefont {P.}~\bibnamefont {Bambhaniya}},\ }\href {\doibase 10.1103/PhysRevD.102.024022} {\bibfield  {journal} {\bibinfo  {journal} {Phys. Rev. D}\ }\textbf {\bibinfo {volume} {102}},\ \bibinfo {pages} {024022} (\bibinfo {year} {2020})},\ \Eprint {http://arxiv.org/abs/2004.06525} {arXiv:2004.06525 [gr-qc]} \BibitemShut {NoStop}%
\bibitem [{\citenamefont {Gyulchev}\ \emph {et~al.}(2018)\citenamefont {Gyulchev}, \citenamefont {Nedkova}, \citenamefont {Tinchev},\ and\ \citenamefont {Yazadjiev}}]{Gyulchev:2018fmd}%
  \BibitemOpen
  \bibfield  {author} {\bibinfo {author} {\bibfnamefont {G.}~\bibnamefont {Gyulchev}}, \bibinfo {author} {\bibfnamefont {P.}~\bibnamefont {Nedkova}}, \bibinfo {author} {\bibfnamefont {V.}~\bibnamefont {Tinchev}}, \ and\ \bibinfo {author} {\bibfnamefont {S.}~\bibnamefont {Yazadjiev}},\ }\href {\doibase 10.1140/epjc/s10052-018-6012-9} {\bibfield  {journal} {\bibinfo  {journal} {Eur. Phys. J. C}\ }\textbf {\bibinfo {volume} {78}},\ \bibinfo {pages} {544} (\bibinfo {year} {2018})},\ \Eprint {http://arxiv.org/abs/1805.11591} {arXiv:1805.11591 [gr-qc]} \BibitemShut {NoStop}%
\bibitem [{\citenamefont {Atamurotov}\ \emph {et~al.}(2013)\citenamefont {Atamurotov}, \citenamefont {Abdujabbarov},\ and\ \citenamefont {Ahmedov}}]{Atamurotov:2013sca}%
  \BibitemOpen
  \bibfield  {author} {\bibinfo {author} {\bibfnamefont {F.}~\bibnamefont {Atamurotov}}, \bibinfo {author} {\bibfnamefont {A.}~\bibnamefont {Abdujabbarov}}, \ and\ \bibinfo {author} {\bibfnamefont {B.}~\bibnamefont {Ahmedov}},\ }\href {\doibase 10.1103/PhysRevD.88.064004} {\bibfield  {journal} {\bibinfo  {journal} {Phys. Rev. D}\ }\textbf {\bibinfo {volume} {88}},\ \bibinfo {pages} {064004} (\bibinfo {year} {2013})}\BibitemShut {NoStop}%
\bibitem [{\citenamefont {Zeng}\ \emph {et~al.}(2020)\citenamefont {Zeng}, \citenamefont {Zhang},\ and\ \citenamefont {Zhang}}]{Zeng:2020dco}%
  \BibitemOpen
  \bibfield  {author} {\bibinfo {author} {\bibfnamefont {X.-X.}\ \bibnamefont {Zeng}}, \bibinfo {author} {\bibfnamefont {H.-Q.}\ \bibnamefont {Zhang}}, \ and\ \bibinfo {author} {\bibfnamefont {H.}~\bibnamefont {Zhang}},\ }\href {\doibase 10.1140/epjc/s10052-020-08449-y} {\bibfield  {journal} {\bibinfo  {journal} {Eur. Phys. J. C}\ }\textbf {\bibinfo {volume} {80}},\ \bibinfo {pages} {872} (\bibinfo {year} {2020})},\ \Eprint {http://arxiv.org/abs/2004.12074} {arXiv:2004.12074 [gr-qc]} \BibitemShut {NoStop}%
\bibitem [{\citenamefont {Cunha}\ and\ \citenamefont {Herdeiro}(2020)}]{Cunha:2020azh}%
  \BibitemOpen
  \bibfield  {author} {\bibinfo {author} {\bibfnamefont {P.~V.~P.}\ \bibnamefont {Cunha}}\ and\ \bibinfo {author} {\bibfnamefont {C.~A.~R.}\ \bibnamefont {Herdeiro}},\ }\href {\doibase 10.1103/PhysRevLett.124.181101} {\bibfield  {journal} {\bibinfo  {journal} {Phys. Rev. Lett.}\ }\textbf {\bibinfo {volume} {124}},\ \bibinfo {pages} {181101} (\bibinfo {year} {2020})},\ \Eprint {http://arxiv.org/abs/2003.06445} {arXiv:2003.06445 [gr-qc]} \BibitemShut {NoStop}%
\bibitem [{\citenamefont {Cunha}\ \emph {et~al.}(2017)\citenamefont {Cunha}, \citenamefont {Herdeiro}, \citenamefont {Kleihaus}, \citenamefont {Kunz},\ and\ \citenamefont {Radu}}]{Cunha:2016wzk}%
  \BibitemOpen
  \bibfield  {author} {\bibinfo {author} {\bibfnamefont {P.~V.~P.}\ \bibnamefont {Cunha}}, \bibinfo {author} {\bibfnamefont {C.~A.~R.}\ \bibnamefont {Herdeiro}}, \bibinfo {author} {\bibfnamefont {B.}~\bibnamefont {Kleihaus}}, \bibinfo {author} {\bibfnamefont {J.}~\bibnamefont {Kunz}}, \ and\ \bibinfo {author} {\bibfnamefont {E.}~\bibnamefont {Radu}},\ }\href {\doibase 10.1016/j.physletb.2017.03.020} {\bibfield  {journal} {\bibinfo  {journal} {Phys. Lett. B}\ }\textbf {\bibinfo {volume} {768}},\ \bibinfo {pages} {373} (\bibinfo {year} {2017})},\ \Eprint {http://arxiv.org/abs/1701.00079} {arXiv:1701.00079 [gr-qc]} \BibitemShut {NoStop}%
\bibitem [{\citenamefont {Wei}\ \emph {et~al.}(2019)\citenamefont {Wei}, \citenamefont {Liu},\ and\ \citenamefont {Mann}}]{Wei:2018xks}%
  \BibitemOpen
  \bibfield  {author} {\bibinfo {author} {\bibfnamefont {S.-W.}\ \bibnamefont {Wei}}, \bibinfo {author} {\bibfnamefont {Y.-X.}\ \bibnamefont {Liu}}, \ and\ \bibinfo {author} {\bibfnamefont {R.~B.}\ \bibnamefont {Mann}},\ }\href {\doibase 10.1103/PhysRevD.99.041303} {\bibfield  {journal} {\bibinfo  {journal} {Phys. Rev. D}\ }\textbf {\bibinfo {volume} {99}},\ \bibinfo {pages} {041303} (\bibinfo {year} {2019})},\ \Eprint {http://arxiv.org/abs/1811.00047} {arXiv:1811.00047 [gr-qc]} \BibitemShut {NoStop}%
\bibitem [{\citenamefont {Jafarzade}\ \emph {et~al.}(2021)\citenamefont {Jafarzade}, \citenamefont {Kord~Zangeneh},\ and\ \citenamefont {Lobo}}]{Jafarzade:2020ova}%
  \BibitemOpen
  \bibfield  {author} {\bibinfo {author} {\bibfnamefont {K.}~\bibnamefont {Jafarzade}}, \bibinfo {author} {\bibfnamefont {M.}~\bibnamefont {Kord~Zangeneh}}, \ and\ \bibinfo {author} {\bibfnamefont {F.~S.~N.}\ \bibnamefont {Lobo}},\ }\href {\doibase 10.1088/1475-7516/2021/04/008} {\bibfield  {journal} {\bibinfo  {journal} {JCAP}\ }\textbf {\bibinfo {volume} {04}},\ \bibinfo {pages} {008} (\bibinfo {year} {2021})},\ \Eprint {http://arxiv.org/abs/2010.05755} {arXiv:2010.05755 [gr-qc]} \BibitemShut {NoStop}%
\bibitem [{\citenamefont {Anacleto}\ \emph {et~al.}(2021)\citenamefont {Anacleto}, \citenamefont {Campos}, \citenamefont {Brito},\ and\ \citenamefont {Passos}}]{Anacleto:2021qoe}%
  \BibitemOpen
  \bibfield  {author} {\bibinfo {author} {\bibfnamefont {M.~A.}\ \bibnamefont {Anacleto}}, \bibinfo {author} {\bibfnamefont {J.~A.~V.}\ \bibnamefont {Campos}}, \bibinfo {author} {\bibfnamefont {F.~A.}\ \bibnamefont {Brito}}, \ and\ \bibinfo {author} {\bibfnamefont {E.}~\bibnamefont {Passos}},\ }\href {\doibase 10.1016/j.aop.2021.168662} {\bibfield  {journal} {\bibinfo  {journal} {Annals Phys.}\ }\textbf {\bibinfo {volume} {434}},\ \bibinfo {pages} {168662} (\bibinfo {year} {2021})},\ \Eprint {http://arxiv.org/abs/2108.04998} {arXiv:2108.04998 [gr-qc]} \BibitemShut {NoStop}%
\bibitem [{\citenamefont {Gogoi}\ \emph {et~al.}(2024)\citenamefont {Gogoi}, \citenamefont {Heidari}, \citenamefont {Kriz},\ and\ \citenamefont {Hassanabadi}}]{Gogoi:2023lvw}%
  \BibitemOpen
  \bibfield  {author} {\bibinfo {author} {\bibfnamefont {D.~J.}\ \bibnamefont {Gogoi}}, \bibinfo {author} {\bibfnamefont {N.}~\bibnamefont {Heidari}}, \bibinfo {author} {\bibfnamefont {J.}~\bibnamefont {Kriz}}, \ and\ \bibinfo {author} {\bibfnamefont {H.}~\bibnamefont {Hassanabadi}},\ }\href {\doibase 10.1002/prop.202300245} {\bibfield  {journal} {\bibinfo  {journal} {Fortsch. Phys.}\ }\textbf {\bibinfo {volume} {72}},\ \bibinfo {pages} {2300245} (\bibinfo {year} {2024})},\ \Eprint {http://arxiv.org/abs/2307.09976} {arXiv:2307.09976 [gr-qc]} \BibitemShut {NoStop}%
\bibitem [{\citenamefont {Ara\'ujo~Filho}(2025)}]{AraujoFilho:2024lsi}%
  \BibitemOpen
  \bibfield  {author} {\bibinfo {author} {\bibfnamefont {A.~A.}\ \bibnamefont {Ara\'ujo~Filho}},\ }\href {\doibase 10.1088/1475-7516/2025/01/072} {\bibfield  {journal} {\bibinfo  {journal} {JCAP}\ }\textbf {\bibinfo {volume} {01}},\ \bibinfo {pages} {072} (\bibinfo {year} {2025})},\ \Eprint {http://arxiv.org/abs/2410.23165} {arXiv:2410.23165 [gr-qc]} \BibitemShut {NoStop}%
\bibitem [{\citenamefont {Wei}\ \emph {et~al.}(2015)\citenamefont {Wei}, \citenamefont {Cheng}, \citenamefont {Zhong},\ and\ \citenamefont {Zhou}}]{Wei:2015dua}%
  \BibitemOpen
  \bibfield  {author} {\bibinfo {author} {\bibfnamefont {S.-W.}\ \bibnamefont {Wei}}, \bibinfo {author} {\bibfnamefont {P.}~\bibnamefont {Cheng}}, \bibinfo {author} {\bibfnamefont {Y.}~\bibnamefont {Zhong}}, \ and\ \bibinfo {author} {\bibfnamefont {X.-N.}\ \bibnamefont {Zhou}},\ }\href {\doibase 10.1088/1475-7516/2015/08/004} {\bibfield  {journal} {\bibinfo  {journal} {JCAP}\ }\textbf {\bibinfo {volume} {08}},\ \bibinfo {pages} {004} (\bibinfo {year} {2015})},\ \Eprint {http://arxiv.org/abs/1501.06298} {arXiv:1501.06298 [gr-qc]} \BibitemShut {NoStop}%
\bibitem [{\citenamefont {Konoplya}(2019)}]{Konoplya:2019sns}%
  \BibitemOpen
  \bibfield  {author} {\bibinfo {author} {\bibfnamefont {R.~A.}\ \bibnamefont {Konoplya}},\ }\href {\doibase 10.1016/j.physletb.2019.05.043} {\bibfield  {journal} {\bibinfo  {journal} {Phys. Lett. B}\ }\textbf {\bibinfo {volume} {795}},\ \bibinfo {pages} {1} (\bibinfo {year} {2019})},\ \Eprint {http://arxiv.org/abs/1905.00064} {arXiv:1905.00064 [gr-qc]} \BibitemShut {NoStop}%
\bibitem [{\citenamefont {{Koivisto}}(2023)}]{2023IJGMM..2050040K}%
  \BibitemOpen
  \bibfield  {author} {\bibinfo {author} {\bibfnamefont {T.}~\bibnamefont {{Koivisto}}},\ }\href {\doibase 10.1142/S0219887824500403} {\bibfield  {journal} {\bibinfo  {journal} {International Journal of Geometric Methods in Modern Physics}\ }\textbf {\bibinfo {volume} {20}},\ \bibinfo {eid} {2450040-362} (\bibinfo {year} {2023})},\ \Eprint {http://arxiv.org/abs/2306.00963} {arXiv:2306.00963 [gr-qc]} \BibitemShut {NoStop}%
\bibitem [{\citenamefont {{Wiesendanger}}(2019)}]{2019CQGra..36f5015W}%
  \BibitemOpen
  \bibfield  {author} {\bibinfo {author} {\bibfnamefont {C.}~\bibnamefont {{Wiesendanger}}},\ }\href {\doibase 10.1088/1361-6382/ab04e9} {\bibfield  {journal} {\bibinfo  {journal} {Classical and Quantum Gravity}\ }\textbf {\bibinfo {volume} {36}},\ \bibinfo {eid} {065015} (\bibinfo {year} {2019})},\ \Eprint {http://arxiv.org/abs/1806.05037} {arXiv:1806.05037 [gr-qc]} \BibitemShut {NoStop}%
\bibitem [{\citenamefont {Abdelqader}\ and\ \citenamefont {Lake}(2015)}]{Abdelqader:2014vaa}%
  \BibitemOpen
  \bibfield  {author} {\bibinfo {author} {\bibfnamefont {M.}~\bibnamefont {Abdelqader}}\ and\ \bibinfo {author} {\bibfnamefont {K.}~\bibnamefont {Lake}},\ }\href {\doibase 10.1103/PhysRevD.91.084017} {\bibfield  {journal} {\bibinfo  {journal} {Phys. Rev. D}\ }\textbf {\bibinfo {volume} {91}},\ \bibinfo {pages} {084017} (\bibinfo {year} {2015})},\ \Eprint {http://arxiv.org/abs/1412.8757} {arXiv:1412.8757 [gr-qc]} \BibitemShut {NoStop}%
\bibitem [{\citenamefont {Tavlayan}\ and\ \citenamefont {Tekin}(2020)}]{Tavlayan:2020chf}%
  \BibitemOpen
  \bibfield  {author} {\bibinfo {author} {\bibfnamefont {A.}~\bibnamefont {Tavlayan}}\ and\ \bibinfo {author} {\bibfnamefont {B.}~\bibnamefont {Tekin}},\ }\href {\doibase 10.1103/PhysRevD.101.084034} {\bibfield  {journal} {\bibinfo  {journal} {Phys. Rev. D}\ }\textbf {\bibinfo {volume} {101}},\ \bibinfo {pages} {084034} (\bibinfo {year} {2020})},\ \Eprint {http://arxiv.org/abs/2002.01135} {arXiv:2002.01135 [hep-th]} \BibitemShut {NoStop}%
\bibitem [{\citenamefont {Pantig}\ and\ \citenamefont {\"Ovg\"un}(2022)}]{Pantig:2022sjb}%
  \BibitemOpen
  \bibfield  {author} {\bibinfo {author} {\bibfnamefont {R.~C.}\ \bibnamefont {Pantig}}\ and\ \bibinfo {author} {\bibfnamefont {A.}~\bibnamefont {\"Ovg\"un}},\ }\href {\doibase 10.1002/prop.202200164} {\bibfield  {journal} {\bibinfo  {journal} {Fortsch. Phys.}\ }\textbf {\bibinfo {volume} {2022}},\ \bibinfo {pages} {2200164} (\bibinfo {year} {2022})},\ \Eprint {http://arxiv.org/abs/2210.00523} {arXiv:2210.00523 [gr-qc]} \BibitemShut {NoStop}%
\bibitem [{\citenamefont {Akhmedov}\ \emph {et~al.}(2006)\citenamefont {Akhmedov}, \citenamefont {Akhmedova},\ and\ \citenamefont {Singleton}}]{Akhmedov:2006pg}%
  \BibitemOpen
  \bibfield  {author} {\bibinfo {author} {\bibfnamefont {E.~T.}\ \bibnamefont {Akhmedov}}, \bibinfo {author} {\bibfnamefont {V.}~\bibnamefont {Akhmedova}}, \ and\ \bibinfo {author} {\bibfnamefont {D.}~\bibnamefont {Singleton}},\ }\href {\doibase 10.1016/j.physletb.2006.09.028} {\bibfield  {journal} {\bibinfo  {journal} {Phys. Lett. B}\ }\textbf {\bibinfo {volume} {642}},\ \bibinfo {pages} {124} (\bibinfo {year} {2006})},\ \Eprint {http://arxiv.org/abs/hep-th/0608098} {arXiv:hep-th/0608098} \BibitemShut {NoStop}%
\bibitem [{\citenamefont {Parikh}\ and\ \citenamefont {Wilczek}(2000)}]{Parikh:1999mf}%
  \BibitemOpen
  \bibfield  {author} {\bibinfo {author} {\bibfnamefont {M.~K.}\ \bibnamefont {Parikh}}\ and\ \bibinfo {author} {\bibfnamefont {F.}~\bibnamefont {Wilczek}},\ }\href {\doibase 10.1103/PhysRevLett.85.5042} {\bibfield  {journal} {\bibinfo  {journal} {Phys. Rev. Lett.}\ }\textbf {\bibinfo {volume} {85}},\ \bibinfo {pages} {5042} (\bibinfo {year} {2000})},\ \Eprint {http://arxiv.org/abs/hep-th/9907001} {arXiv:hep-th/9907001} \BibitemShut {NoStop}%
\bibitem [{\citenamefont {Shankaranarayanan}\ \emph {et~al.}(2002)\citenamefont {Shankaranarayanan}, \citenamefont {Padmanabhan},\ and\ \citenamefont {Srinivasan}}]{Shankaranarayanan:2000qv}%
  \BibitemOpen
  \bibfield  {author} {\bibinfo {author} {\bibfnamefont {S.}~\bibnamefont {Shankaranarayanan}}, \bibinfo {author} {\bibfnamefont {T.}~\bibnamefont {Padmanabhan}}, \ and\ \bibinfo {author} {\bibfnamefont {K.}~\bibnamefont {Srinivasan}},\ }\href {\doibase 10.1088/0264-9381/19/10/310} {\bibfield  {journal} {\bibinfo  {journal} {Class. Quant. Grav.}\ }\textbf {\bibinfo {volume} {19}},\ \bibinfo {pages} {2671} (\bibinfo {year} {2002})},\ \Eprint {http://arxiv.org/abs/gr-qc/0010042} {arXiv:gr-qc/0010042} \BibitemShut {NoStop}%
\bibitem [{\citenamefont {Angheben}\ \emph {et~al.}(2005)\citenamefont {Angheben}, \citenamefont {Nadalini}, \citenamefont {Vanzo},\ and\ \citenamefont {Zerbini}}]{Angheben:2005rm}%
  \BibitemOpen
  \bibfield  {author} {\bibinfo {author} {\bibfnamefont {M.}~\bibnamefont {Angheben}}, \bibinfo {author} {\bibfnamefont {M.}~\bibnamefont {Nadalini}}, \bibinfo {author} {\bibfnamefont {L.}~\bibnamefont {Vanzo}}, \ and\ \bibinfo {author} {\bibfnamefont {S.}~\bibnamefont {Zerbini}},\ }\href {\doibase 10.1088/1126-6708/2005/05/014} {\bibfield  {journal} {\bibinfo  {journal} {JHEP}\ }\textbf {\bibinfo {volume} {05}},\ \bibinfo {pages} {014} (\bibinfo {year} {2005})},\ \Eprint {http://arxiv.org/abs/hep-th/0503081} {arXiv:hep-th/0503081} \BibitemShut {NoStop}%
\bibitem [{\citenamefont {Bera}\ \emph {et~al.}(2020)\citenamefont {Bera}, \citenamefont {Ghosh},\ and\ \citenamefont {Majhi}}]{Bera:2019oxg}%
  \BibitemOpen
  \bibfield  {author} {\bibinfo {author} {\bibfnamefont {A.}~\bibnamefont {Bera}}, \bibinfo {author} {\bibfnamefont {S.}~\bibnamefont {Ghosh}}, \ and\ \bibinfo {author} {\bibfnamefont {B.~R.}\ \bibnamefont {Majhi}},\ }\href {\doibase 10.1140/epjp/s13360-020-00693-1} {\bibfield  {journal} {\bibinfo  {journal} {Eur. Phys. J. Plus}\ }\textbf {\bibinfo {volume} {135}},\ \bibinfo {pages} {670} (\bibinfo {year} {2020})},\ \Eprint {http://arxiv.org/abs/1909.12607} {arXiv:1909.12607 [gr-qc]} \BibitemShut {NoStop}%
\bibitem [{\citenamefont {Kerner}\ and\ \citenamefont {Mann}(2008{\natexlab{a}})}]{Kerner:2007rr}%
  \BibitemOpen
  \bibfield  {author} {\bibinfo {author} {\bibfnamefont {R.}~\bibnamefont {Kerner}}\ and\ \bibinfo {author} {\bibfnamefont {R.~B.}\ \bibnamefont {Mann}},\ }\href {\doibase 10.1088/0264-9381/25/9/095014} {\bibfield  {journal} {\bibinfo  {journal} {Class. Quant. Grav.}\ }\textbf {\bibinfo {volume} {25}},\ \bibinfo {pages} {095014} (\bibinfo {year} {2008}{\natexlab{a}})},\ \Eprint {http://arxiv.org/abs/0710.0612} {arXiv:0710.0612 [hep-th]} \BibitemShut {NoStop}%
\bibitem [{\citenamefont {Kerner}\ and\ \citenamefont {Mann}(2006)}]{Kerner:2006vu}%
  \BibitemOpen
  \bibfield  {author} {\bibinfo {author} {\bibfnamefont {R.}~\bibnamefont {Kerner}}\ and\ \bibinfo {author} {\bibfnamefont {R.~B.}\ \bibnamefont {Mann}},\ }\href {\doibase 10.1103/PhysRevD.73.104010} {\bibfield  {journal} {\bibinfo  {journal} {Phys. Rev. D}\ }\textbf {\bibinfo {volume} {73}},\ \bibinfo {pages} {104010} (\bibinfo {year} {2006})},\ \Eprint {http://arxiv.org/abs/gr-qc/0603019} {arXiv:gr-qc/0603019} \BibitemShut {NoStop}%
\bibitem [{\citenamefont {Kerner}\ and\ \citenamefont {Mann}(2008{\natexlab{b}})}]{Kerner:2008qv}%
  \BibitemOpen
  \bibfield  {author} {\bibinfo {author} {\bibfnamefont {R.}~\bibnamefont {Kerner}}\ and\ \bibinfo {author} {\bibfnamefont {R.~B.}\ \bibnamefont {Mann}},\ }\href {\doibase 10.1016/j.physletb.2008.06.012} {\bibfield  {journal} {\bibinfo  {journal} {Phys. Lett. B}\ }\textbf {\bibinfo {volume} {665}},\ \bibinfo {pages} {277} (\bibinfo {year} {2008}{\natexlab{b}})},\ \Eprint {http://arxiv.org/abs/0803.2246} {arXiv:0803.2246 [hep-th]} \BibitemShut {NoStop}%
\bibitem [{\citenamefont {Yale}\ and\ \citenamefont {Mann}(2009)}]{Yale:2008kx}%
  \BibitemOpen
  \bibfield  {author} {\bibinfo {author} {\bibfnamefont {A.}~\bibnamefont {Yale}}\ and\ \bibinfo {author} {\bibfnamefont {R.~B.}\ \bibnamefont {Mann}},\ }\href {\doibase 10.1016/j.physletb.2009.02.019} {\bibfield  {journal} {\bibinfo  {journal} {Phys. Lett. B}\ }\textbf {\bibinfo {volume} {673}},\ \bibinfo {pages} {168} (\bibinfo {year} {2009})},\ \Eprint {http://arxiv.org/abs/0808.2820} {arXiv:0808.2820 [gr-qc]} \BibitemShut {NoStop}%
\bibitem [{\citenamefont {Gibbons}(2016)}]{Gibbons:2015qja}%
  \BibitemOpen
  \bibfield  {author} {\bibinfo {author} {\bibfnamefont {G.~W.}\ \bibnamefont {Gibbons}},\ }\href {\doibase 10.1088/0264-9381/33/2/025004} {\bibfield  {journal} {\bibinfo  {journal} {Class. Quant. Grav.}\ }\textbf {\bibinfo {volume} {33}},\ \bibinfo {pages} {025004} (\bibinfo {year} {2016})},\ \Eprint {http://arxiv.org/abs/1508.06755} {arXiv:1508.06755 [gr-qc]} \BibitemShut {NoStop}%
\bibitem [{\citenamefont {Chanda}\ \emph {et~al.}(2017)\citenamefont {Chanda}, \citenamefont {Gibbons},\ and\ \citenamefont {Guha}}]{Chanda:2016aph}%
  \BibitemOpen
  \bibfield  {author} {\bibinfo {author} {\bibfnamefont {S.}~\bibnamefont {Chanda}}, \bibinfo {author} {\bibfnamefont {G.~W.}\ \bibnamefont {Gibbons}}, \ and\ \bibinfo {author} {\bibfnamefont {P.}~\bibnamefont {Guha}},\ }\href {\doibase 10.1063/1.4978333} {\bibfield  {journal} {\bibinfo  {journal} {J. Math. Phys.}\ }\textbf {\bibinfo {volume} {58}},\ \bibinfo {pages} {032503} (\bibinfo {year} {2017})},\ \Eprint {http://arxiv.org/abs/1612.00375} {arXiv:1612.00375 [math-ph]} \BibitemShut {NoStop}%
\bibitem [{\citenamefont {Das}\ \emph {et~al.}(2017)\citenamefont {Das}, \citenamefont {Sk},\ and\ \citenamefont {Ghosh}}]{Das:2016opi}%
  \BibitemOpen
  \bibfield  {author} {\bibinfo {author} {\bibfnamefont {P.}~\bibnamefont {Das}}, \bibinfo {author} {\bibfnamefont {R.}~\bibnamefont {Sk}}, \ and\ \bibinfo {author} {\bibfnamefont {S.}~\bibnamefont {Ghosh}},\ }\href {\doibase 10.1140/epjc/s10052-017-5295-6} {\bibfield  {journal} {\bibinfo  {journal} {Eur. Phys. J. C}\ }\textbf {\bibinfo {volume} {77}},\ \bibinfo {pages} {735} (\bibinfo {year} {2017})},\ \Eprint {http://arxiv.org/abs/1609.04577} {arXiv:1609.04577 [gr-qc]} \BibitemShut {NoStop}%
\bibitem [{\citenamefont {Srinivasan}\ and\ \citenamefont {Padmanabhan}(1999)}]{Srinivasan:1998ty}%
  \BibitemOpen
  \bibfield  {author} {\bibinfo {author} {\bibfnamefont {K.}~\bibnamefont {Srinivasan}}\ and\ \bibinfo {author} {\bibfnamefont {T.}~\bibnamefont {Padmanabhan}},\ }\href {\doibase 10.1103/PhysRevD.60.024007} {\bibfield  {journal} {\bibinfo  {journal} {Phys. Rev. D}\ }\textbf {\bibinfo {volume} {60}},\ \bibinfo {pages} {024007} (\bibinfo {year} {1999})},\ \Eprint {http://arxiv.org/abs/gr-qc/9812028} {arXiv:gr-qc/9812028} \BibitemShut {NoStop}%
\bibitem [{\citenamefont {Robinson}\ and\ \citenamefont {Wilczek}(2005)}]{Robinson:2005pd}%
  \BibitemOpen
  \bibfield  {author} {\bibinfo {author} {\bibfnamefont {S.~P.}\ \bibnamefont {Robinson}}\ and\ \bibinfo {author} {\bibfnamefont {F.}~\bibnamefont {Wilczek}},\ }\href {\doibase 10.1103/PhysRevLett.95.011303} {\bibfield  {journal} {\bibinfo  {journal} {Phys. Rev. Lett.}\ }\textbf {\bibinfo {volume} {95}},\ \bibinfo {pages} {011303} (\bibinfo {year} {2005})},\ \Eprint {http://arxiv.org/abs/gr-qc/0502074} {arXiv:gr-qc/0502074} \BibitemShut {NoStop}%
\bibitem [{\citenamefont {Iso}\ \emph {et~al.}(2006)\citenamefont {Iso}, \citenamefont {Umetsu},\ and\ \citenamefont {Wilczek}}]{Iso:2006wa}%
  \BibitemOpen
  \bibfield  {author} {\bibinfo {author} {\bibfnamefont {S.}~\bibnamefont {Iso}}, \bibinfo {author} {\bibfnamefont {H.}~\bibnamefont {Umetsu}}, \ and\ \bibinfo {author} {\bibfnamefont {F.}~\bibnamefont {Wilczek}},\ }\href {\doibase 10.1103/PhysRevLett.96.151302} {\bibfield  {journal} {\bibinfo  {journal} {Phys. Rev. Lett.}\ }\textbf {\bibinfo {volume} {96}},\ \bibinfo {pages} {151302} (\bibinfo {year} {2006})},\ \Eprint {http://arxiv.org/abs/hep-th/0602146} {arXiv:hep-th/0602146} \BibitemShut {NoStop}%
\bibitem [{\citenamefont {Majhi}(2010)}]{Majhi:2010onr}%
  \BibitemOpen
  \bibfield  {author} {\bibinfo {author} {\bibfnamefont {B.~R.}\ \bibnamefont {Majhi}},\ }\emph {\bibinfo {title} {{Quantum Tunneling in Black Holes}}},\ \href@noop {} {Ph.D. thesis},\ \bibinfo  {school} {Calcutta U.} (\bibinfo {year} {2010}),\ \Eprint {http://arxiv.org/abs/1110.6008} {arXiv:1110.6008 [gr-qc]} \BibitemShut {NoStop}%
\bibitem [{\citenamefont {Gibbons}\ and\ \citenamefont {Werner}(2008)}]{Gibbons:2008rj}%
  \BibitemOpen
  \bibfield  {author} {\bibinfo {author} {\bibfnamefont {G.~W.}\ \bibnamefont {Gibbons}}\ and\ \bibinfo {author} {\bibfnamefont {M.~C.}\ \bibnamefont {Werner}},\ }\href {\doibase 10.1088/0264-9381/25/23/235009} {\bibfield  {journal} {\bibinfo  {journal} {Class. Quant. Grav.}\ }\textbf {\bibinfo {volume} {25}},\ \bibinfo {pages} {235009} (\bibinfo {year} {2008})},\ \Eprint {http://arxiv.org/abs/0807.0854} {arXiv:0807.0854 [gr-qc]} \BibitemShut {NoStop}%
\bibitem [{\citenamefont {Lambert}\ and\ \citenamefont {Le~Poncin-Lafitte}(2011)}]{Lambert:2011fk}%
  \BibitemOpen
  \bibfield  {author} {\bibinfo {author} {\bibfnamefont {S.~B.}\ \bibnamefont {Lambert}}\ and\ \bibinfo {author} {\bibfnamefont {C.}~\bibnamefont {Le~Poncin-Lafitte}},\ }\href {\doibase 10.1051/0004-6361/201016370} {\bibfield  {journal} {\bibinfo  {journal} {A\&A}\ }\textbf {\bibinfo {volume} {529}},\ \bibinfo {pages} {A70} (\bibinfo {year} {2011})}\BibitemShut {NoStop}%
\bibitem [{\citenamefont {Dolan}(2023)}]{dolan2023einstein}%
  \BibitemOpen
  \bibfield  {author} {\bibinfo {author} {\bibfnamefont {B.~P.}\ \bibnamefont {Dolan}},\ }\href@noop {} {\emph {\bibinfo {title} {Einstein's General Theory of Relativity: A Concise Introduction}}}\ (\bibinfo  {publisher} {Cambridge University Press},\ \bibinfo {year} {2023})\BibitemShut {NoStop}%
\bibitem [{\citenamefont {Synge}(1966)}]{Synge:1966}%
  \BibitemOpen
  \bibfield  {author} {\bibinfo {author} {\bibfnamefont {J.~L.}\ \bibnamefont {Synge}},\ }\href@noop {} {\bibfield  {journal} {\bibinfo  {journal} {Monthly Notices of the Royal Astronomical Society}\ }\textbf {\bibinfo {volume} {131}},\ \bibinfo {pages} {463} (\bibinfo {year} {1966})}\BibitemShut {NoStop}%
\bibitem [{\citenamefont {Cunningham}\ and\ \citenamefont {Bardeen}(1972)}]{Cunningham:1972}%
  \BibitemOpen
  \bibfield  {author} {\bibinfo {author} {\bibfnamefont {C.~T.}\ \bibnamefont {Cunningham}}\ and\ \bibinfo {author} {\bibfnamefont {J.~M.}\ \bibnamefont {Bardeen}},\ }\href@noop {} {\bibfield  {journal} {\bibinfo  {journal} {Astrophysical Journal Letters}\ }\textbf {\bibinfo {volume} {173}},\ \bibinfo {pages} {L137} (\bibinfo {year} {1972})}\BibitemShut {NoStop}%
\bibitem [{\citenamefont {{Bardeen}}\ \emph {et~al.}(1972)\citenamefont {{Bardeen}}, \citenamefont {{Press}},\ and\ \citenamefont {{Teukolsky}}}]{Bardeen:1973a}%
  \BibitemOpen
  \bibfield  {author} {\bibinfo {author} {\bibfnamefont {J.~M.}\ \bibnamefont {{Bardeen}}}, \bibinfo {author} {\bibfnamefont {W.~H.}\ \bibnamefont {{Press}}}, \ and\ \bibinfo {author} {\bibfnamefont {S.~A.}\ \bibnamefont {{Teukolsky}}},\ }\href {\doibase 10.1086/151796} {\bibfield  {journal} {\bibinfo  {journal} {Astrophys. J}\ }\textbf {\bibinfo {volume} {178}},\ \bibinfo {pages} {347} (\bibinfo {year} {1972})}\BibitemShut {NoStop}%
\bibitem [{\citenamefont {Luminet}(1979)}]{Luminet:1979nyg}%
  \BibitemOpen
  \bibfield  {author} {\bibinfo {author} {\bibfnamefont {J.~P.}\ \bibnamefont {Luminet}},\ }\href {https://ui.adsabs.harvard.edu/abs/1979A&A....75..228L} {\bibfield  {journal} {\bibinfo  {journal} {Astron. Astrophys.}\ }\textbf {\bibinfo {volume} {75}},\ \bibinfo {pages} {228} (\bibinfo {year} {1979})}\BibitemShut {NoStop}%
\bibitem [{\citenamefont {Zeldovich}\ and\ \citenamefont {Novikov}(1971)}]{zeldovich_relativistic_1966}%
  \BibitemOpen
  \bibfield  {author} {\bibinfo {author} {\bibfnamefont {Y.~B.}\ \bibnamefont {Zeldovich}}\ and\ \bibinfo {author} {\bibfnamefont {I.~D.}\ \bibnamefont {Novikov}},\ }\href@noop {} {\emph {\bibinfo {title} {Relativistic Astrophysics, Volume 1: Stars and Relativity}}}\ (\bibinfo  {publisher} {University of Chicago Press},\ \bibinfo {year} {1971})\ \bibinfo {note} {original Russian Edition: 1966}\BibitemShut {NoStop}%
\bibitem [{\citenamefont {Bardeen}(1973)}]{Bardeen:1973b}%
  \BibitemOpen
  \bibfield  {author} {\bibinfo {author} {\bibfnamefont {J.~M.}\ \bibnamefont {Bardeen}},\ }in\ \href@noop {} {\emph {\bibinfo {booktitle} {Proceedings of the Les Houches Summer School}}},\ \bibinfo {editor} {edited by\ \bibinfo {editor} {\bibfnamefont {C.}~\bibnamefont {DeWitt}}\ and\ \bibinfo {editor} {\bibfnamefont {B.~S.}\ \bibnamefont {DeWitt}}}\ (\bibinfo  {publisher} {Gordon and Breach},\ \bibinfo {year} {1973})\ pp.\ \bibinfo {pages} {215--239}\BibitemShut {NoStop}%
\bibitem [{\citenamefont {Chandrasekhar}(1998)}]{Chandrasekhar:1998}%
  \BibitemOpen
  \bibfield  {author} {\bibinfo {author} {\bibfnamefont {S.}~\bibnamefont {Chandrasekhar}},\ }\href {https://cds.cern.ch/record/579245} {\emph {\bibinfo {title} {The mathematical theory of black holes}}},\ Oxford classic texts in the physical sciences\ (\bibinfo  {publisher} {Oxford University Press},\ \bibinfo {year} {1998})\BibitemShut {NoStop}%
\bibitem [{\citenamefont {Luminet}(2018)}]{luminet_seeing_2018}%
  \BibitemOpen
  \bibfield  {author} {\bibinfo {author} {\bibfnamefont {J.-P.}\ \bibnamefont {Luminet}},\ }\href@noop {} {\bibfield  {journal} {\bibinfo  {journal} {The European Physical Journal H}\ }\textbf {\bibinfo {volume} {43}},\ \bibinfo {pages} {293} (\bibinfo {year} {2018})}\BibitemShut {NoStop}%
\bibitem [{\citenamefont {Falcke}\ \emph {et~al.}(2000)\citenamefont {Falcke}, \citenamefont {Melia},\ and\ \citenamefont {Agol}}]{Falcke_2000}%
  \BibitemOpen
  \bibfield  {author} {\bibinfo {author} {\bibfnamefont {H.}~\bibnamefont {Falcke}}, \bibinfo {author} {\bibfnamefont {F.}~\bibnamefont {Melia}}, \ and\ \bibinfo {author} {\bibfnamefont {E.}~\bibnamefont {Agol}},\ }\href@noop {} {\bibfield  {journal} {\bibinfo  {journal} {Astrophysical Journal Letters}\ }\textbf {\bibinfo {volume} {528}},\ \bibinfo {pages} {L13} (\bibinfo {year} {2000})}\BibitemShut {NoStop}%
\bibitem [{\citenamefont {Johnson}\ \emph {et~al.}(2020)\citenamefont {Johnson} \emph {et~al.}}]{johnson_universal_2020}%
  \BibitemOpen
  \bibfield  {author} {\bibinfo {author} {\bibfnamefont {M.~D.}\ \bibnamefont {Johnson}} \emph {et~al.},\ }\href {\doibase 10.1126/sciadv.aaz1310} {\bibfield  {journal} {\bibinfo  {journal} {Sci. Adv.}\ }\textbf {\bibinfo {volume} {6}},\ \bibinfo {pages} {eaaz1310} (\bibinfo {year} {2020})},\ \Eprint {http://arxiv.org/abs/1907.04329} {arXiv:1907.04329 [astro-ph.IM]} \BibitemShut {NoStop}%
\bibitem [{\citenamefont {Tsukamoto}(2018{\natexlab{b}})}]{Tsukamoto:2017fxq}%
  \BibitemOpen
  \bibfield  {author} {\bibinfo {author} {\bibfnamefont {N.}~\bibnamefont {Tsukamoto}},\ }\href {\doibase 10.1103/PhysRevD.97.064021} {\bibfield  {journal} {\bibinfo  {journal} {Phys. Rev. D}\ }\textbf {\bibinfo {volume} {97}},\ \bibinfo {pages} {064021} (\bibinfo {year} {2018}{\natexlab{b}})},\ \Eprint {http://arxiv.org/abs/1708.07427} {arXiv:1708.07427 [gr-qc]} \BibitemShut {NoStop}%
\bibitem [{\citenamefont {Ali}\ \emph {et~al.}(2025)\citenamefont {Ali}, \citenamefont {Islam}, \citenamefont {Ghosh},\ and\ \citenamefont {Ramasamya}}]{ali_testing_2025}%
  \BibitemOpen
  \bibfield  {author} {\bibinfo {author} {\bibfnamefont {A.}~\bibnamefont {Ali}}, \bibinfo {author} {\bibfnamefont {S.~U.}\ \bibnamefont {Islam}}, \bibinfo {author} {\bibfnamefont {S.~G.}\ \bibnamefont {Ghosh}}, \ and\ \bibinfo {author} {\bibfnamefont {A.}~\bibnamefont {Ramasamya}},\ }\href {\doibase 10.1016/j.dark.2024.101768} {\bibfield  {journal} {\bibinfo  {journal} {Physics of the Dark Universe}\ }\textbf {\bibinfo {volume} {47}},\ \bibinfo {pages} {101768} (\bibinfo {year} {2025})}\BibitemShut {NoStop}%
\bibitem [{\citenamefont {Bozza}\ \emph {et~al.}(2001)\citenamefont {Bozza}, \citenamefont {Capozziello}, \citenamefont {Iovane},\ and\ \citenamefont {Scarpetta}}]{Bozza:2001xd}%
  \BibitemOpen
  \bibfield  {author} {\bibinfo {author} {\bibfnamefont {V.}~\bibnamefont {Bozza}}, \bibinfo {author} {\bibfnamefont {S.}~\bibnamefont {Capozziello}}, \bibinfo {author} {\bibfnamefont {G.}~\bibnamefont {Iovane}}, \ and\ \bibinfo {author} {\bibfnamefont {G.}~\bibnamefont {Scarpetta}},\ }\href {\doibase 10.1023/A:1012292927358} {\bibfield  {journal} {\bibinfo  {journal} {Gen. Rel. Grav.}\ }\textbf {\bibinfo {volume} {33}},\ \bibinfo {pages} {1535} (\bibinfo {year} {2001})},\ \Eprint {http://arxiv.org/abs/gr-qc/0102068} {arXiv:gr-qc/0102068} \BibitemShut {NoStop}%
\bibitem [{\citenamefont {Einstein}(1936)}]{einstein_lens-like_1936}%
  \BibitemOpen
  \bibfield  {author} {\bibinfo {author} {\bibfnamefont {A.}~\bibnamefont {Einstein}},\ }\href {\doibase 10.1126/science.84.2188.506} {\bibfield  {journal} {\bibinfo  {journal} {Science}\ }\textbf {\bibinfo {volume} {84}},\ \bibinfo {pages} {506} (\bibinfo {year} {1936})}\BibitemShut {NoStop}%
\bibitem [{\citenamefont {Mellier}(1999)}]{mellier_probing_1999}%
  \BibitemOpen
  \bibfield  {author} {\bibinfo {author} {\bibfnamefont {Y.}~\bibnamefont {Mellier}},\ }\href {\doibase 10.1146/annurev.astro.37.1.127} {\bibfield  {journal} {\bibinfo  {journal} {Annual Review of Astronomy and Astrophysics}\ }\textbf {\bibinfo {volume} {37}},\ \bibinfo {pages} {127} (\bibinfo {year} {1999})}\BibitemShut {NoStop}%
\bibitem [{\citenamefont {Bartelmann}\ and\ \citenamefont {Schneider}(2001)}]{bartelmann_weak_2001}%
  \BibitemOpen
  \bibfield  {author} {\bibinfo {author} {\bibfnamefont {M.}~\bibnamefont {Bartelmann}}\ and\ \bibinfo {author} {\bibfnamefont {P.}~\bibnamefont {Schneider}},\ }\href {\doibase 10.1016/S0370-1573(00)00082-X} {\bibfield  {journal} {\bibinfo  {journal} {Physics Reports}\ }\textbf {\bibinfo {volume} {340}},\ \bibinfo {pages} {291} (\bibinfo {year} {2001})}\BibitemShut {NoStop}%
\bibitem [{\citenamefont {Schmidt}(2008)}]{schmidt_weak_2008}%
  \BibitemOpen
  \bibfield  {author} {\bibinfo {author} {\bibfnamefont {F.}~\bibnamefont {Schmidt}},\ }\href {\doibase 10.1103/PhysRevD.78.043002} {\bibfield  {journal} {\bibinfo  {journal} {Physical Review D}\ }\textbf {\bibinfo {volume} {78}},\ \bibinfo {pages} {043002} (\bibinfo {year} {2008})}\BibitemShut {NoStop}%
\bibitem [{\citenamefont {Guzik}\ \emph {et~al.}(2010)\citenamefont {Guzik}, \citenamefont {Jain},\ and\ \citenamefont {Takada}}]{guzik_tests_2010}%
  \BibitemOpen
  \bibfield  {author} {\bibinfo {author} {\bibfnamefont {J.}~\bibnamefont {Guzik}}, \bibinfo {author} {\bibfnamefont {B.}~\bibnamefont {Jain}}, \ and\ \bibinfo {author} {\bibfnamefont {M.}~\bibnamefont {Takada}},\ }\href {\doibase 10.1103/PhysRevD.81.023503} {\bibfield  {journal} {\bibinfo  {journal} {Physical Review D}\ }\textbf {\bibinfo {volume} {81}},\ \bibinfo {pages} {023503} (\bibinfo {year} {2010})}\BibitemShut {NoStop}%
\bibitem [{\citenamefont {Bozza}\ and\ \citenamefont {Mancini}(2004)}]{bozza_time_2004}%
  \BibitemOpen
  \bibfield  {author} {\bibinfo {author} {\bibfnamefont {V.}~\bibnamefont {Bozza}}\ and\ \bibinfo {author} {\bibfnamefont {L.}~\bibnamefont {Mancini}},\ }\href {\doibase 10.1023/B:GERG.0000010486.58026.4f} {\bibfield  {journal} {\bibinfo  {journal} {General Relativity and Gravitation}\ }\textbf {\bibinfo {volume} {36}},\ \bibinfo {pages} {435} (\bibinfo {year} {2004})}\BibitemShut {NoStop}%
\end{thebibliography}%
\end{document}